\documentclass[a4paper,12pt,oneside]{book}
\usepackage[nottoc]{tocbibind}
\usepackage[top=1.5 in,bottom=1.3 in,left=1.5 in,right=0.8 in]{geometry}
\usepackage{latexsym} 
\usepackage{fancyhdr}
\def\be{\begin{equation}}
\def\ee{\end{equation}}
\def\pa{\partial}
\def\eps{\epsilon}
\def\th{\theta}
\newcommand{\beq}{\begin{equation}}
\newcommand{\eeq}{\end{equation}}
\newcommand{\pr}{\partial}
\newcommand{\lam}{\lambda}
\newcommand{\sg}{\sigma}
\newcommand{\mn}{\mu\nu}
\newcommand{\mnl}{\mu\nu\lambda}
\newcommand{\til}{\tilde}
\newcommand{\al}{\alpha}

\begin{document}
\frontmatter
\begin{titlepage}
\begin{center}
\begin{huge}
\textbf{Study of Planar Models In \\Quantum Mechanics, Field Theory \\\vspace {0.75cm} And Gravity}
\end{huge}
\end{center}
\vskip 4 cm
\begin{center} 
\begin{Large}\bf{Thesis submitted for the degree of\\
Doctor  of Philosophy(Science)\\
in\\
Physics(Theoretical)\\
by\\
SARMISHTHA  KUMAR
\vskip 3cm
Department of Theoretical Sciences\\
%\hspace{5cm}
\,\,UNIVERSITY OF CALCUTTA\\
\,\,\,\,\,2013}  
\end{Large}                                       
                                                                                 
\end{center}
\end{titlepage}
%%%%%%%%%%%%%%%%%%%%%%%%5
%\include{dedi}
\pagestyle{empty}
\begin{Huge} \noindent To \\my beloved Parents\end{Huge}
% % % % % % % % %
%\include{new-acknow}
\newpage
\textbf{\begin{Huge}Acknowledgement :\end{Huge}}
\\\\
It is the time to convey my utmost gratitude to those without whom the dream of being successful
in this journey could not be possible. So this is my golden opportunity to thank you all.

To name first, I am deeply indebted   to  \textbf{``\textit{Satayandranath Bose National Centre For Basic Sciences }"} (Salt lake, Sector-III. Block-JD, Kol-98) where  I have carried my entire Ph-D work, and to the Director Prof. Arup Roychowdhury for his kind approval to let me get an entry in this respected institute as an external part-time Ph-D candidate providing me all the necessary facilities to pursue my thesis work.\\
I am very much grateful to the academic authority of the Centre for setting a suitable time table required to continue the ``Ph-D course work programme"  for working candidates like me. On the other side I am indebted to the members of the administrative office, the Library officials for their keen support and their continual help.

It is my privilege having such a tough and hard taskmaster as my thesis supervisor Prof. Rabin Banerjee. His cheerful and energetic guidance, constant endeavour to the problem has led me to finalise my work at this stage. I consider myself to be fortunate enough to come in contact with such a scientist who, with his encyclopaedic knowledge has introduced me to diverse fields of science during the tenure of my work. His valuable suggestions, criticism and immense encouragement has enriched my journey to reach my goal. I owe to him for the rest part of my life.

I would like to express my sincere thanks to all my fellow mates in this institute for lending me their helping hands whenever needed. Thank you Friends !

I pay my gratitude to my work place \textit{``Camellia Institute Of Technology"}, (Madhyamgram, Kolkata-129) and my heartfelt thanks goes to the Director, Prof. Subhasis Sengupta for his kindness to pursue this Ph-D programme. My heartiest thanks to all my well-wishers and specially my colleagues of Physics Dept. of this college for their constant support and help. 

This is high time to pay tribute to my teachers I have met since my school days. Down the memory lane I can still recall the vivid moments of my school ``\textit{Bally Bangasishu(Balika) vidyalaya}", our favourite teachers(\textit{Mina di,Swapna di,Sumita di, Rina di, Rekha di, Subha di, Sobhona di , Shikha di} and so on..!) and very much their genuine loving personality which deserve  to be adored. They were my true inspiration to choose teaching as my profession.

My warm regards to the Professors ( Late Prof. Amal Kr. Som, Prof. Shukla Mukherjee, Prof. Keka Basu Chaudhuri and so many ..) of my Undergraduate College ``\textit{Brahmananda Keshabchandra College, Bonhoogly}" from where I had earned my Bachelors degree in Physics. I pay respect to my private tutor Prof. Basudeb Ghosh (the then professor in the dept.of Physics, \textit{Vidyamandir, Belurmath}) who specially instilled the aspirations for achieving higher education in this field.

It was the \textit{Dept. of Physics of Calcutta University}(Rajabazar science college campus)where the spell really came from the awesome teaching of the eminent Professors. Their relentless effort to ease diverse topics to us was truly amazing and unparalleled. And I did aim at Ph-D degree then.

Last but not the least is my \textit{\textbf{Family}}. My homage to our close family friends Sri Debashish Majumdar and his wife Smt. Mousumi Majumdar for sharing their views , valuable suggestions and showing me the track to opt correct decision at that primary stage. I feel honoured to materialise the wish of my parents
Sri Jatindranath Kumar and my mother Smt. Nilima Kumar. I owe very much to my elder sister and brothers for their steady support to finish this mammoth task. I am privileged with the heartful blessings from my in-laws.

At the end of this golden opportunity I am left with none but my son Anubhab and my husband Indranath.
It is only their tremendous mental support, passion and love that boosted my work to the ultimate destination. So not to utter any word for them, and they know why.
\\
Lastly \textit{Thank to the Almighty}
%%%%%%%%%%%%%%%%%%%%%%%
%\include{publist}
\newpage
\textbf{\begin{LARGE}
\underline{List of Publications}:
\end{LARGE}}
\vskip 1.5 cm
\begin{flushleft}
1. \begin{large}R. Banerjee and S. Kumar\end{large}, \\
\textbf{Phys. Rev. D60}(1999) 085005\\
\textit{Title: Self-duality and soldering in odd dimensions}
\vskip 0.5 cm
2. \begin{large} Rabin Banerjee and Sarmishtha Kumar\end{large}, \\
\textbf{Phys. Rev. D63} (2001) 125008\\
\textit{Title: Self-dual models and mass generation in planar field theory}
 \\e-print  ar{X}ive no: hep-th/0007148
\vskip 0.5 cm
3.  \begin{large}Sarmishtha Kumar\end{large}, \\
\textbf{ IJMPA}. vol.18 , No. 9(2003)1613-1622\\
\textit{Title: Lagrangian and Hamiltonian formulations of higher order Chern-Simons theories}\\
e-print ar{X}ive no: hep-th/0112121
\vskip 0.5 cm
4.  \begin{large} Sarmistha Kumar(Chaudhuri) and Saurav Samanta\end{large}, \\
{\bf IJMPA},vol.25,No.16(2010)3221-3233\\
\textit{Title: Study on the noncommutative representations of Galilean generators}
 \\e-print ar{X}ive no: hep-th/0909.2527
\vskip 0.5 cm
5. \begin{large} Rabin Banerjee and Sarmishtha Kumar(Chaudhuri)\end{large},\\
\textbf{Phys. Rev. D85},(2012) 125002\\
\textit{Title: Dual composition of odd dimensional models}
\\e-print ar{X}ive no: hep-th/1203.6229
\vskip 0.5 cm
My thesis is entirely based on these papers [1--5].
\end{flushleft}
%%%%%%%%%%%%%%%%%%%%%%%%%
\tableofcontents
%%%%%%%%%%%%%%%
\mainmatter
%%%%%%%%%%%%%%%%%%%%%% Main text starts   %%%%%%%%%%%%%%
\chapter{Introduction}
\label{intro}
Self duality is a powerful notion in classical mechanics, classical field theory, quantum mechanics as well in quantum field theory. In these theories, the interactions have particular forms and special strength so that 
the second order equations of motion reduce to first order equations which are simpler to analyse. The minimization of energy or action leads to the ``self dual point" at which the interactions and coupling strengths take their special self dual effects. This signifies the self dual theories physically. For example the self dual Yang -Mills  equations have minimum action solutions known as instantons, the Bogomol'nyi equations of self dual Yang-Mills Higgs theory have minimum solutions known as 't Hooft-Polyakov monopoles, the Planar Abelian Higgs model has minimum energy self dual solutions known as Nielsen-Olesen vortices. Thus instantons, monopoles and vortices have become paradigms of topological structures in field theory and quantum mechanics, with important applications in particle physics, astrophysics, condensed matter physics and mathematics.

We have discussed here the self-dual Chern-Simons theory specially in (2+1) dimensions (i.e two spatial dimensions). The physical context in which these self-dual Chern-Simons models arise is that of anyonic quantum field theory \cite{forte,frolich,sousa}, with direct applications to such planar models as the quantum Hall effect \cite{frad1,PG,stone2}, anyonic superconductivity  \cite{lyk} and Aharonov- Bohm scattering. Self dual models in (2+1)dimensions have certain distinct features which are essentially connected with the presence of the Chern-Simons term.

The possibility of describing gauge theories with a Chern-Simons(CS) term  is a special feature of odd dimensional space time. More on that, the (2+1)dimensional case is distinguished in the sense that the derivative part of CS lagrangian is quadratic in gauge fields.

To review the significant properties of the CS lagrangian density let:
$$
{\cal {L}}_ {CS}= \epsilon^{\mu\nu\rho}tr(\pa_{\mu}A_{\nu}A_{\rho} + 
\frac{2}{3}A_{\mu}A_{\nu}A_{\rho})
\nonumber
$$
where the gauge field is $ A_\mu $ and $ \epsilon $ -symbol stands for anti-symmetric tensor and normalised with $ \epsilon^{012} =+1$ In an Abelian theory gauge fields commute and so the trilinear term in the above expression vanishes. Also the action,
$ S= \int d^3x \cal {L}_ {CS} $ is gauge invariant and so we expect that a sensible gauge theory may be formulated. Another important feature of CS theories is that the CS term describes a topological gauge field theory in the sense that there is no explicit dependence on the space time metric. Thus the action is independent of the space time metric and the Cher-Simons lagrangian density $\cal {L}_{CS}$  does not contribute to the energy momentum tensor. Moreover the special feature of the first order  in space time derivative of $\cal {L}_{CS}$ modifies the structure of the theory leading to many interesting features in CS theories.

The CS lagrangian density when coupled to an external matter current $ J^\mu $  as 
$$ {\cal{L}}= \frac{k}{2}{\cal {L}}_{CS} - tr(A_\mu J^\mu)\nonumber $$
(where $ k $ is CS coupling coefficient)yields crucial features when applied to condensed matter systems such as the quantum Hall effect. %\cite{frad1,PG,stone2}.

Most prominent aspect of CS theory evolves when above lagrangian is coupled with Maxwell term to form a gauge model which describe a massive dynamical gauge mode, with  mass determined by the CS coupling parameter $ k $ and with spin $ \pm 1 $ given by the sign of $ k $. This system has been termed ``topologically massive gauge theory" \cite{DJT}.
The equations of motion when expressed in terms of the dual to the field tensor manifest a self duality. An equivalent version of this model also exists, where the self duality is revealed in the equations of motion for the basic field \cite{TPN, DJ, BR}. 
More recently, another possibility has been considered where instead of the first derivative CS term a parity violating third derivative term is added to the Maxwell term \cite{DJ1}.

An intriguing fact first observed in \cite{DJT} and briefly discussed in \cite{D, BW, BK1} is that topologically massive CS-doublets, with identical mass parameters having opposite sign, are equivalent to a parity preserving vector theory with an explicit mass term. This is the Proca model. The invariance of the CS doublet under the combined parity and field interchanges is thereby easily understood from the equivalent theory. These equivalent descriptions of the same physical theory become useful and play a significant role in expanding our understanding. Aspects of a theory that are hidden in one formulations become transparent in some other formulation. Most common example is bosonization technique in (1+1)dimension in this context \cite{CO}.

In recent times the role of duality as a qualitative tool in the investigation of physical systems is being gradually realised in different contexts \cite{gz}. Several technical aspects of duality symmetric actions have been explored \cite{Wotz}. Specially the technique able to work with distinct manifestations of duality symmetry proposed by Stone \cite{stone} is relevant to our thesis work, where a soldering technique has been developed and applied to different models. This technique for fusing together opposite aspects of duality symmetries provides a new formalism that includes the quantum interference effects between the independent components. This leads to a unique way of obtaining physical results.

\vskip 1.0 cm
%%%%%%%%%%%%%%%%%%%%%%%%%%%%%%%%%%%%%%%%5
\section{\large{Outline of the thesis :}}

We will restrict our study within planar models i.e models in (2+1)dimension. As we know quantum models in (0+1)dimension may be interpreted as toy models useful for studying higher dimensional field theoretic examples. Pursuing this feature we start typically with a relevant topological quantum mechanical model (such as Landau problem consisting of two basic chiral oscillators) and extrapolate the analysis to (2+1)dimensional vector field theory. Also from a contemporary view we will consider the aspects of selfdual symmetry in topologically massive gravity model. We will consider here three different approaches to analyse this selfdual doublet structure of a composite theory. The first one is soldering which is solely a lagrangian formulation. The idea is to solder the distinct lagrangians through a contact term \cite{BW,BK1}.
In the hamiltonian approach, on the other hand, there is a canonical transformation which diagonalises the composite hamiltonian into two independent pieces \cite{DJTr,BG,BK2}. 
The third method is based on the exploitation of equations of motion. Simply redefining the old variables by
the new ones does the trick.We have organized the chapters as follows.

In Chap. \ref{quan-model} we will focus on the quantum mechanical models. We will first demonstrate how duality symmetric (or chiral) actions are already present in the quantum mechanical examples such as in usual harmonic oscillator. Using the chiral oscillator form,  we will briefly develop the key concepts of the soldering mechanism. This is purely a lagrangian formulation where schematic  addition of two such independent lagrangians yields a final effective theory. 
Alternately, a canonical transformation based on hamiltonian analysis demonstrates the splitting of a composite hamiltonian into its constituent basic components. We will briefly discuss the consequence of the factorisability property of the final equation of motion of the soldered laqrangian.
Next we introduce another form of the chiral oscillator model where the equation of motion approach
is applicable. Necessitating very simple field redefinitions that are generic to a wide variety of models,
 the results of the lagrangian soldering formalism are reproduced.

We have also discussed the non commutative property of such quantum models.
Since we know that non commutativity may be present in both position and momenta co-ordinates, 
we incorporate these features in non-commutative(NC) quantum mechanics.
Some Galilean generators will be constructed out of these NC features of the phase space coordinates. 
A simple dynamical model  will be presented that displays these aspects.

We know the compatibility of the (0+1)dimensional quantum mechanical extracts with the (2+1)dimensional field theoretic models. So we will establish the  correlation between the chiral oscillator(CO) models with that of the self(antiself) dual vector models and the outcome of the lagrangian or the hamiltonian formulations regarding such quantum theories. In Chap. \ref{vecmodel} thus we have extended the results obtained in Chap. \ref{quan-model} to the case of spin $1$ vector models in $(2+1)$ dimensions. We will first exploit the equations of motion technique for the fusion of vector models. The results obtained here will be reproduced using other techniques(soldering formalism) leading to fresh insights. 
 We will also discuss the equivalence of the Maxwell-Chern-Simons(MCS) theory with that of the selfdual model from the aspect of soldering as well as from the path integral interpretations. Subsequently the hamiltonian reduction will be executed. A non-trivial canonical mapping will be introduced to decompose the composite hamiltonian into constituting hamiltonians of the net effective theory. 
 An elaborate description of the spin content of the respective vector theories comes from the analysis of the  energy momentum tensor. The calculation of the spin of the excitations follows by taking not just the angular momentum operator, but by including also boosts.  

In Chap. \ref{TCS} we have considered models involving higher order derivative of Abelian CS-term in (2+1)dimensions, specially the leading third order derivative Chern-Simons term).
Inclusion of this term with usual Maxwell term or with CS term or to both of these terms
 reveals many interesting observations. For example polarisation vectors in these models possess an identical
 structure with corresponding expressions for usual Maxwell-CS, Proca or the MCS-Proca 
model. We know that higher derivative model poses problem in hamiltonian formulation. So we have considered only the Maxwell-Third order Chern-Simons  model for convenience. 
Due to the presence of third order time derivative term the hamiltonian formulation is very tricky and proper care has been taken to identify appropriate canonical pairs. These issues were bypassed by adopting Ostrogradski's formalism for higher order lagrangian and successively constructing the momentum as well as hamiltonian. We have also illustrated the constrained features of the model and computed relevant Dirac brackets. 

In Chap. 5  emphasis is  on the  spin $2$ tensor models which appear in discussions \cite{DJT,BH} of linearised gravity in $(2+1)$ dimensions. Taking a doublet of self dual massive spin $2$ models
considered earlier in \cite{AK}, we show that the effective theory is a new type of generalised
self dual model that has a Fierz-Pauli term, a first order Chern-Simons term and the Einstein-Hilbert
term. Subject to a specific condition, it reduces to the model taken in \cite{ DM}. 
\\
General conclusions and final remarks are relegated to Chap. 6.
%%%%%%%%%%%%%%%%%%%%%%%%%%%%%%%%%%%
%%%%%%%%%%%%%%%%%%%%%%%%    CHAPTER 2    %%%%%%%%%%%%%%%%%%%%%%5
\chapter{Study of quantum mechanical models}
\label{quan-model}
Quantum field theories in (2+1) space time dimension are now in vogue not only for mathematical reasons but also for studies in planar condensed matter (i.e Hall effect, high $T_c$ superconductivity) and cosmological
(string) settings. These models are interesting due to special structure of Chern-Simons terms, available in three dimensions (and in any odd dimensions) which give rise to topologically intricate phenomenon without even dimensional analogs. We know that self-dual models in odd-dimensions, characterised by the presence of Chern-Simons term have been focussed for their relevance in higher dimensional bosonization \cite{R}. Some new
 results in this connection were reported in \cite{BW} by using the concept of soldering \cite{MS}. 
 Interestingly, several facets of self dual field theories in odd dimensions may be better appreciated 
and understood by looking at their one dimensional counterpart -- the so called topological quantum mechanics \cite{DJTr}.
It is well known that the harmonic oscillator(HO) model pervades our understanding of quantum mechanics as well as field theoretical models in various contexts. 
In some instances, the chiral oscillator (CO) instead of the usual HO captures the essence of the problem \cite{BG}. This is tied to the fact that the chiral oscillator simulates the left-right symmetry and takes a decisive roles where this symmetry is significant. Nonetheless the `chirality' property in (0+1) dimension resemblance the effect of self-duality (anti-selfduality) in higher odd dimensions  as is prominent from the various examples known so far having topological term.
We have started with this fact, exploited it in standard quantum mechanical models before extending it to quantum field theory.

We have organised this chapter as follows. In Sec.2.1 we will discuss the systematic derivation of CO from the HO, where the decomposition of the HO leads to a pair of left-right symmetric chiral oscillators. Issue of symmetries will be clarified. 

In Sec.2.2 the method of soldering which is purely a lagrangian approach is introduced. The basic idea of the soldering procedure is to raise a global Noether symmetry of the self and anti-self dual constituents into a local one, but for an effective composite system, consisting of the dual components and an interference term.
In this formalism it combines two distinct lagrangians manifesting  dual aspects of some symmetry (like left-right chirality or self and antiself duality etc) to yield a new effective lagrangian which is devoid of that symmetry. Here specifically it will demonstrate the fusion of the chiral oscillators to regain the HO.

 Sec.2.3 is based on a topological quantum mechanical model. Analysis of this model enables us to explore different aspects of chiral symmetry. In this section other than soldering formalism, a technique using the equations of motion is also adopted to provide an alternative view point.  Apart from these lagrangian approaches a qualitative discussion on the canonical transformation
 based on the hamiltonian formulation, is involved. Such transformations are used to decompose a composite hamiltonian into two distinct pieces. An example \cite{BK2} is the decomposition of the hamiltonian of a particle in two dimensions, moving in a constant magnetic field and quadratic potential. It finally decomposes into two one dimensional oscillators, rotating in a clockwise and anti-clockwise direction respectively 
\cite{BG,RG2}. We have also discussed the compatibility of the lagrangian and hamiltonian approach.

In Sec.2.4 we have revealed how  the noncommutative structures may appear in planar quantum mechanics. A simple dynamical model is considered along with the study of certain deformation characters with respect to several symmetry transformations and its generators.\\
Sec.2.5 goes for the conclusion.
%%%%%%%%%%%%%%%%%%%%%%%%%%%%%%%%555
%%%%%%%%%%%%%%%%%%%%%%%%%%%%%%
\section{Chiral Oscillator Model}
\label{CO}
The basic feature of duality symmetric actions are already present in the quantum mechanical examples. 
This present analysis will illustrate using basic harmonic oscillator model.
Let us consider the lagrangian of an one-dimensional oscillator,
\begin{equation}
L=\frac{1}{2}(\dot{q}^2-q^2)\label{eq1}
\end{equation}
leading to equation of motion $\ddot{q} + q = 0$. Thus to obtain chiral oscillator(CO) form, 
the basic step is to convert (\ref{eq1}) in a first order form by introducing an auxiliary variable, such that
\begin{equation}
L=\frac{1}{2}(p\dot{q}-q\dot{p}-p^2-q^2)\label{eq2}
\end{equation}
Here a symmetrisation has been performed. There are now two possible classes for relabelling these variables
corresponding to proper and improper rotations generated by the matrices $R^{+}(\theta)$ and $R^{-}(\phi)$
with determinant $+1$ and $-1$.
\begin{eqnarray}
\left(\matrix{
{q}\cr
{p}}\right)=\left(\matrix{
{\cos\theta}&{\sin\theta}\cr
{-\sin\theta}&{\cos\theta}}\right)\left(\matrix{
{x_1}\cr
{x_2}}\right)
\label{eq3}
\end{eqnarray}
 \begin{eqnarray}
\left(\matrix{
{q}\cr
{p}}\right)=\left(\matrix{
{\sin\phi}&{\cos\phi}\cr
{\cos\phi}&{-\sin\phi}}\right)\left(\matrix{
{x_1}\cr
{x_2}}\right)
\label{equ4}
\end{eqnarray}
leading to the distinct lagrangian,
\begin{equation}
L_{\pm}=\frac{1}{2}(\pm x_{\alpha}\epsilon_{\alpha\beta}\dot{x}_\beta - {x_\alpha}^2)
\label{eq5}
\end{equation}
By these change of variables an index $\alpha=(1,2)$ has been introduced that characterises a
symmetry in this internal space. So setting the angle $\theta=0$ or $\phi=0$ in the rotation matrices
in (\ref{eq3}) and (\ref{equ4}) it follows
\begin{eqnarray}
&&q=x_{1}\quad\quad;\quad\quad p=x_{2} \nonumber\\
&&q=x_{2}\quad\quad;\quad\quad p=x_{1}
\label{equ6}
\end{eqnarray}
Correspondingly, the lagrangian (\ref{eq2}) goes over to (\ref{eq5}). We can easily observe that lagrangian
in (\ref{eq5}) are manifestly invariant under the continuous duality transformation,
\begin{equation}
x_{\alpha}\rightarrow{R}^{+}_{\alpha\beta}x_{\beta}
\label{eq7}
\end{equation}
where, ${R}^{+}_{\alpha\beta}x_{\beta}$ is the usual SO(2) rotation matrix in (\ref{eq3}).
It is easy to verify the basic symplectic bracket obtained from (\ref{eq5}),
\begin{equation}
\{x_{\alpha},x_{\beta}\}= \pm \epsilon_{\alpha\beta}
\label{equ8}
\end{equation}
It is important to note that, there are actually two and not one, duality symmetric actions
$L_\pm$ in (\ref{eq5}), corresponding to the signatures in the determinant of the transformation matrices.
 In the co-ordinate language these lagrangian correspond to two 
bi-dimensional chiral oscillator(CO) rotating in opposite directions. This is easily verified either from 
the classical equation of motion or by examining the spectrum of the angular momentum operator,
\begin{equation}
J_{\pm}= \pm \epsilon_{ij}x_{i}p_{j}=\pm H 
\label{equ9}
\end{equation}
where, H is the hamiltonian of the usual harmonic oscillator.
In other words the two lagrangian manifest the dual aspects of rotational symmetry in the two-dimensional
internal space.
%%%%%%%%%%%%%%%%%%%%%
\section{Introduction to Soldering Formalism}
\label{sold}
The role of duality as a qualitative tool in the investigation of physical systems is getting explored
 at different context and dimensions. In this respect emphasis has been given on several technical aspects of selfdual  symmetric actions. The technique able to work with distinct manifestations of the duality symmetric actions was proposed by Stone \cite{stone}. His mechanism for fusing together opposite aspects of duality symmetries provides a new formalism. The effect is intrinsically quantal without any classical analogue. 
 This can be easily explained by the observation that a simple addition of two independent classical lagrangians is a trivial operation without leading to anything significant. On the contrary the direct sum of classical actions depending on different fields would not give anything new. So only this soldering process leads to a new and nontrivial result. 
 \\
The basic idea of the soldering procedure is to raise a global Noether symmetry of the self and
anti-selfdual constituents into a local one, but for an effective composite system , consisting of the dual components and an interference term. In this formalism it combines two distinct lagrangians manifesting
 dual aspects of some symmetry (like left-right chirality or self and antiself duality etc) to yield a
 new lagrangian which is devoid of that symmetry. From the previous section we have obtained two such chiral oscillators (of opposite chirality) and it is now possible to solder these two lagrangians to produce
a composite Lagrangian. This is achieved by soldering technique \cite{ADW,ABW}. 
\\
The point is that the Lagrangian in (\ref{eq5}) are now considered as functions of independent variables, namely $L_{+}(x)$ and $L_{-}(y)$ instead of same $x$.
\\
It follows
\begin{equation}
L_{+}=\frac{1}{2}(+ x_{\alpha}\epsilon_{\alpha\beta}\dot{x}_\beta - {x_\alpha}^2)
%\label{eq5a}
\nonumber
\end{equation}
\begin{equation}
L_{-}=\frac{1}{2}(- y_{\alpha}\epsilon_{\alpha\beta}\dot{y}_\beta - {y_\alpha}^2)
\label{eq5b}
\end{equation}

Consider the gauging of the lagrangian under the following gauge transformations,
\begin{equation}
\delta{x}_{\alpha}=\delta{y}_{\alpha}=\eta_\alpha
\label{101}
\end{equation}
Then the gauge variations are given by,
\begin{equation}
\delta{L_\pm}(z)=\epsilon_{\alpha\beta}\eta_{\alpha}{J^{\pm}}_{\beta}(z)\quad\quad;
z=x,y
\label{102}
\end{equation}
where the currents are defined by,
\begin{equation}
J^{\pm}_{\beta}(z)=\pm\dot{z_\alpha}+\epsilon_{\alpha\beta}z_{\beta}
\label{103}
\end{equation}
Introducing a new field $B_\alpha$ transforming as 
\begin{equation}
\delta{B}_{\alpha}=\epsilon_{\alpha\beta}\eta_\beta
\label{103s}
\end{equation}
which will affect the soldering such a way that a new lagrangian,
\begin{equation}
L=L_{+}(x)+L_{-}(y)-B_{\alpha}(J^{+}_{\alpha}(x)+J^{-}_{\alpha}(y))-B^{2}_{\alpha}
\label{104}
\end{equation}
It is easy to show 
\begin{equation}
\delta{L}=0
\end{equation}
Eliminating $B_{\alpha}$ by the equation of motion, we obtain the final soldered lagrangian,
\begin{equation}
L(w)=\frac{1}{4}({\dot{w}}^{2}_\alpha-{w}^{2}_{\alpha})
\label{105}
\end{equation}
which is no longer a function of $x$ and $y$ independently, but only on their gauge invariant
combination,
\begin{equation}
w_{\alpha}=(x_{\alpha}-y_{\alpha})\label{106}
\end{equation}
The soldered lagrangian just corresponds to a simple bi-dimensional oscillator. Thus by starting from
two lagrangian which contained the opposite aspects of a duality symmetry, it is feasible to combine 
them into a single lagrangian which has a richer symmetry. 
%%%%%%%%%%%%%%%%%%%%%%%%%%%%%%%%%%%%%Application of Soldering%%%%%%
\section{Study of planar models in quantum mechanics}

Self dual models in odd dimensions, characterised by the presence of Chern-Simons terms \cite{dunne}, have been in vogue for quite sometime \cite{TPN,DJ}. Interest in these models has been rekindled by noting
 their relevance in higher dimensional bosonisation \cite{R}. In this section we discuss the concepts of self duality and soldering in the context of topological quantum mechanics. 

The quantum mechanical topological models are governed by the Lagrangian \cite{DJTr},
\begin{equation}
{\cal {L}} = \frac{m}{2}{\dot{\vec{x}}}^2 + e\dot{\vec{x}}.\vec{A}(\vec{x})-eV(\vec{x})
\label{lag}
\end{equation}
implying the motion of a particle of mass $m$ and charge $e$ in the external electric $(-\partial_iV)$ and magnetic $(\partial_iA_j-\partial_jA_i)$ fields. For the simplest explicitly solvable model, the motion is two
dimensional $(i=1,2)$ and rotationally symmetric in a constant magnetic field(B) and a quadratically scalar potential so that, 
\begin{eqnarray}
A_i &=& -\frac{1}{2}\epsilon_{ij}x_jB \nonumber \\
V   &=& \frac{k}{2}{x_i}^2  \nonumber
\end{eqnarray}
%%%%%%%%%%%%% Landau model %%%%%%%%%%%%%%%%
The lagrangian (\ref{lag}) therefore simplifies to (setting e=1),
\begin{equation}
 {\cal {L}} = \frac{m}{2}{\dot{x_i}}^2 + \frac{B}{2}\epsilon_{ij}x_i
\dot{x_j} - \frac{k}{2}{x_i}^2
\label{lag1}
\end{equation}
There are some interesting features of this lagrangian. If the magnetic field is switched off (B=0), the model represents a bi-dimensional harmonic oscillator,
\begin{equation}
 {\cal{L}}_{H.O} = \frac{m}{2}{\dot{x_i}}^2-\frac{k}{2}{x_i}^2
\label{ho}
\end{equation}
Now consider the motion of the particle in the absence of the electric field so that we have, 
\begin{equation}
 {\cal {L}}_+ = \frac{m}{2}{\dot{x_i}}^2 + \frac{B}{2}\epsilon_{ij}x_i\dot{x_j}
\label{l+}
\end{equation}
Let us next illustrate the connection between (\ref{l+}) and (\ref{ho}). Together with (\ref{l+}), consider the Lagrangian (\ref{l-}) with  an independent set of coordinates $y_i$ and where the direction of 
the magnetic field is reversed,
\begin{equation}
 {\cal{L}}_- = \frac{m}{2}{\dot{y_i}}^2 - \frac{B}{2}\epsilon_{ij}y_i\dot{y_j}
\label{l-}
\end{equation}
It is now possible to combine (\ref{l+}) and (\ref{l-}) by the soldering formalism \cite{BW,RG2}. Consider the following transformation, 
\begin{equation}
\delta x_i = \delta y_i = \eta_i
\label{varq}
\end{equation}
which effects the changes,
\begin{equation}
\delta{\cal L}_{\pm} = J_{\pm i}\dot{\eta_i}
\label{varl}
\end{equation}
where, 
\begin{equation}
J_{{\pm}{i}}(z) = m\dot{z_i} \pm B\epsilon_{ji}z_j 
\label{cur1}
\end{equation}
and $z_i = x_i ,y_i$. Introduce the soldering field $W_i$ transforming as 
\begin{equation}
\delta{W}_i = \dot{\eta_i}
\label{ax1}
\end{equation}
Then the first iterated lagrangian, 
\begin{equation}
{\cal L}^{(1)} = {\cal L}_+ + {\cal L}_- - (J_{+i}(x) + J_{-i}(y))W_i
\label{itl1}
\end{equation}
transforms as, 
\begin{equation}
\delta{\cal L}^{(1)} = -2m\dot{\eta_i}W_i
\label{varl1}
\end{equation}
Including the term ${W_i}^2$ now yields an invariant lagrangian,
\begin{equation}
{\cal L} = {\cal L}^{(1)} + m{W_i}^2  \quad\quad;\quad\quad  
 \delta{\cal L} = 0
\label{finl1}
\end{equation}
Since ${W_i}$ is an auxiliary variable, it is possible to eliminate it by using the equation of motion,
\begin{equation}
W_i = \frac{1}{2m}(J_{+i} + J_{-i})
\label{solax}
\end{equation}
The solution is compatible with the variations (\ref{varq}) and (\ref{ax1}). Inserting (\ref {solax}) into (\ref{finl1}), the final soldered Lagrangian is obtained,
\begin{equation}
{\cal L} = \frac{m}{2}{\dot {q_i}}^2 - \frac{B^2}{2m}{q_i}^2   \quad\quad ;\quad
q_i = \frac{1}{\sqrt 2}(x_i - y_i)
\label{soldlag}
\end{equation}
which is no longer a function of $x$ and $y$ independently, but only on their difference. Identifying $k$ with $\frac{B^2}{m}$, it is found that (\ref{soldlag}) exactly maps on to (\ref{ho}). 
This exercise shows how two identical particles moving in the presence of magnetic fields  
with same magnitudes but opposite directions simulate the effect of a single particle moving 
in the presence of a quadratic scalar potential. We can supplement the above Lagrangian analysis by the familiar hamiltonian formulation \cite {DJTr}. The hamiltonian corresponding to (\ref{l+}) is given by, 
%%%%%%%%%
\begin{equation}
H_+ = \frac{1}{2m}{(p_i + \frac{B}{2}\epsilon_{ij}x_j)}^2
\label{h+}
\end{equation}
where $p_i$ is the conjugate momentum,
%%%%%%%%%%%%%
\begin{equation}
p_i = \frac{\partial {\cal L}_+}{\partial{\dot{ x_i}}} = m{\dot{ x_i}} - \frac{B}{2}\epsilon_{ij}x_j
\label{picon}
\end{equation}
\\
Making a canonical transformation, 
\begin{equation}
p_{\pm} = p_1 \pm \frac{B}{2}x_2
\label{p+}
\end{equation}
\begin{equation}
x_{\pm} = \frac{1}{2}x_1 \mp \frac{1}{B}p_2
\label{x+}
\end{equation}
we obtain, in the new canonical variables,
\begin{equation}
H_+ = \frac{1}{2m}{p_+}^2 + \frac{1}{2}{B}^2{x_+}^2 
\label{newh}
\end{equation}
The hamiltonian is that of the usual HO. It is however expressed only in terms of $(x_+,p_+)$ while the other canonical pair $(x_-,p_-)$ gets eliminated. The fact that the two dimensional Lagrangian (\ref{l+}) simplifies to a one dimensional oscillator (\ref{newh}) is essentially tied to its symplectic structure. Likewise (\ref{l-}) yields the hamiltonian for the H.O expressed only in terms of the canonical set $(x_-,p_-)$. Thus the combination of (\ref{l+}) and (\ref{l-}) should yield a two dimensional HO which is precisely shown by the soldering mechanism leading to (\ref{soldlag}).
%%%%%%%%%%%%%%%%%%%%%
%%%%%%%%%%%%%%%%%%%%%%5
\subsection{One dimensional Harmonic Oscillator - an alternate form}
\label{alterform}
%%%%%%%
It is worthwhile to mention that the massless version of (\ref{lag1}),
\begin{equation}
{\cal L}_0 = \frac{B}{2}\epsilon_{ij}x_i\dot{x_j }-\frac{k}{2}{x_i}^2
\label{l0}
\end{equation}
also yields a one dimensional HO. The simplest way to realise
 this is by eliminating either $x_1$ or $x_2$ in favour of the other.
  In the beginning of this section  the feature of this type of model 
 have been demonstrated and utilised this form to the soldering scheme.In this sense it 
 is similar to (\ref{l+}).
 
 Going back to the original Lagrangian (\ref{lag1}), it is well known \cite{DJTr}
from a hamiltonian analysis that the model corresponds to two decoupled 
one-dimensional oscillators described by the canonical pairs $(x_{\pm},p_{\pm})$and frequencies $\omega_\pm$ where,
%%%%%
\begin{eqnarray}
p_{\pm} & =& {\sqrt {\frac{\omega_\pm}{2m{\Omega}}}}p_1 \pm {\sqrt {\frac{{\omega_\pm}m\Omega}{2}} }x_2       \nonumber \\
x_{\pm} & =& {\sqrt {\frac{m{\Omega}}{2{\omega_\pm}} }}x_1 \mp \frac{1}{\sqrt 
{{\omega_\pm}m\Omega}}p_2    \nonumber \\
\omega_{\pm} & =& \Omega \pm \frac{B}{2m} \quad\quad   ;    \quad\quad \Omega = \sqrt { \frac{B^2}{4m^2} + \frac{k}{m}  }    
\label{R} 
\end{eqnarray}
%%%%%%%%%%
These are the analogous of (\ref{p+}), (\ref{x+}). While the hamiltonian analysis reveals the 
decoupling of (\ref{lag1}) into the two one-dimensional oscillator, the soldering formalism will 
explicitly demonstrate the reverse process. Let us therefore consider the following 
{\it independent} lagrangian : 
%%%%%%%%%%%%%%%
\begin{equation}
{\cal L}_+ = \frac{1}{2}(\omega_+\epsilon_{ij}x_i{\dot {x_j}} - {\omega_+}^2{x_i}^2)
\label{L}
\end{equation}
\begin{equation}
{\cal L}_- = \frac{1}{2}(-\omega_-\epsilon_{ij}y_i{\dot {y_j}} - {\omega_-}^2{y_i}^2)
\label{M}
\end{equation}
%%%%%
These Lagrangian are similar to the previous cases (see, for instance 
(\ref{l0})), except that the frequencies are different $\omega_{\pm}$.
 As stated before both these represent one dimensional harmonic 
oscillators but there are two points which ought be stressed. The equations of motion are given by, 
%%%%%
\begin{equation}
x_i = \frac{1}{\omega _+}\epsilon_{ij}{\dot{x_j}}
\label{A}
\end{equation}
\begin{equation}
y_i = - \frac{1}{\omega _-}\epsilon_{ij}{\dot{y_j}}
\label{B1}
\end{equation}
Define a dual field as, 
$$ {\tilde {x_i}} = \frac{1}{\omega}\epsilon_{ij}\dot {x_j}$$
The duality property is only on-shell because, 
%%%%%%%%%%
\begin{eqnarray}
{\tilde {\tilde{x_i}} }  = \frac{1}{\omega}\epsilon_{ij}{\tilde {\dot {x_j}} }
 = x_i   \nonumber 
\end{eqnarray}
requires the use of the equation of motion. In this sense, therefore,  equations (\ref {A}) and (\ref{B1}) characterise self and antiself dual solutions, respectively. Moreover, as discussed in \cite{BW, RG2}, 
it is possible to interpret the lagrangians (\ref{L}) and (\ref{M}) as chiral oscillators with varying frequencies $\omega_\mp$ rotating in clockwise and anticlockwise directions. Thus the ubiquitous role of self duality and chriality becomes apparent in these models. The process of soldering will combine the dual 
aspects of these symmetries to yield a new model expressed in terms of the composite variable $(x-y)$.
%%%%%%%%%%%%%%%%%%%%%%%%%%%%%%%%%%%555 
%%%%%%%%%%%%%%%%%%%%%%%%%%%%%%%%%%%%
\subsection{The approach based on Equations of motion}
\label{eqofmotion}

Before going into the intricacy of usual soldering formalism which sometime becomes technically involved
requiring arcane field redefinition, in this part we have adopted a method \cite{D} necessitating very simple field redefinitions that are generic to a wide variety of models.
%%%%%%%
Let us consider the doublet as in (\ref{L}) and (\ref{M}),
\begin{eqnarray}
{\cal {L}_+}& =&  \frac{1}{2\omega_+}\epsilon_{ij}x_i\dot{x_j} - \frac{1}{2}{x_i}^2     
\label{1a} \\
{\cal{L}_-} &= & - \frac{1}{2\omega_-}\epsilon_{ij}y_i\dot{y_j}-\frac{1}{2}{y_i}^2
\label{1b}
\end{eqnarray}
where independent set of coordinates $x_i$ and $y_i$ have been used. The equations of motion
are given by,
\begin {equation}
\frac{1}{{\omega}_+}\epsilon_{ik}\dot{x_k} - x_i= 0
\label{2a}
\end{equation}
\begin{equation}
-\frac{1}{{\omega}_-}\epsilon_{ik}\dot{y_k} - y_i= 0
\label{2b}
\end{equation}
These may be put in the form,
$$\ddot{x_i} = -{{\omega_+}^2}{x_i}\quad\quad,\quad\quad \ddot{y_i} = -{{\omega_-}^2}{y_i} \nonumber$$
which are just the standard oscillator equations. Consequently (\ref{1a}) and (\ref{1b}) represent
two chiral oscillators (with frequencies $\omega_+ $ and $\omega_-$) having opposite chirality 
which is manifested by the different signs of the first term in $\cal{L}_\pm$.\\
Let us introduce a new set of variables $(f_i, g_i)$ by combining chiral ones as,
\be
 x_i+y_i = f_i, \,\,\,\,\, x_i-y_i = g_i
\label{2c}
\ee
Subtracting (\ref{2b}) from (\ref{2a}) and using the definition of $g_i$ we obtain,
\begin{equation}
\frac{1}{\omega_+}\epsilon_{ik}\dot{g_k} + \Omega\epsilon_{ik}\dot{y_k} - g_i= 0 \quad;
\quad\quad\quad\quad \Omega=\frac{1}{\omega_+} +\frac{1}{\omega_-}
\label{3a}
\end{equation}
%where $\Omega=\frac{1}{\omega_+} +\frac{1}{\omega_-}$.\\
Contracting (\ref{3a}) by ${\frac{1}{\omega_-}}\epsilon_{ip}$ yields,
\be
\frac{1}{{\omega_+}{\omega_-}}\ddot{g_p}+\frac{1}{\omega_-}\epsilon_{pi}\dot{g_i}+
\frac{\Omega}{\omega_-}\ddot{y_p}=0
\label{3b}
\ee
Taking the difference of (\ref{3a}) and (\ref{3b}) and exploiting the oscillator equation of motion 
for $y$ yields, 
\begin{equation}
\ddot{g_i} + ({\omega_+}-{\omega_-})\epsilon_{ik}\dot{g_k}+{\omega_+}{\omega_-}{g_i}=0
\label{4amot}
\end{equation}
Adopting a similar analysis we can show that the other variable $f_i$ also satisfies
an identical equation,
\begin{equation}
\ddot{f_i} + ({\omega_+}-{\omega_-})\epsilon_{ik}\dot{f_k}+{\omega_+}{\omega_-}{f_i}=0
\label{4bmot}
\end{equation}
These equations of motion factorise in terms of their dual(chiral) components as \cite{BK1},
\be
(\epsilon_{ji}\partial_t + \omega_-\delta_{ji})(\epsilon_{ik}\partial_t - \omega_+\delta_{ik})X_k = 0 \quad\quad;
\quad\quad \quad{X_k}=({f_k},{g_k})
\label{4b.1}
\ee
Observe that the above equations of motion can be obtained from the Lagrangian,
\be
{\cal L} = \frac{1}{2} {\dot{X_i}}^2 - \frac{1}{2}(\omega_+ - \omega_-)\epsilon
_{ij}X_i\dot{X_j} - \frac{1}{2}{\omega_+}{\omega_-}{X_i}^2
\label{4cmot}
\ee
It is now straightforward to identify this Lagrangian with (\ref{lag1}) for a unit mass(m=1) by taking,
${\omega_-} - {\omega_+}=B$  and  $ {\omega_+}{\omega_-}=k $.
This shows how two chiral oscillators with distinct frequencies moving in clockwise and anticlockwise
directions simulate the motion of a charged particle in the presence of electric and magnetic fields.
Moreover the magnetic part is a consequence of the different angular frequencies. For 
${\omega_+}={\omega_-}$, this term just drops out and only the effect of the electric field is retained.
Finally, note that the obtention of (\ref{4cmot}) is effected by the change of variables (\ref{2c}).
Such a change will also play a crucial role in both field theory and gravity to be discussed in the subsequent
sections. 
%%%%%                       method of soldering %%%%%%%%%%%%%%
\\
However following the process of soldering also yield the similar effect coupling these above chiral lagrangians. To demonstrate briefly let us consider the transformations,
$$
\delta x_i = \delta y_i = \eta_i
$$
the lagrarians undergo the variations, 
\begin{eqnarray}
\delta{\cal L}_{\pm} = \epsilon_{ij}J_{\pm j }{\eta_i}\quad\quad ;
 z  = x,y  \nonumber \quad\quad ;
J_{\pm i }(z)  = \omega_\pm(\pm \dot{z_i} + \omega_\pm\epsilon_{ij}z_j)
\nonumber
\end{eqnarray}
Inserting the auxiliary variable $W_i$ transforming as, 
$$
{\delta W}_i = \epsilon_{ij}\eta_j
$$
it is possible to construct, in analogy with (\ref{finl1}), the following lagrangian,
$$
{\cal L} = {\cal L} _+(x) + {\cal L} _-(y) + W_i({J_i}^+(x) + {J_i}^-(y))
- \frac{1}{2}({\omega_+}^2 +{\omega_-}^2){W_i}^2
\nonumber 
$$
This expression is on shell invariant. Eliminating $W_i$ by using the equation
of motion, the above Lagrangian is recast in the manifestly invariant form,
$$
{\cal L} = \frac{1}{2} {\dot{X_i}}^2 - \frac{1}{2}(\omega_+ - \omega_-)\epsilon
_{ij}X_i\dot{X_j} - \frac{1}{2}\omega_+\omega_-{X_i}^2
\nonumber
$$
\begin{equation}
X_i = {\sqrt {\frac{\omega_+\omega_-}{{\omega_+}^2 +{\omega_-}^2}}} \quad(x_i - y_i)
\label{S}
\end{equation}
where use has been made of the on-shell conditions (\ref{A}, \ref{B1}). Identifying the frequencies $\omega_\pm$ with those occurring in (\ref{R}) we find,
\begin{eqnarray}
\omega_- - \omega_+ &=& \frac{B}{m} \nonumber \\
 {\omega_-}{\omega_+} &=& \frac{k}{m}
\label{T}
\end{eqnarray}
After a suitable scaling it is now simple to observe that the Lagrangian in 
(\ref{S}) exactly reproduces (\ref{lag1}).
\\
%%%%%%%%%%%%%%%%%%5
The above exercise therefore shows, in a precise manner, how the self and anti-self dual (or, alternatively, the left and right chiral) oscillators combine to yield the model (\ref{lag1}). For identical frequencies $(\omega_+
=\omega_- =\omega)$, the epsilon term in (\ref{S}) vanishes so that the Lagrangian (\ref{ho}) is obtained, a result found earlier \cite{BW,RG2} in a different context. This is also expected since (\ref{ho}) was 
derived directly from a soldering of (\ref{l+}) and(\ref{l-}), models which are equivalent to (\ref{L}) and 
(\ref{M}) with identical frequencies.

We conclude our discussion on the topological quantum mechanics by pointing out that the equation of motion obtained from (\ref{S}) factorises into its dual (chiral) components as follows,
$$
(\omega_+\delta_{ij} + \epsilon_{ij}\partial_t)(\omega_-\delta_{jk} - \epsilon_{jk}\partial_t)X_k = 0
\nonumber
$$
The possibility of this factorisation is ingrained in the soldering of (\ref{L}) and (\ref{M}) [with equations of motions (\ref{A}) and (\ref{B1}), respectively] to yield the final structure (\ref{S}).
Such a phenomenon illuminates the dual alternately the chiral composition of the models. Actually the symmetry gets hidden in the effective lagrangian but only manifests after the factorisation.
The above exercise therefore shows, in a precise manner, how the left and right chiral oscillators combine to yield the final model (\ref{lag1}). 
%%%%%%%%%%%%%%%%%5
%%%%%%%%%%%%%%%%%%%%%
\section{Study of noncommutative features }
\label{NC}
It is generally believed that the measurement of space time coordinates at small scale involves unavoidable effects of quantum gravity. This effect, as suggested in the work of Doplicher {\it et. al.}\cite{doplicher}, can be incorporated in a physical theory by making the space time coordinates noncommutative. Without going into any detail one can write a general commutator among the space time coordinates as,
\begin{eqnarray}
[\hat{y}_{\mu}, \hat{y}_{\nu}]=i\theta_{\mu\nu}(\hat{y},\hat{q})\label{non}
\end{eqnarray}
Here $y,q$ are phase space variables.  The studies which are built on a structure like (\ref{non}) are called noncommutative physics \cite{review}. In the simplest nontrivial case one takes the noncommutative parameter $\Theta(=\theta_{\mu\nu})$ to be a constant antisymmetric matrix which is commonly named as canonical noncommutativity. Even in that case the commutator relation (\ref{non}) violates the Poincare symmetries\cite{kuldeep}.

In the last few years an interesting study has been found\cite{wess} where appropriate deformations of the representations of Poincare generators lead to different symmetry transformations which leave the basic commutator algebra covariant. In this way original Poincare algebra is preserved at the expense of modified coproduct rules. Quantum group theoretic approach following from the twist functions also give the identical results\cite{chaichian}. Construction of field theory based on these ideas and their possible consequences in field theory have been discussed in \cite{kuldeep,gonera}.

In nonrelativistic quantum mechanics, unlike space coordinates time is treated as a parameter instead of an operator. In that case though $\theta_{0i}=0$, remaining nonvanishing $\theta_{ij}$ breaks the Galilean invariance even for the canonical (constant $\theta$) case. But once again deforming the generators one can keep the theory consistent with the noncommutating algebra among space coordinates. This has been shown in \cite{RB1} for the larger Schroedinger group, a subgroup of which is the Galilean group.

However, in all these analysis, the basic noncommutative brackets taken were somewhat restricted in the sense that noncommutativity among momenta coordinates were always taken to be zero. Interestingly, in the planar Landau problem which is frequently referred for the physical realization of canonical noncommutativity, it was shown in \cite{RB2,SS1} that noncommutativity among position coordinates and momenta coordinates has a dual nature. In the semiclassical treatment of Bloch electrons under magnetic field, a nonzero Berry curvature leads to a modification of the commutator algebra\cite{xiao}. When both the magnetic field and Berry curvature are constant the commutator brackets take a simple form and even in that case none of them is zero. On top of it even the standard position-momentum ($x-p$) algebra gets modified. 

In this Sec. \ref{NC} we consider both position position and momentum momentum noncommutativity in 2+1 dimension and study the invariance of Galilean group. In Sec. \ref{NCphase-space} we give a general 
mapping between the commutating (which satisfy Heisenberg algebra) and noncommutating phase space variables. By a systematic method the values of different coefficients in this map are fixed. An inverse mapping is then obtained. In Sec. \ref{jacobi}, starting from a general noncommutative phase space algebra, we show how Jacobi identities lead to different brackets studied in earlier papers. Using the inverse map found in 
Sec. \ref{NCphase-space}, the deformed representations of the generators of the  Galilean group are obtained in the Sec. \ref{galilean}. These generators satisfy the usual closure algebra on the noncommutative plane. In Sec. \ref{symmetry}, using each deformed generator we calculate the symmetry transformations of the phase space coordinates. A dynamical model is then proposed in Sec. \ref{dynamic}. Constraint analysis of this model leads to nonzero Dirac brackets among position coordinates as well as momenta coordinates. These bracket structures are classical analogues of the quantum commutators considered in the earlier sections. Noether analysis is performed in Sec. \ref{noether} for the same model to get the classical version of the deformed Galilean generators. Finally we conclude.
 
The model we have so far considered in expression (\ref{eq5})or in (\ref{L}) shows noncommutativity.
But the model in (\ref{lag1}), known as planar Landau model, frequently referred for physical 
realization of canonical noncommutativity. Here the position coordinates and momenta coordinates has a dual
nature. It has been observed that above commutator relation violets the Poincare symmetries. 
We have discussed some features of noncommutativity as well as the deformed algebra in this regard.

%%%%%%%%%%%%%%%%%%%     sub: 2.4.1  %%%%%%%%%%%%%%%%%%%%%
\subsection{Noncommutative Phase-Space}
\label{NCphase-space}

In this section we show how  noncommutativity  can be introduced by suitable mapping
 of phase space variables obeying the commutative algebra. We have the standard Heisenberg algebra
 in ( D=2+1 ) dimensional space as,
\begin{eqnarray}
&&[\hat{x}_i,\hat{x}_j]=0\nonumber\\
&&[\hat{p}_i,\hat{p}_j]=0\label{nce1}\\
&&[\hat{x}_i,\hat{p}_j]=i\hbar\delta_{ij}\quad\quad(i=1,2)\nonumber
\end{eqnarray}
Here a quantum mechanical operator ($\hat{O}$) is denoted by putting a hat on its classical counterpart ($O$). Now we define two sets of variables $\hat{y}_i$ and $\hat{q}_i$ in terms of the commutative phase space variables in the following way
\begin{equation}
\hat{y}_i=\hat{x}_i+\alpha_1\epsilon_{ij}\hat{p}_j+\alpha_2\epsilon_{ij}\hat{x}_j\label{e2}
\end{equation}
\begin{equation}
\hat{q}_i=\hat{p}_i+\beta_1\epsilon_{ij}\hat{x}_j+\beta_2\epsilon_{ij}\hat{p}_j\label{e3}
\end{equation}
where $\alpha$ ($\alpha_1$, $\alpha_2$) and $\beta$ ($\beta_1$, $\beta_2$) are some arbitrary constants. Since for small values of $\alpha$ and $\beta$, ($\hat{y}, \hat{q}$) reduces to ($\hat{x}, \hat{p}$) we interpret $\hat{y}$ and $\hat{q}$ as modified coordinates and momenta. Making use of (\ref{nce1}) one finds that the new phase--space variables defined in the above two equations satisfy the algebra
\begin{equation}
[\hat{y}_i,\hat{y}_j]=-2i\hbar\alpha_1\epsilon_{ij}\label{eq4}
\end{equation}
\begin{equation}
[\hat{q}_i,\hat{q}_j]=2i\hbar\beta_1\epsilon_{ij}\label{e5}
\end{equation}
\begin{equation}
[\hat{y}_i,\hat{q}_j]=i\hbar(1+\alpha_2\beta_2-\alpha_1\beta_1)\delta_{ij}\label{eq6}
\end{equation}
Evidently the new brackets show the non-commutative nature of newly defined coordinates ($\hat{y}$) and momenta ($\hat{q}$). Henceforth they will be called non-commutative phase--space variables.
Note that a certain amount of flexibility is there due to different values of the constants $\alpha$ and $\beta$. We keep the bracket (\ref{eq6}) to its simplest undeformed form (\ref{nce1}). This gives the condition
\begin{eqnarray}
\alpha_2\beta_2=\alpha_1\beta_1
\end{eqnarray}
Now without any loss of generality we can take $\alpha_2=\beta_2$ which fixes the constants $\alpha_2$ and $\beta_2$ in terms of the other two constants
\begin{eqnarray}
\alpha_2=\beta_2=\sqrt{\alpha_1\beta_1}\label{n1}
\end{eqnarray}
Next to give (\ref{eq4}) and (\ref{e5}) a neat form we set the following values of $\alpha_1$ and $\beta_1$
\begin{eqnarray}
&&\alpha_1=-\frac{\theta}{2}\label{n2}\\
&&\beta_1=\frac{\eta}{2}\label{n3}
%&&\alpha_2=\beta_2=\frac{1}{2}\sqrt{-\theta\eta}\label{eq7}
\end{eqnarray}
where $\theta$ and $\eta$ are non commutative parameters which in the present study are assumed to be non zero. The choice of constants (\ref{n2}), (\ref{n3}) together with (\ref{n1}) yield the required non-commutative algebra
\begin{eqnarray}
&&[\hat{y}_i,\hat{y}_j]=i\hbar\theta\epsilon_{ij}\nonumber\\
&&[\hat{q}_i,\hat{q}_j]=i\hbar\eta\epsilon_{ij}\nonumber\\
&&[\hat{y}_i,\hat{q}_j]=i\hbar\delta_{ij}\label{eq8}
\end{eqnarray}
Such noncommutative structures appears in the chiral oscillator problem and the Landau model 
where a charged particle moves on a plane subjected to a strong perpendicular magnetic field. Phenomenological discussion of this structure was given in \cite{bertolami}.
 The inverse phase--space transformations of (\ref{eq8}) is given by,
\begin{eqnarray}
&&\hat{x}_i=A\hat{y}_i+B\epsilon_{ij}\hat{y}_j+C\hat{q}_i+D\epsilon_{ij}\hat{q}_j\nonumber\\
&&\hat{p}_i=E\hat{y}_i+F\epsilon_{ij}\hat{y}_j+A \hat{q}_i+B\epsilon_{ij}\hat{q}_j\label{eq9}
\end{eqnarray}
Here the various constants are,
\begin{eqnarray*}
&&A=\frac{2-\theta\eta}{2(1-\theta\eta)}\quad\quad\quad
B=\frac{-\sqrt{-\theta\eta}}{2(1-\theta\eta)}\\
&&C=\frac{\theta\sqrt{-\theta\eta}}{2(1-\theta\eta)}\quad\quad\quad
D=\frac{\theta}{2(1-\theta\eta)}\\
&&E=\frac{-\eta\sqrt{-\theta\eta}}{2(1-\theta\eta)}\quad\quad\quad
F=\frac{-\eta}{2(1-\theta\eta)}
\end{eqnarray*}
Note that the hermiticity of physical operators $x, p$ and 
$\hat{y}$ , $\hat{q}$ can be restored by demanding different signs of $\theta$ and $\eta$ which will keep the various co-efficients  real and well defined.
%%%%%%%%%%%%%%%%%%%%%%%%%%%%%%%%%%%%%%%%%  sub:2.4.2   %%%%%%%%%%%%%%%%%%
\subsection{Role of Jacobi Identities in Planar Non-commutativity}
\label{jacobi}
Jacobi identities are known to play an important role in fixing the structure of the non-commutative algebra. For instance in \cite{magg} the algebra of Kappa-deformed space was obtained in this manner. In this section we discuss the obtention of planar non-commutative algebra by exploiting Jacobi identities.

Consider a plane where the noncommutative parameters are not constants. They are taken to be arbitrary functions of the position coordinates. Since Jacobi identities must be satisfied for the phase space commutator algebra, the functions appearing in the brackets cannot all be independent. The relations among these functions will enable us to generate different types of noncommutative structures studied in earlier papers. 

We take the noncommutative structure in the following form
\begin{eqnarray}
&&[\hat{y}_i,\hat{y}_j]=i\hbar\Omega f(x)\epsilon_{ij}\label{1001}\\
&&[\hat{q}_i,\hat{q}_j]=i\hbar Bg(x)\epsilon_{ij}\label{1002}\\
&&[\hat{y}_i,\hat{q}_j]=i\hbar s(x)\delta_{ij}\label{1003}
\end{eqnarray}
where $\Omega, \ B$ are constants and $f,g,s$ are some functions of coordinates.
The Jacobi identity for $y_i$--$q_j$--$q_k$ is
\begin{eqnarray}
[y_i,[q_j,q_k]]+[q_j,[q_k,y_i]]+[q_k,[y_i,q_j]]=0
\end{eqnarray}
Using (\ref{1002}) and (\ref{1003}) in the above equation we find
\begin{eqnarray}
(B\Omega f\partial_kg-s\partial_k s)\delta_{ij}-(B\Omega f\partial_jg-s\partial_js)\delta_{ik}=0\label{1004}
\end{eqnarray}
where $\partial_k=\frac{\partial}{\partial x_k}$ in the above equation. Similarly the Jacobi identity for  $y_i$--$y_j$--$q_k$ give
\begin{eqnarray}
%&&h\partial_i g\epsilon_{jk}+h\partial_jg\epsilon_{ki}+h\partial_kg\epsilon_{ij}=0\label{1005}\\
&&f\partial_ms(\epsilon_{im}\delta_{jk}-\epsilon_{jm}\delta_{ki})-s\partial_kf\epsilon_{ij}=0\label{1006}
\end{eqnarray}
Other two Jacobi identities are identically zero. In the equations (\ref{1004}) and (\ref{1006}), $i,j,k$ take values only 1 and 2. Now we take specific choices for $i,j,k$ in these equations to get simplified equations. For $i=1,j=2,k=2$ (\ref{1006}) gives
\begin{eqnarray}
f\partial_2s-s\partial_2f=0\label{001}
\end{eqnarray}
And the choice $i=2,j=1,k=1$ in the same equation gives
\begin{eqnarray}
f\partial_1s-s\partial_1f=0\label{001a}
\end{eqnarray}
Equations (\ref{001}) and (\ref{001a}) are written in a covariant way
\begin{eqnarray}
f\partial_is-s\partial_if=0\label{AA}
\end{eqnarray}
Equation (\ref{1004}) under the choices $i=1,j=1,k=2$ and  $i=2,j=1,k=2$ gives following two equations
\begin{eqnarray}
&&B\Omega f\partial_2g-s\partial_2s=0\label{002}\\
&&B\Omega f\partial_1g-s\partial_1s=0\label{003}
\end{eqnarray}
These two equations are written in a covariant manner
\begin{eqnarray}
B\Omega f\partial_ig-s\partial_is=0\label{BB}
\end{eqnarray}
Equation (\ref{AA}) immediately implies
\begin{eqnarray}
\partial_i\big(\frac{s}{f}\big)=0
\end{eqnarray}
Or equivalently,
\begin{eqnarray}
s(x)=\xi f(x)\label{CC}
\end{eqnarray}
where $\xi$ is some arbitrary constant. Replacing $f$ in terms of $s$ in the commutator algebra (\ref{1001}), we see the constant $\frac{1}{\xi}$ can be absorbed in $\Omega$. So without any loss of generality we set $\xi=1$ in (\ref{CC}) to get
\begin{eqnarray}
f(x)=s(x)\label{DD}
\end{eqnarray}
Substituting (\ref{DD}) in (\ref{BB}) we find
\begin{eqnarray}
s\partial_i(B\Omega g-s)=0
\end{eqnarray}
Thus the term within the parentheses is a constant and we write it as $-\lambda$ {\it i.e.}
%\begin{eqnarray}
%h(x)-B\Omega g(x)=\lambda
%\end{eqnarray}
\begin{eqnarray}
g(x)=\frac{1}{B\Omega}(s(x)-\lambda)\label{EE}
\end{eqnarray}
Equations (\ref{DD}) and (\ref{EE}) give severe restrictions on the structure of the noncommutative algebra (\ref{1001})--(\ref{1003}). For example, if we set $s=1$, then from (\ref{DD}) and (\ref{EE}) we get
\begin{eqnarray}
&&f=1\\
&&g=\frac{1-\lambda}{B\Omega}
\end{eqnarray}
These when substituted in (\ref{1001})--(\ref{1003}) give
\begin{eqnarray}
&&[\hat{y}_i,\hat{y}_j]=i\hbar\Omega\epsilon_{ij}\label{1001'}\\
&&[\hat{q}_i,\hat{q}_j]=i\hbar \frac{1-\lambda}{\Omega}\epsilon_{ij}\label{1002'}\\
&&[\hat{y}_i,\hat{q}_j]=i\hbar\delta_{ij}\label{1003'}
\end{eqnarray}
This noncommutative structure is nothing but the algebra (\ref{eq8}) under the identification $\Omega=\theta$ and $\frac{1-\lambda}{\Omega}=\eta$.

It is interesting to take a different choice of $s$
\begin{eqnarray}
s(x)=g(x)
\end{eqnarray}
Then from (\ref{EE})
\begin{eqnarray}
g=\frac{\lambda}{1-B\Omega}\label{FF}
\end{eqnarray}
Equations (\ref{DD}), (\ref{EE}) and (\ref{FF}) show that $f,g$ and $s$ are constant and
\begin{eqnarray}
f=g=s=\frac{\lambda}{1-B\Omega}\label{FF1}
\end{eqnarray}
Using the above equation in (\ref{1001})--(\ref{1003}) we get
\begin{eqnarray}
&&[\hat{y}_i,\hat{y}_j]=i\hbar\frac{\Omega}{1-B\Omega}\epsilon_{ij}\label{1001''}\\
&&[\hat{q}_i,\hat{q}_j]=i\hbar \frac{B}{1-B\Omega}\epsilon_{ij}\label{1002''}\\
&&[\hat{y}_i,\hat{q}_j]=i\hbar\frac{1}{1-B\Omega}\delta_{ij}\label{1003''}
\end{eqnarray}
where we set $\lambda=1$ which appeared as an overall scaling. Above noncommutative structure appears when an electron is subjected to a uniform magnetic field ($B$) and constant Berry curvature ($\Omega$)\cite{xiao}. 
%%%%%%%%%%%%%%%%%%%%%%%%  sub:   2.4.3  %%%%%%%%%%%%%%%%%%%5
\subsection{Realization of Galilean Generators in noncommutative phase space}
\label{galilean}
In this section we will study the deformations of the generators of Galilean group in the noncommutative plane characterized by (\ref{eq8}){\footnote{Symmetry analysis for particular type of nonconstant noncommutative parameter may be found in \cite{Luki2,RSS}.}}.This group consists of  Angular momentum ($\hat{J}$),
  Translation ($\hat{\bf{P}}$), and Boost ($\hat{\bf{G}}$) which
 in (2+1)dimensional commutative space take the form
\begin{eqnarray}  
&&\hat{J}=\epsilon_{ij}\hat{x}_i\hat{p}_j\nonumber\\
%&&\hat{H}=\frac{\hat{p}^2}{2m}\nonumber\\
&&\hat{P}_i=\hat{p}_i\nonumber\\
&&\hat{G}_i=m\hat{x}_i-t\hat{p}_i\label{eq10}
\end{eqnarray}
They are known to satisfy the following closure properties
 \begin{eqnarray}
 &&[\hat{P}_i,\hat{P}_j]=0\quad\quad\quad   [\hat{G}_i,\hat{J}]=-i\hbar\epsilon_{ij}\hat{G}_k\nonumber\\
&&[\hat{G}_i,\hat{G}_j]=0\quad\quad\quad[\hat{J},\hat{J}]=0\nonumber\\
 &&[\hat{P}_i,\hat{J}]=-i\hbar\epsilon_{ij}\hat{P}_k\quad\quad\quad
[\hat{P}_i,\hat{G}_j]=-im\hbar\delta_{ij}\label{eq11}
 %&&[\hat{H},\hat{G}_i]=-i\hbar\hat{P}_i\quad\quad\quad[\hat{H},\hat{J}]=0\nonumber\\
 %&& [\hat{H},\hat{P}_j]=0     \quad\quad\quad[\hat{H},\hat{P}_j]=0
 \end{eqnarray}
 In a 2+1 dimensional noncommutative space (\ref{eq8}), mere substitution of old phase space variables ($x$, $p$) by new variables ($y, q$) would not preserve the closure algebra. Thus, it is necessary to deform the generators in an appropriate manner so that the symmetry remains invariant. The deformed generators are readily obtained by using the mapping (\ref{eq9}) in the expressions (\ref{eq10}). These are given by,
\begin{eqnarray}
\hat{J}&=&\frac{E}{2}\epsilon_{ij}\hat{y}_i\hat{y}_j+\frac{C}{2}\epsilon_{ij}\hat{q}_i\hat{q}_j+
\frac{1}{1-\theta\eta}\epsilon_{ij}\hat{y}_i\hat{q}_j+D{\hat{q}_j}^2-F\hat{y}_j^2
\label{j}\\
%%%\end{equation}
%\hat{H}&=&\frac{1}{2m}(E\hat{y}_i+F\epsilon_{ij}\hat{y}_j+A \hat{q}_i+B\epsilon_{ij}\hat{q}_j)^2 \nonumber\\
\hat{P}_i&=&E\hat{y}_i+F\epsilon_{ij}\hat{y}_j+A \hat{q}_i+B\epsilon_{ij}\hat{q}_j
\label{pi}\\
\hat{G}_i&=&m(A\hat{y}_i+B\epsilon_{ij}\hat{y}_j+C\hat{q}_i+D\epsilon_{ij}\hat{q}_j)\nonumber\\
&&-t(E\hat{y}_i+F\epsilon_{ij}\hat{y}_j+A \hat{q}_i+B\epsilon_{ij}\hat{q}_j)
\label{eq12}
\end{eqnarray}
It can be easily verified that in the noncommutative space (\ref{eq8}) the deformed generators (\ref{eq12}) satisfy the undeformed algebra (\ref{eq11}). Quite naturally in the limit of vanishing noncommutative parameters ($\theta,\eta\rightarrow 0$) (\ref{eq12}) reduces to the primitive form of the generators (\ref{eq10}) under the identification $(\hat{y},\hat{q})\rightarrow(\hat{x},\hat{p})$.
%%%%%%%%%%%%%%%%%%%%%%%%%%%%%%%%%%%%%%%%%%%%%%%%%%%%%%%%%%%%%%%%%%%%%%%%%%%%%%%%%
%%%%%%%%%%%%%%%%%%%%  sub:2.4.4  %%%%%%%%%%%%%%%%%%%%%%%%%%
\subsection{Infinitesimal Symmetry Transformations}
\label{symmetry}
The explicit presence of non-commutative parameters in the phase space algebra hints at possible 
deformations of the symmetry transformations. These are obtained by calculating the commutator of noncommutative phase space variables and the deformed generators listed in (\ref{eq12}). The results are given separately for each generators. 

{\bf Translation}

We first consider the translation generator ($P$) which gives the following transformation rule for the noncommutative coordinate $\hat{y}$
\begin{eqnarray}
\delta\hat{y}_i&=&\frac{i}{\hbar}[\hat{P},\hat{y}_i]\nonumber\\
%&=&\frac{i}{\hbar}a_k[p_k,\hat{y}_i]\\
&=&\frac{i}{\hbar}a_k[E\hat{y}_k+F\epsilon_{kj}\hat{y}_j+A\hat{q}_k+B\epsilon_{kj}\hat{q}_j,\hat{y}_i]\nonumber\\
&=&{a_i}+\frac{1}{2}\sqrt{-\theta\eta}\epsilon_{ij}a_j\label{100}.
\end{eqnarray}
The transformation rule for momentum variable is obtained in a likewise manner
\begin{equation}
\delta{\hat{q}_i}=\frac{\eta}{2}\epsilon_{ij}a_j\label{eq14.1}
\end{equation}
Expectedly, $(\theta,\eta)\rightarrow 0$ gives the correct commutative space results for the expressions (\ref{100}) and (\ref{eq14.1}).

{\bf Rotation}

An identical treatment for the deformed rotation generator gives the following transformation rules for the phase space variables
\begin{eqnarray}
\delta\hat{y}_i&=&\frac{i}{\hbar}\alpha[\hat{J},\hat{y}_i]\nonumber\\
&=&-\alpha\epsilon_{ik}\hat{y}_k\nonumber\\
\delta{\hat{q}_i}&=&\frac{i}{\hbar}\alpha[\hat{J},\hat{q}_i]\nonumber\\
&=&-\alpha\epsilon_{ik}\hat{q}_k\label{eq14.2}
\end{eqnarray}
Interestingly, these expressions are identical with the corresponding transformation rules for the commutative space. This is a very special property which holds only in 2+1 dimension.

{\bf Boost}

Similarly for the boost generator the transformation rules are found to be
\begin{eqnarray}
\delta\hat{y}_i&=&\frac{m\theta}{2}\epsilon_{ik}a_k-t(\frac{\sqrt{-\theta\eta}}{2}\epsilon_{ik}a_k
+{a_i})\nonumber\\
\textrm{and}\quad\quad\delta\hat{q}_i&=&-m({a_i}+\frac{\sqrt{-\theta\eta}}{2}\epsilon_{il}a_l)-
t(\frac{\eta}{2}\epsilon_{il}a_l)\label{eq14.3}
\end{eqnarray}
These expressions also have the smooth commutative limit $(\theta,\eta)\rightarrow 0$.
%%%%%%%%%%%%%%%%%%%%%%%%%%%%%%%%%%%%%%%%%%%%%%%%%%%%%%%%%%%%%%%%%%%%%%%%%%%%%%
%%%%%%%%%%%%%%%%%%%  sub: 2.4.5 
\subsection{Construction of a Dynamical Model}
\label{dynamic}
 In order to generate the noncommutative algebra (\ref{eq8}) in a natural way from a model, we consider the first order form of the non relativistic free particle lagrangian 
\begin{equation}
L= p_i\dot{x}_i-\frac{p^2}{2m}\label{eq15}
\end{equation}
and use the classical version of the transformation (\ref{eq9}) to get the following form of the lagrangian
\begin{eqnarray}
L&=&[\frac{E}{2}y_k\dot{y}_k+F\epsilon_{kl}\dot{y}_ky_l+Mq_k\dot{y}_k
  +\frac{C}{2}q_k\dot{q}_k+D\epsilon_{kl}q_k\dot{q}_l]\nonumber\\
&&-\frac{1}{2m}[(E^2+F^2){y_i}^2+(A^2+B^2)q_i^2+
Ey_iq_i-2F\epsilon_{ij}y_iq_j]\label{eq16}
\end{eqnarray}
where $M=1/(1-\theta\eta)$ in the above equation. For $\eta=0$, physical application of this model was studied in \cite{Duval} and theoretical discussion, notably lagrangian involving second order time derivative was given in \cite{LSZ,Luki1}. The relation between the Chern-Simons field theory and the Landau problem in the 
noncommutative plane has been studied in \cite{plyuschay}. In the above model we interpret $y$ and $q$ as the configuration space variables in an extended space.
The canonical momenta conjugate to $y$ and $q$ are,
\begin{eqnarray}
&&{\pi_i}^y=\frac{E}{2}y_i+F\epsilon_{ik}y_k+Mq_i\label{eq17a}\\
&&{\pi_i}^q=\frac{C}{2}q_i-D\epsilon_{ik}q_k\label{eq17}
\end{eqnarray}
These momenta ($\pi_i^y,\pi_i^q$) together with the configuration space variables ($y_i,q_i$) give the following Poisson algebra
\begin{eqnarray}
&&\{y_i,\pi_j^y\}=\delta_{ij}\\
&&\{q_i,\pi_j^q\}=\delta_{ij}
\end{eqnarray}
All other brackets are zero. Since none of the momenta ((\ref{eq17a}) and (\ref{eq17})) involve velocities these are interpreted as primary constraints. These are given by,
\begin{eqnarray}
\Omega_{1,i}&=&{\pi_i}^y-[\frac{E}{2}y_i+F\epsilon_{ik}y_k+Mq_i] \approx 0\nonumber\\
\Omega_{2,i}&=&{\pi_i}^q-[\frac{C}{2}q_i-D\epsilon_{ik}q_k]\approx 0\label{eq18}
\end{eqnarray}
They satisfy the Poisson algebra
\begin{eqnarray}
&&\{\Omega_{1,i},\Omega_{1,j}\}=-2F\epsilon_{ij}\nonumber\\
&&\{\Omega_{1,i},\Omega_{2,j}\}=-M\delta_{ij}=-\{\Omega_{2,i},\Omega_{1,j}\}\nonumber\\
%&&=M\delta_{ij}\nonumber\\
&&\{\Omega_{2,i},\Omega_{2,j}\}=2D\epsilon_{ij}\label{eq19}
\end{eqnarray}

Evidently $\Omega_{1,i}$ and $\Omega_{2,i}$ do not close among themselves so they are the second class constraints according to Dirac's classification\cite{Dirac}. This set can be eliminated by computing Dirac brackets. For that we write the constraint matrix. 
 \begin{eqnarray}
 \Lambda_{ij} =\big( \Lambda_{ij}^{mn}\big)&=&\left(\matrix{
{\{\Omega_{1,i},\Omega_{1,j}\}}&{\{\Omega_{1,i},\Omega_{2,j}\}}\cr
{\{\Omega_{2,i},\Omega_{1,j}\}} & {\{\Omega_{2,i},\Omega_{2,j}\}}}\right)\nonumber\\&=&\left(\matrix{
{-2F\epsilon_{ij}}&{-M\delta_{ij}}\cr
{M\delta_{ij}} & {2D\epsilon_{ij}}}\right)
\label{eq20}
\end{eqnarray}
We write the inverse of $\Lambda_{ij}$ as $\Lambda^{(-1)}_{ij}$ such that ${\Lambda_{ij}}^{mn}\Lambda_{jk}^{(-1)ns}=
{\delta^{ms}}_{ik}$ $(\textrm{for} \  i,j,k=1,2)$.
It is given by,
\begin{eqnarray}
\Lambda_{ij}^{(-1)ns}=\left(\matrix{
%\left(\begin{array}{11}
{\theta}\epsilon_{ij} & \delta_{ij}\cr
-\delta_{ij} & {\eta}\epsilon_{ij}}\right)
%\end{array}\right)
\label{eq21}
\end{eqnarray}

Using the definition of Dirac bracket\cite{Dirac}
\begin{displaymath}
\{f,g\}_{DB}=\{f,g\}-\{f,\Phi_{i,m}\}{\Lambda_{ij}}^{(-1)ms}\{\Phi_{j,s},g\}
\end{displaymath}
our model yields the following Dirac brackets in configuration space 
\begin{eqnarray}
\{y_i,y_j\}&=&{\theta}\epsilon_{ij}\nonumber\\
\{q_i,q_j\}&=&{\eta}\epsilon_{ij}\nonumber\\
\{y_i,q_j\}&=&\delta_{ij}\label{eq22}
\end{eqnarray}
This algebra manifests the classical analog of the noncommutative algebra ({\ref{eq8}).
%%%%%%%%%%%%%%
%%%%%%%%%%%%%%%%%%%%%%%% sub: 2.4.7
\subsection{ Noether's theorem and generators}
\label{noether}
In this section we reproduce the deformed Galilean generators from a Noether analysis of the lagrangian ({\ref{eq16}}). The generators thus obtained from a knowledge of infinitesimal symmetry transformation will be shown to be identical with those found in  Sec. \ref{galilean}. This clearly shows the consistency between the dynamical approach of previous section and the algebraic approach of Sec. \ref{galilean}.

The invariance of an action $S$ under an infinitesimal symmetry transformation
\begin{equation}
\delta{Q_i}=\{G,Q_i\}\nonumber\\
\label{eq23}
\end{equation}
is given by,
\begin{equation}
\delta{S} =\int{dt} \delta L=\int{dt}{\frac{d}{dt}}(\delta{Q_i}P_i-G)\label{eq24}   
\end{equation}
where G is the generator of the transformation and $P_i$ is the canonical momenta
conjugate to $Q_i$. If we denote the quantity inside the parentheses by $B(Q,P)$, then
\begin{equation}
G=\delta{Q_i}{P_i}-B\nonumber\\   
\end{equation}
Then this can be taken as the definition of the generator $G$. For the model (\ref{eq16}) both $y$ and $q$ have been interpreted as 
configuration space variables so we write the above equation as,
\begin{equation}
G=\delta{q_i}{\pi_i}^q+\delta{y_i}{\pi_i}^y-B\label{eq25} 
\end{equation}
Using the expressions of the canonical momenta (\ref{eq17a},\ref{eq17}), above equation is written in an explicit form as, 
\begin{eqnarray}
  G=\delta{y_i}\big(\frac{E}{2}y_i+F\epsilon_{ik}{y_k}+Mq_i\big)
+\delta{q_i}\big(\frac{C}{2}q_i-D\epsilon_{ik}q_k\big)-B
\label{eq26}
\end{eqnarray}
Knowing the deformed transformation rules, this relation is now used to find the deformed generators one by one.

{\bf{Translations}}\\
It is obvious from (\ref{eq14.1}) that the time derivatives of the variations
$\delta{y}_i$ and $\delta{q}_i$ are zero. So from (\ref{eq16}) we write $\delta{L}$ as,
\begin{eqnarray}
\delta{L}&=&[\frac{E}{2}\delta{y}_k+F\epsilon_{kl}\delta{y}_l+M\delta{q}_k]\dot{y}_k+[\frac{C}{2}\delta{q}_k
+D\epsilon_{lk}\delta{q}_l]\dot{q}_k\nonumber\\
&&-\frac{1}{2m}[((E^2+F^2)2y_i+Eq_i-2F\epsilon_{ij}q_j)\delta{y}_i+\nonumber\\
&&((A^2+B^2)2q_i+Ey_i-2F\epsilon_{ki}y_k)\delta{q}_i]\nonumber\\
&&=\frac{d}{dt}[\frac{E}{4}\sqrt{-\theta\eta}\epsilon_{ks}y_{k}a_{s}+\frac{D\eta}{2}a_kq_k+\frac{C\eta}{4}\epsilon_{ks}q_ka_s]\label{B}
\end{eqnarray}
Using (\ref{B}) and the phase space transformation rules (\ref{100}, \ref{eq14.1}) in (\ref{eq25})
we get the Translational generator ,
\begin{eqnarray}
G_{Tr}=a_i(E\hat{y}_i+F\epsilon_{ij}\hat{y}_j+A \hat{q}_i+B\epsilon_{ij}\hat{q}_j)\label{eq27}
\end{eqnarray}
%where $p_i$ is given by Eq.(\ref{eq9}).
This result matches exactly with the expression of the translation generator obtained in (\ref{pi}).
\\
{\bf{Rotation}}\\
\vskip 0.05cm
Under rotation the transformations (\ref{eq14.2}) gives the following variation of the lagrangian 
%\pagebreak
\begin{eqnarray*}
\delta{L}&=&[\frac{E}{2}(\delta{y}_k\dot{y}_k+y_k{\delta{\dot {y}}_k})+F\epsilon_{kl}({\delta{\dot{y}}}_ky_l
+y_l\dot{\delta{y}_k})\nonumber\\
&&+\frac{C}{2}(\delta{q}_k\dot{q}_k+q_k{\delta{\dot{q}}_k})+
D\epsilon_{kl}(\delta{q}_k\dot{q}_l+q_k{\delta{\dot{q}}_l})
+M(\delta{q}_k\dot{y}_k+q_k\delta{\dot{y}}_k)]\nonumber\\
&&-\frac{1}{2m}[(E^2+F^2)2y_i+Eq_i-2F\epsilon_{ij}q_j)\delta{y}_i\nonumber\\&&+
((A^2+B^2)2q_i+Ey_i-2F\epsilon_{ki}y_k)\delta{q}_i]\nonumber\\
&=&\frac{d}{dt}(0)\nonumber\\
\end{eqnarray*}

Since $B=0$ for the lagrangian (\ref{eq16}) under rotation we obtain from (\ref{eq26}) and (\ref{eq14.2})
 the desired form of rotational generator 
\begin{equation}
G_{Rot}=\alpha({\frac{E}{2}\epsilon_{ij}\hat{y}_i\hat{y}_j+\frac{C}{2}\epsilon_{ij}\hat{q}_i\hat{q}_j+
\frac{1}{1-\theta\eta}\epsilon_{ij}\hat{y}_i\hat{q}_j+D{\hat{q}_j}^2-F\hat{y}_j^2}) %\quad\quad\quad\textrm{where J is given by Eq.(12)}
\label{eq28}
\end{equation}
Here also the rotational generator is same as (\ref{j}).
\\
{\bf{Boost}}\\
Following similar approach for the lagrangian (\ref{eq16}) in case of Boost symmetry (\ref{eq14.3}) we find,
\begin{eqnarray*}
\delta{L}&=&\frac{d}{dt}[a_iy_i\{-m(M+\frac{\theta}{2}F\}+a_k\epsilon_{ki}y_im(-\frac{E\theta}{4}+\frac{M}{2}
\sqrt{-\theta\eta})\nonumber\\
&&+a_iq_i(-mC)+a_k\epsilon_{ki}q_im(\frac{C}{4}\sqrt{-\theta\eta}-D)]+
\frac{d}{dt}a_i\frac{F\theta}{2}{q_i}t
+a_k\epsilon_{ki}\frac{\sqrt{-\theta\eta}}{2}\frac{E}{2}y_it\nonumber\\
&&+a_k\epsilon_{ki}q_i\frac{C\eta}{4}t]\nonumber
\end{eqnarray*}

Thus we can identify $B$ from the above result and using this in (\ref{eq26}) together with (\ref{eq14.3}) we achieve the expression for the Boost generator 
\begin{eqnarray}
G_{Boost}&=&a_i[m(A\hat{y}_i+B\epsilon_{ij}\hat{y}_j+C\hat{q}_i+D\epsilon_{ij}\hat{q}_j)\nonumber\\&&-
t(E\hat{y}_i+F\epsilon_{ij}\hat{y}_j+A \hat{q}_i+B\epsilon_{ij}\hat{q}_j)\nonumber
%&&=a_i(m\hat{x}_i-t\hat{p}_i)\nonumber
\end{eqnarray}
Thus again the Boost generator is identical with (\ref{eq12}) found from algebraic approach. 
%%%%%%%%%%%%%%%%
%%%%%%%%%%%%%%%%%%%Final conclusion of Chap-2  
\section{Conclusion}
\label{con1}
The present analysis clearly revealed the possibility of obtaining new results from combining two apparently
independent theories into a single effective theory. The essential requirement was that these theories must posses the dual(chiral) aspects of the same symmetry. We have used thus soldering mechanism to abstract 
different quantum mechanical models starting from the basic harmonic oscillator.
It is now clear that there are two approaches to visualize the doublet structure of a composite theory
- one based on lagrangian formulation while the other involves the hamiltonian formulation. 
In the first approach the standard viewpoint is to solder the distinct lagrangians through a  contact term
while in the latter (hamiltonian) case, a canonical transformation is found that diagonalizes the hamiltonian
into independent pieces. Moreover the explicit demonstration was done at the level of equations of motion
with an appropriate change of variables. Specifically the generalised Landau problem with electric and magnetic fields was shown to be composed of such oscillators.\\
It appears therefore, the soldering formalism which fuses the symmetries while canonical transformation that
decomposes the symmetries are complementary to each other. This has been illustrated by
considering a particular symmetry, namely `chirality'. This is known to play a pivotal role in discussing 
different aspects of two dimensional field theories with left movers and right movers.
\\Since all these models show noncommutative features regarding phase space variables 
we have considered a system where not only position variables but also momentum variables are intrinsically noncommutating. This is an important departure from earlier studies in this context where noncommutativity appeared only in position position coordinates. Imposing the Jacobi identities among the various variables we were able to reproduce from general arguments the specific structure of noncommutativity discussed in the context of physical models. We obtained the deformed Galilean generators in an algebraic approach which are compatible with this space. We also constructed a dynamical model invariant under deformed transformation rules of phase space variables. Constraint analysis of this model allowed us to identify the second class constraints which finally lead to the noncommutative Dirac brackets. These brackets are the classical analogues of the noncommutative algebra.  Finally Noether's theorem was applied to this dynamical model to obtain the classical deformed generators. These generators were identical with those found by the previous method. In this way consistency between the algebraic approach and the dynamical approach was established.
%%%%%%%%%%%%%
%%%%%%%%%%%%%%%%%%%%%%%%%%%%%%%%%%%   VECTOR MODEL   %%%%%%%%%%%%%%%%%%
%%%%%%%%%%%%%%%%%%%%%  CHAPTER 3 %%%%%%%%%%%%%%%%%%%%
\chapter{Analysis of Vector Models in Field Theory}
\label{vecmodel}
We have used soldering formalism to  abstract different quantum mechanical models starting from the 
basic harmonic oscillator. Specifically, self and anti-self dual or alternately,
the left and right chiral oscillators combine to yield the model in (\ref{lag1}). It shows that the 
motion of a charged particle in the presence of electric and magnetic fields is simulated by a doublet of 
chiral oscillators, one moving in the clockwise direction and the other in the anti-clockwise direction.
Now we can interpret the results analysed in detail for the quantum mechanical model as a field theory in
(0+1)dimension which is precursor to field theory in (2+1)dimensions. Naturally the prior discussion can be extended to spin-1 vector models in (2+1) dimensions. At one point we will briefly show the extension of the analysis to arbitrary $ (4k-1) $dimensions.

In this context we recall that (\ref{lag1}) had been regarded \cite{DJT} analogous to the 
lagrangian density for three dimensional topologically massive electrodynamics (Maxwell-Chern-Simons theories)
in the Weyl $(A_0=0)$ gauge,
$$
{\cal L} = \frac{1}{2} {\dot{\vec A}}^2 + \frac{\mu}{2}{\dot{\vec A}}\times
{\vec A} - \frac{1}{2}{(\vec\nabla \times \vec A)}^2
\nonumber
$$
However, we should interpret (\ref{lag1}) to be the analogue of the topologically massive electrodynamics augmented by the usual mass term,
\begin{equation}
{\cal L}_{S} = \frac {1}{2}{ A_\mu}{ A^\mu}- \frac {\theta}{2m^2}
\epsilon_ {\mu \nu \sigma}{\partial ^\mu}{ A^\nu}{A^\sigma}
- \frac {1}{4m^2}{ A _{\mu \nu}}{A^{\mu \nu}}
\label{V}
\end{equation}
where $A_{\mu \nu}=\partial_{[\mu}A_{\nu]}$ and in the limit where all spatial derivatives are neglected \cite{dunne}. Correspondingly, (\ref{L}) and (\ref{M}) would be interpreted as the analogues of the self and anti-self dual models \cite{DJ}whose dynamics is governed, respectively, by the following lagrangian densities,

\be
{\cal L}_{SD} = {\cal L} _{-}(g) = \frac{m_{-}}{2} g_{\mu} g^{\mu} - \frac{1}{2} 
\eps_{\mu \nu\lambda} g^{\mu} \pa ^{\nu} g ^{\lambda}
\label{anew}
\ee
\be
{\cal L}_{ASD} = {\cal L} _{+}(f) = \frac{m_{+}}{2} f_{\mu} f^{\mu} + \frac{1}{2}
\eps_{\mu \nu\lambda} f^{\mu} \pa ^{\nu} f ^{\lambda}
\label{bnew}
\ee
once again in the limit where all spatial derivatives are ignored.\\ 
Self dual models in three space time dimensions, have certain distinct features which are 
essentially connected with the presence of the Chern-Simons term which is both metric and gauge independent.
In this section we will make a detailed analysis of a doublet of topologically massive theories
with distinct mass parameters. Precisely there are two approaches to visualize this doublet structure of a composite theory -- one based on the lagrangian formulation while the other involves the hamiltonian formulation. In the first method combining the distinct lagrangians either by soldering or using the equation of motion approach leads to a resultant theory. Here it is a parity violating non-gauge vector
theory with explicit as well as topological mass terms. Also a hamiltonian reduction of the effective theory, based on canonical transformations, is performed. The diagonalization of the hamiltonian in this context reveals the presence of two massive modes, which are a combination of topological and explicit mass parameters.

We have organised this chapter as follows. In Sec.3.1 we shall concentrate on the lagrangian version.
To generalise our view source terms are included in above doublet structure. Sec.3.2 will consider a similar analysis adopting the soldering mechanism. These results are then interpreted in Sec.3.3 in the path integral approach. A brief discussion of the equivalence of self-dual(SD) and Maxwell-Chern-Simons(MCS)model is given in Sec.3.4. 
A complementary viewpoint will be presented in the hamiltonian formulation in Sec.3.5. By solving the constraint, the hamiltonian of the model is expressed in term of a reduced set of variables.
Next, by means of a canonical transformation, the hamiltonian gets decomposed into two distinct pieces, which correspond to the hamiltonians of the Maxwell-Chern-Simons doublet. This technique of using canonical transformations to diagonalise a hamiltonian is of course well known and appears in different versions and different situations as observed in the previous chapter. Sec.3.6 elaborates the diagonalization of the energy-momentum tensor. The spin of the excitations is also calculated. 
The helicity states are $\pm 1$, corresponding to the two modes of the theory. Conclusive remarks are left for the Sec.3.7.
%%%%%%%%%%%%%%%%%%%%%%%%%%%%%%%%%%%
%%%%%%%%%%%%%%%%%%%%%%%%%%%%%%%%%%%
\vskip 0.5cm
\section{Approach based on equations of motion}
\label{eqmot1}

We will use here the expressions for selfdual and antiself-dual lagrangian densities in a slightly
alternate form with the metric $g_{\mu\nu}=(+1 -1 -1)$. Let us consider therefore the following doublet of models,
\be
{\cal L}_{SD} = {\cal L} _{-}(g) = \frac{1}{2} g_{\mu} g^{\mu} - \frac{1}{2m_-}
\eps_{\mu \nu\lambda}g^{\mu}\pa ^{\nu} g^\lambda + \frac{1}{2}{g_\mu}{J^\mu}
\label{2.1}
\ee 
%%%%%%%%%%%
\be
{\cal L}_{ASD} = {\cal L} _{+}(f) = \frac{1}{2} f_{\mu} f^{\mu} + \frac{1}{2m_+} 
\eps_{\mu \nu\lambda} f^{\mu} \pa ^{\nu} f ^{\lambda} + \frac{1}{2}{f_\mu}{J^\mu}
\label{2.2} 
\ee
The different signs in the Chern-Simons piece show that they may be regarded as a set of selfdual and antiself-dual models \cite{TPN,DJ,BW} where inclusion of source terms generalises our view. The property of self (or anti-self) duality follows on exploiting the equations of motion . \\
Now the equations of motion are given by,
 %%%%%%%%%%%%%%%%%%%%%%   EqN of Motion   %%%%%%%%%%%%%%%%%%55
\be
g_{\mu}  - \frac{1}{m_-}\eps_{\mu \nu\lambda}\pa ^{\nu}g ^{\lambda} + \frac{1}{2}J^{\mu} = 0
\label{2.3}
\ee
and
\be
f_{\mu}  + \frac{1}{m_+}\eps_{\mu \nu\lambda}\pa ^{\nu}f^{\lambda} + \frac{1}{2}J^{\mu} = 0
\label{2.4}
\ee
Proceeding as before we introduce a new set of fields which are the analogues of (\ref{2c}),
\be
 g^{\mu} +f^{\mu} = k^\mu ;\quad\quad\quad g^{\mu} -f^{\mu} = h^\mu 
\label{2.4a} 
\ee
Now adding (\ref{2.3}) and (\ref{2.4}) and substituting the old fields by the new one $k_\mu$(say)
leads to
\be
k_{\mu}  - \frac{1}{m_-}\eps_{\mu \nu\lambda}\pa ^{\nu}k ^{\lambda} +
 M\eps_{\mu \nu\lambda}\pa ^{\nu}g ^{\lambda} + J^{\mu} = 0
\label{2.5}
\ee 
where $M=(\frac{1}{m_-} + \frac{1}{m_+}) $.\\
Multiplying (\ref{2.5}) by $\frac{1}{m_+}\eps^{\sigma \rho\mu}\pa_{\rho}$ yields,
\be
\frac{1}{m_+}\eps^{\sigma \rho\mu}\pa_{\rho}{k_\mu} + \frac{1}{{m_-}{m_+}}\pa_{\rho}{k^{\rho\sigma}}
 - \frac{M}{m_+}\pa_{\rho}f^{\rho\sigma} + \frac{1}{m_+}\eps^{\sigma \rho\mu}\pa_{\rho}{J_\mu} = 0
\label{2.6}
\ee
where, $k_{\mu \nu}= \partial_{[\mu}k_{\nu]}$ and  $f_{\mu \nu}= \partial_{[\mu}f_{\nu]}$ are antisymmetric
combinations that may be interpreted as field tensors associated with the respective fields $k_\mu $
 and $g_\mu $. \\
In order to eliminate the $f_\mu$ variable from (\ref{2.6}), we add (\ref{2.5}) to it and then exploit (\ref{2.4}). One finally obtains
%%%%%%%%%
\be
\pa_{\rho}k^{\rho\sigma} - ({m_+} - {m_-}) \eps^{\sigma \rho\lambda}\pa_{\rho}{k_\lambda}  
+ {m_+}{m_-}{k^\sigma} 
=  - {m_+}{m_-}{J^\sigma} - \frac{1}{2}({m_-} - {m_+})\eps^{\sigma \rho\lambda}\pa_{\rho}{J_\lambda}
\label{2.7}
\ee
The expression here is purely in terms of the new variable $k_\mu$.

Following similar steps an equation involving only the $h_\mu$ variable is obtained,
\be
\pa_{\rho}h^{\rho\sigma} -  ({m_+} - {m_-}) \eps^{\sigma \rho\lambda}\pa_{\rho}{h_\lambda}  
+ {m_+}{m_-}{h^\sigma} 
=  - \frac{1}{2}({m_-} + {m_+})\eps^{\sigma\rho\lambda}\pa_{\rho}{J_\lambda}
\label{2.7a}
\ee
%%%%%%%%%%%%%%  factorisation %%%%%%%%%5
In the absence of sources the above equations display a factorisation property analogous to (\ref{4b.1}),
\begin{equation}
[g^{\rho\mu} + \frac{1}{m_+}\epsilon^{\rho\sigma\mu}\partial_{\sigma} ]
[g_{\mu\lambda} - \frac{1}{m_-}\epsilon_{\mu\nu\lambda}\partial ^{\nu} ] X^{\lambda} =0  \quad\quad;
\quad\quad\quad(X^\lambda = k^\lambda,h^\lambda)
\label{4b.2}
\end{equation}
The lagrangian which lead to (\ref{2.7}) and (\ref{2.7a}) are given by,
\begin{eqnarray}
{{\cal L}_k} = - \frac {1}{4}k_{\mu \nu}k^{\mu \nu}  - \frac {1}{2}({m_+} - {m_-})
 \eps_{\mu \nu \sigma}\partial ^{\mu}{ k^\nu}k^{\sigma} + \frac{1}{2}{m_-}{m_+}{ k_\mu}{ k^\mu}
 + {m_-}{m_+}{ k_\mu}{ J^\mu} \cr
+ \frac{1}{2}({m_+} - {m_-})\eps_{\mu \nu \sigma}{k^\mu}\partial ^{\nu}J^{\sigma}
\label{2.8}
\end{eqnarray}
and
\be
{{\cal L}_h}  = - \frac {1}{4}{ h_{\mu \nu}}{h^{\mu \nu}} - \frac {1}{2}({m_+} - {m_-})
 \eps_ {\mu \nu \sigma}\partial ^{\mu} h^{\nu}{h^\sigma}+ \frac{1}{2}{m_-}{m_+}{h_\mu}{h^\mu}
 + \frac{1}{2}({m_+}+{m_-})\eps_{\mu \nu \sigma}h^{\mu}\partial ^{\nu}J^{\sigma}
\label{2.9}
\ee
It is notable that in the absence of the sources both the effective lagrangian in
(\ref{2.8}) and (\ref{2.9}) are identical.
The first term represents the ordinary Maxwell term, the second one involving epsilon 
specifies the Chern-Simons term and the last one is a mass term.
Therefore the effective lagrangian density ${{\cal L}_h} $ or ${\cal L}_k $ gets identified 
with the Maxwell-Chern Simons-Proca(MCSP) model. 
This result was obtained earlier using various approaches ranging from the soldering of actions
\cite{BW,BK1} to path integral methods \cite{BK2} based on master actions. Within the hamiltonian
formalism this was achieved by using the canonical transformations \cite{BG,BK2}. Compared to these,
the present analysis is very economical and follows as a natural extension of the quantum mechanical
analysis presented earlier. Furthermore, although both $ k_\mu $ and  $ h_\mu $ yield the same free
theory, their roles are quite distinct in the presence of interactions as may be evidenced
from the different source contributions appearing in (\ref{2.8}) and (\ref{2.9}) respectively.

The dual composition is succinctly expressed by the following maps:
\be
{\cal L}_{2SD}(f,g) \Longleftrightarrow {\cal L}_{MCSP}(g \pm f)
\label{2.10}
\ee
where the left hand side indicates a doublet comprising self and anti-self dual models (\ref{2.1},\ref{2.2})
while the right hand side depicts the composite model that is expressed either in terms of the $(g+f)$
variable (\ref{2.8}) or the $(g-f)$ variable (\ref{2.9})
Let us now look into the case when ${m_-}={m_+}={m}$ . Then the epsilon term vanishes reducing the expression
for ${\cal L}_k $ or ${\cal L}_h $ to the usual Proca model. Also the source term gets considerably
simplified.
\vskip 0.5cm
%%%%%%%%%
%%%%%%%%%%%%%%%%%%  Method of Soldering %%%%%%%%%%%%%%%%%%
\section{An operator approach- Soldering formalism}
\label{opapp}
 
In this section we will implement the soldering formalism which is based on \cite{MS}. 
The idea is to construct an effective lagrangian that will characterise the doublet as in (\ref{anew}) and (\ref{bnew}). \\
Consider the variation of the lagrangian under the local transformation,
\be
\delta f_{\mu}=\delta g_{\mu}=\Lambda_{\mu}(x)
\label{1}
\ee
The requisite variations are given by,
\be
\delta{\cal L}_{\mp} = J_{\mp}^{\mu}\Lambda_{\mu}
\label{2}
\ee
where the currents are defined as,
\be
J^{\mu}_{\mp} = m_{\mp}h^{\mu} \mp \eps^{\mu\alpha\beta}\pa_{\alpha}h_{\beta}
 \,  ;\,\,\,  h = f,g
\label{3}
\ee
Next we introduce the soldering field $W_\mu$ transforming as,
\be
\delta W_{\mu} = \Lambda _{\mu}
\label{4}
\ee
It is now simple to check that the following lagrangian,
\be
{\cal L}={\cal L}_{-}(g)+{\cal L}_{+}(f) - W_{\mu}(J_+^{\mu}(f) +
J_-^{\mu}(g)) + \frac{1}{2}(m_+ + m_-)W_{\mu}W^{\mu}
\label{511}
\ee
is invariant under the transformations introduced earlier. The field $W_\mu$ plays the role of an auxiliary variable that can be eliminated by using the equation of motion,
\be
W_{\mu} = \frac{1}{m_+ + m_-}(J_{\mu}^+(f) + J_{\mu}^-(g))
\label{6}
\ee
The final theory is manifestly invariant under the transformations
containing only the difference of the original fields. It is given by,
\be
{\cal L} = -\frac{1}{4} F_{\mu\nu}F^{\mu\nu}(A) + \frac{1}{2} \eps_{\mu \nu\lambda}
(m_{-} - m_{+})A^{\mu}\pa^{\nu}A^{\lambda} + \frac{1}{2} m_{+}m_{-}A_{\mu}A^{\mu} 
\label{j11}
\ee   
where,
\be
A_\mu = \frac{1}{\sqrt {m_{+} + m_{-}}} (f_\mu - g_\mu ) 
\label{kvec}
\ee 
This is the Maxwell-Chern-Simons theory with an explicit mass term. A word
about the degree of freedom count might be useful. The lagrangian
(\ref{anew}) and (\ref{bnew}) individually correspond to single massive modes. The
composite model (\ref{j11}) corresponds to two massive modes. There is thus
a matching of the degree of freedom count.
 
It is now possible to take a different variation of the fields, but the final result will be the same. To illustrate this consider, instead of (\ref{1}), the following variations, 
\be
\delta f_{\mu} = \delta g_{\mu} = \eps_ {\mu\alpha\beta} \pa^{\alpha}\Lambda^{\beta}
\label{c11}
\ee
The variations in the individual lagrangian can be written in terms of the parameter $ \Lambda$ as,
\be
\delta{\cal L}_{\mp} = J_{\mp}^{\alpha\beta} \pa_{\alpha}\Lambda_{\beta}
\label{d11}
\ee
where,
\be
J_{\mp}^{\alpha\beta} = m_{\mp}\eps^{\alpha\beta\mu} h_{\mu} \hskip .1 cm {\mp}\hskip .1 cm 
h^{\alpha\beta}; \hskip .25 cm h = f, g 
\label{e11}
\ee
and,
\be
h^{\alpha\beta} = \pa^{\alpha} h^{\beta} - \pa^{\beta} h^{\alpha} 
\label{f11}
\ee
Introducing an antisymmetric tensor field $ B_{\alpha\beta} $ transforming as,
\be
\delta B_{\alpha\beta} = \pa_{\alpha} \Lambda_{\beta} - \pa_{\beta}\Lambda_{\alpha}
\label{g11}
\ee
it is possible to write a modified lagrangian,
\be
{\cal L} = {\cal L}_{SD} + {\cal L}_{ASD} - \frac{1}{2} B^{\alpha\beta}
({J_{\alpha\beta}^{+}(f) + J_{\alpha\beta}^{-}(g)}) + \frac{1}{4}
(m_{+} + m_{-}) B_{\alpha\beta}B^{\alpha\beta} 
\label{h11}
\ee
that is invariant under (\ref{c11}) and (\ref{g11}); i.e. $\delta{\cal L} =  0 $.
Since $B_{\alpha\beta}$ is an auxiliary field it is eliminated from (\ref{h11})
by using its solution. The final effective theory is just (\ref{j11}).

A straightforward extension of the above analysis in $d =4k-1$ dimensions would lead to the soldering of the antiself and self dual lagrangians,
 %%%%%%%%%%  %           D =4K - 1 considered 
%%%%%%%%%%%%%%%%%%%%%%%%%%%%%%%%
\begin{eqnarray}
\!\!\!{\cal L}_{+}&=&\frac{1}{2}\frac{1}{2k!} \epsilon_{{\mu_1}\cdots{\mu_{2k-1}}{\lambda_1}\cdots{\lambda_{2k}}}
f^{\mu_1\cdots\mu_{2k-1}}\partial^{[\lambda_1}f^{\lambda_2\cdots\lambda_{2k}]} 
+  \frac{m_+}{2}f_{\mu_1\cdots\mu_{2k-1}}{f}^{\mu_1\cdots\mu_{2k-1}}
%\label{diff ASDN}
%%%%%%%%%%%%%%%%%%%%%%
\nonumber\\
\!\!\!{\cal L}_{-}& =&- \frac{1}{2}\frac{1}{2k!} \epsilon_{{\mu_1}\cdots{\mu_{2k-1}}{\lambda_1}\cdots{\lambda_{2k}}}
 g^{\mu_1\cdots\mu_{2k-1}}\partial^{[\lambda_1}g^{\lambda_2
\cdots\lambda_{2k}]}  
+ {\frac{m_-}{2}}g_{\mu_1\cdots\mu_{2k-1}}g^{\mu_1\cdots\mu_{2k-1}}
\label{diff SDN}
\end{eqnarray}
to yield the new lagrangian
\begin{eqnarray}
{\cal L}_{S} &=& \frac{(m_{+}m_{-})}{2}A^{\mu_1 \cdots \mu_{2k-1}}A_{\mu_1 \cdots \mu_{2k-1}}
 - \frac{1}{2.2k} F_{ \sigma_1 \cdots \sigma_{2k}}F^{ \sigma_1 \cdots \sigma_{2k}}\nonumber \\
 &+& \frac{(m_{-} - m_{+})}{2}\frac{1}{(2k-1)!}\epsilon ^{\mu_1 \cdots \mu_{2k-1} \sigma_1
\cdots \sigma_{2k}}A_{\mu_1 \cdots \mu_{2k-1}}\partial _{\sigma_1}A_{ \sigma_2 \cdots \sigma_{2k}}
\label{final mscp}
\end{eqnarray}
where ,
$$
A_{\mu_1 \cdots \mu_{2k-1}} = \frac{1}{\sqrt {m_{+} + m_{-}}}(f_{\mu_1 \cdots \mu_{2k-1}} 
- g_{\mu_1 \cdots \mu_{2k-1}} )$$
and,
$$
 F^{\sigma_1 \cdots \sigma_{2k}}=\partial^{[\sigma_1}A^{\sigma_2
\cdots \sigma_{2k}]}
$$
where in the latter expression antisymmetrisation is done with respect to all the indices in the square bracket. Note that the basic variables $(f,g)$ are $ (2k-1)$-form fields. For identical masses , $m_+ = m_-$, the generalised Proca model is obtained.
\vskip 0.75cm
%%%%%%%%%%%%%%%%%%%%%%5
%%55555555555555555 Showing the direct sum %%%%%%%%%%%%%%%
\textbf{\underline{Direct sum of lagrangians}}
\\\\
The above manipulations have shown that it is possible to glue the two
lagrangians by introducing an auxiliary variable. We could adopt this method
to glue any two lagrangians; however the final result would not be local.
The local expression follows precisely because the self and anti-self dual
nature of the lagrangians engage in a cancelling act. Note that the variations
considered here lead to the combination $f_\mu -g_\mu$ in the effective theory.
By considering the variations with opposite signatures we would have been led
to the same effective theory but with the combination $f_\mu +g_\mu$.

As announced earlier we now show how the above approach enables one to 
directly obtain the effective theory by adding the two lagrangians,
\be
{\cal L}={\cal L}_+(f) + {\cal L}_-(g)
\label{7}
\ee
Introducing the combination (\ref{kvec}), we find,
\begin{eqnarray}
{\cal L} &=& {\cal L}_+(\sqrt{m_+ + m_-}A + g) + {\cal L}_-(g)
\nonumber\\
& = &\frac{m_+}{2}(m_+ + m_-)A^{\mu}A_{\mu} + \frac{1}{2}(m_+ +m_-)
g^{\mu}g_{\mu} + \sqrt{m_++m_-}\eps_{\mu \nu \lambda}g^{\mu}\pa^{\nu}
A^{\lambda}  \nonumber\\&+&  m_+\sqrt{m_++m_-}A_{\mu}g^{\mu}  
+ \frac{m_++m_-}{2}\eps_{\mu \nu\lambda}A^{\mu}\pa^{\nu}A^{\lambda}
\label{8}
\end{eqnarray}
Now $g_\mu$ behaves as an auxiliary variable. It is eliminated in favour
of the other variable by using the equation of motion. The end result reproduces (\ref{j11}).
\\\\
%%%%%%%%%%%%%%%%%% Compatibility of eqn of motion %%%%%%%%%%%%%%%%%
\textbf{\underline{Compatibility of equation of motion}}
\\
\\
Let us discuss the compatibility of the equations of motion of the doublet and the effective theory.
From (\ref{anew}) and (\ref{bnew}) the following equation are obtained,
\be
g_{\mu} = \frac{1}{m_{-}} \eps_{\mu\nu\lambda} \pa^{\nu}g^{\lambda} 
\label{a1}
\ee
\be 
\pa_{\beta} g^{\mu\beta} =  m_{-}\eps^{\mu\alpha\beta}\pa_{\alpha}g_{\beta}
\label{a2}
\ee
and,
\be
f_{\mu} = - \frac{1}{m_{+}}\eps_{\mu\nu\lambda}\pa^{\nu}f^{\lambda}\label{a3}
\ee
\be
\pa_{\beta}f^{\mu\beta} = - m_{+}\eps^{\mu\alpha\beta}\pa_{\alpha}f_{\beta}
\label{a4}
\ee
Using the above sets of equations it follows that,
\be
- \pa^{\nu} (f_{\mu\nu} - g_{\mu\nu}) + (m_{-} - m_{+}) \eps_{\mu\nu\lambda}
\pa^{\nu} (f^{\lambda} - g^{\lambda}) + m_{+}m_{-}(f_{\mu} - g_{\mu}) = 0
\label{a5}
\ee
which is just the equation of motion for the effective lagrangian (\ref{j11}) with the identification (\ref{kvec}).
\\\\
%%%%%%%%%%%%%%%%%%%%%%%%%%%FACTORISATION%%%%%%%%%%%%%%%%%%
%%%%%%%%%%%%%%%%%%%
\textbf{\underline{Factorizability property}}
\\
\\
Let us discuss the factorisability property. As noted 
in \cite{PK}\footnote{there is a sign error in this ref} the equation of motion following from (\ref{V}) factorises as,
\begin{equation}
[g^{\mu}_\sigma \mp (\frac{1}{m_\pm}){\epsilon_{\sigma}}^{\lambda
\mu}\partial_\lambda ]
[g^{\rho}_\mu \pm (\frac{1}{m_\mp}){\epsilon_{\mu}}^{\nu \rho}
\partial _\nu ] A_{\rho} =0
\label{factor}
\end{equation}
For identical masses $(m_+ = m_-)$ this reduces to the Proca equation. The 
structure of the factorisation has led to the claim that the massive modes in these models satisfy the self duality condition. That this is not so. Consider, for simplicity, the following generating functional for the Proca lagrangian,
\begin{equation}
 Z_{P}[j,J] = \int DA_{\mu} e^{ -\frac{1}{2}\int
[{\cal L}_P + {\bf A}_\mu j^\mu +\tilde{\bf A}_\mu J^\mu]{d^3}x }
\label{pfn}
\end{equation}
where, the dual has also been introduced,
\begin{equation}
{\tilde A}_\mu = \frac{1}{m} \epsilon_ {\mu \nu \lambda}
\partial^{\nu}{\bf A}^\lambda
\label{dual}
\end{equation}
The result of the Gaussian integration is ,
\begin{equation}
Z_{P}[j,J] = e^{-\frac {1}{2}\int(j_\mu + \frac{1}{m}\epsilon_ {\mu \lambda \sigma}
\partial^{\lambda}J^{\sigma})C^{\mu \nu}(j_\nu + \frac{1}{m}\epsilon_
{\nu \alpha \beta}\partial^{\alpha}J^\beta)}
\label{pfng}
\end{equation}
where,
$$
C^{\mu \nu}(x,y)=\frac{2}{(m^2 + \Box)}[g^{\mu \nu} + \frac{1}{m^2}
\partial^\mu \partial^\nu]\times \delta(x-y)
$$
It is now easy to calculate the relevant correlation functions,
\begin{equation}
< A_\eta(x) A_\xi(y)> = C_{\eta \xi}(x,y)
\label{gfn2}
\end{equation}
\begin{equation}
< A_\eta(x) A_\xi(y)> = <\tilde{A}_\eta(x)\tilde{A}_\xi(y)>
+ \frac{2}{m^2}{\bf g}_{\eta \xi}{\bf \delta} (x-y)
\label{gfn}
\end{equation}
\begin{equation}
 <{A_\eta}(x){\tilde A}_\xi(y)> = \frac{2}{m}. \frac{1}{(m^2+\Box)}
\epsilon_{\eta \sigma \xi }\partial ^\sigma \delta (x-y)
\label{gfn1}
\end{equation}
 It is seen that all the correlation functions cannot be related modulo only local terms. Thus it is not possible to interpret  $A_\mu = \tilde A_\mu $ operatorially. Hence the $A_\mu$ field cannot be regarded 
as self dual.\\
The origin of the structure of the factorisation in (\ref{factor}) is understood from the soldering analysis performed earlier. The two factors correspond to the self dual and anti self dual modes, not in the model 
(\ref{V}), but rather in the models (\ref{anew}) and (\ref{bnew}), respectively. It is the soldering mechanism that has precisely combined these modes from distinct models with fields $f_\mu$ and $g_\mu$ to yield the new model (\ref{V}) with the field $A_\mu = f_\mu - g_\mu$. This new field $A_\mu$ is altogether a separate entity which lacks the original symmetry properties. 

The above manipulations have shown that it is possible to glue the two lagrangian by introducing an auxiliary variable. We could adopt this method to glue any two lagrangian; however the final result would not be local.
The local expression follows precisely because the self and anti-self dual nature of the lagrangian engage in a cancelling act. Note that the variations considered here lead to the combination $f_\mu -g_\mu$ in the effective theory. By considering the variations with opposite signatures we would have been led to the same effective theory but with the combination $f_\mu +g_\mu$.

%%%%%%%%%%%%%%%%%%%%%%%%%%%%5 New method %%%%%%%%%%%%%%%
%%%%%%%%%%%% EQUIVALENCE %%%%%%%%%%%%%%%%%%%%%%%
\section{Equivalence between Selfdual and Maxwell-Chern-Simons model}
\label{equiv}
As we know an important variant of self-dual model is the topologically massive gauge theory where gauge
invariance co-exists with the finite mass, single helicity and parity violating nature of the excitations.
Its dynamics is governed by a lagrangian comprising both the Maxwell and Chern-Simons terms. 
The equations of motion, when expressed in terms of the dual to the field tensor, manifest a self duality.
An equivalent version of this model also exists, where the self duality is revealed in the equations of motion for the basic field %\cite{TPN,DJ,BR}. 
Now the self dual model is known to be equivalent to the Maxwell-Chern-Simons theory \cite{DJ,BR}. 
Consequently the above analysis can be repeated for a doublet of Maxwell-Chern-Simons theories defined by the lagrangian densities,
\be
{\cal L}_{-}(P) = - \frac{1}{4m_{-}}F_{\mu\nu}F^{\mu\nu}(P) + \frac{1}{2}
\eps_{\mu\nu\lambda}P^{\mu}\pa^{\nu}P^{\lambda} 
\label{lequiv} 
\ee
\be
{\cal L}_{+}(Q) = - \frac{1}{4m_{+}}F_{\mu\nu}F^{\mu\nu}(Q) - \frac{1}{2}
\eps_{\mu\nu\lambda}Q^{\mu}\pa^{\nu}Q^{\lambda} 
\label{mequiv}  
\ee
Specifically, the models (\ref{lequiv}) and (\ref{mequiv}) are the analogues of those 
given in (\ref{anew}) and (\ref{bnew}), respectively. For the sake of comparison,
the mass parameters $ m_{\mp} $ are taken to be identical in both cases.  

Now consider the variations of the lagrangian under the following transformations,
\be
\delta P_{\mu} = \delta Q_{\mu} = \Lambda_{\mu} \label{n}
\ee
Then it follows,
\be
\delta{\cal L}_{\mp} = J_{\mu\nu}^{\mp} \pa^{\mu}\Lambda^{\nu} \label{o}
\ee
where,
\be
J_{\mu\nu}^{\mp}(W) = - \frac{1}{m_{\mp}} F_{\mu\nu}(W)  \pm  \eps_{\mu\nu\lambda} 
W^{\lambda};\hskip .25 cm W = P, Q \label{p} 
\ee
Introducing the $B_{\mu\nu}$ field transforming as (\ref{g11}), it is seen that
the following combination,
\be
{\cal L} = {\cal L}_{-}(P) + {\cal L}_{+}(Q) - \frac{1}{2} B_{\mu\nu}
( J_{+}^{\mu\nu} + J_{-}^{\mu\nu} ) - \frac{1}{4} ( \frac{1}{m_{+}} + 
\frac{1}{m_{-}})B_{\mu\nu}B^{\mu\nu} \label{q} 
\ee
is invariant under the relevant transformations (\ref{g11}) and (\ref{n}).
As before, the auxiliary field $B_{\mu\nu}$ is eliminated from (\ref{q}) to yield
the lagrangian (\ref{j11}) in terms of a composite field which is the difference of
the fields in the doublet,
\be
A_\mu = \frac{1}{\sqrt {m_{+} + m_{-}}} (P_\mu - Q_\mu ) \label{r}
\ee
%The other considerations discussed for the self-dual models are all applicable here.

\section{Path Integral Derivation}
\label{pathint}

The above discussion has a natural interpretation in the path integral formalism.
The point is that the analysis related to equations (\ref{7}) and (\ref{8})
shows that it is possible to obtain the effective theory by an addition of the
lagrangian and then identifying an auxiliary variable which is eventually
eliminated. Since the problem is Gaussian it is straightforward to interpret
it in the path integral language. 
The elimination of the auxiliary variable just corresponds to a Gaussian
integration over that variable. 
Let us therefore consider the following generating functional 
{\footnote{Note that the path integral following from the hamiltonian version 
\cite{BR} requires the factor $\delta (f_{0} + \frac{1}{m_{+}} \eps_{ij}\pa_{i}
f_{j} ) \delta (g_{0} - \frac{1}{m_{-}}\eps_{ij}\pa_{i}g_{j}) $ 
in the measure to account for the constraints. Since this is a Gaussian problem the result of the path
integral remains unaltered even if these factors are not included. This is how
we choose to define the basic lagrangian path integral for the self and anti
self dual models.}} 
for the doublet of self and anti-self dual models (\ref{anew}) and (\ref{bnew}),\\
\begin{eqnarray}
{\cal Z } & = & \int df_{\mu} dg_{\mu}\exp i\int d^{3}x [{\cal L}_{-}(g) + {\cal L}_{+}
(f) \nonumber \\
& & + \frac{1}{\sqrt {m_{+} + m_{-}}}(f_{\mu} - g_{\mu}) J^{\mu}]
\end{eqnarray} \label{s}
where a source has been introduced that is coupled to the difference (\ref{kvec}) of
the variables. A relabelling of variables as in (\ref{kvec}) is made for which the 
Jacobian is trivial. 
\newpage
The path integral is now rewritten in terms of the redefined variable $A_{\mu}$ and $g_{\mu}$,
\begin{eqnarray}
{\cal Z } & = & \int dA_{\mu}  dg_{\mu} \exp i \int d^{3}x [ \frac{m_{+}}{2}(\sqrt 
{m_{+} + m_{-}} A_{\mu} + g_{\mu} )^2 \nonumber\\
& & + \frac{1}{2} \eps_{\mu\nu\lambda} (\sqrt {m_{+} + m_{-}} A^{\mu} + g^{\mu} )
\pa^{\nu} (\sqrt {m_{+} + m_{-}}A^{\lambda} + g^{\lambda} ) \nonumber\\ 
& & + \frac{m_{-}}{2} g_{\mu}g^{\mu}- \frac{1}{2} \eps_{\mu\nu\lambda} g^{\mu}
\pa^{\nu}g^{\lambda} + A_{\mu}J^{\mu}]
\label{t}
\end{eqnarray}   
Integrating over the $g_{\mu}$ variable yields,
\begin{eqnarray}
{\cal Z } &  = & \int dA_{\mu} \exp i \int d^{3}x [ - \frac{1}{4} F_{\mu\nu}F^{\mu\nu}
 + \frac{1}{2} (m _{-} - m_ {+} ) \eps_{\mu\nu\lambda}A^{\mu}\pa^{\nu}
A^{\lambda}\nonumber \\
& & + \frac{m_{+}m_{-}}{2} A_{\mu}A^{\mu} + A^{\mu} J_{\mu}]
\label{u}
\end{eqnarray}
In the absence of sources this is just the partition function for the 
Maxwell-Chern-Simons-Proca model (\ref{j11}). Furthermore, the $ A_{\mu} $ 
field in (\ref{u}) is related to the original doublet fields by exactly the same equation
(\ref{kvec}). This shows the equivalence of the results obtained by the two approaches.

It is equally possible to carry out a similar analysis for a doublet of Maxwell-Chern-Simons theories. 
However, a gauge fixing is necessary to account for the gauge
invariance of these theories. As was shown in \cite{BR}, through the use of master 
lagrangians, the basic field in the self dual model can be identified with the basic 
field in the Maxwell-Chern-Simons theory defined in the covariant gauge. We therefore
consider the generating functional obtained from (\ref{lequiv}),(\ref{mequiv});  
\begin{eqnarray} 
{\cal Z } & = & \int dP_{\mu} dQ_{\mu} \delta(\pa_{\mu}P^{\mu}) \delta(\pa_
{\mu}Q^ {\mu}) \exp i \int d^{3}x [{\cal L}_{-}(P) + {\cal L}_{+}(Q) 
\nonumber \\
&  & \hskip 2 cm  + \frac{1}{\sqrt {m_{+} + m_{-}}}(P_{\mu} - Q_{\mu})  J^{\mu}]
\end{eqnarray} \label{v} 
where, as before, a coupling with an external source has been done with the difference
(\ref{r}) of the variables. Because of the gauge invariance of the integrand, 
the source $ J_{\mu} $ 
should be conserved. 

To perform the path integration, a renaming of variables according to (\ref{r}) is done for which
the Jacobian is trivial. 
\pagebreak
Then, 
\begin{eqnarray}
{\cal Z } & = & \int dA_{\mu} dQ_{\mu} \delta(\pa_{\mu}A^{\mu}) \delta(\pa_{\mu}
Q^ {\mu}) \exp i \int d^{3}x [ - \frac{1}{4m_{-}}(m_{+} + m_{-}) F_{\mu\nu}(A)
F^{\mu\nu}(A) \nonumber\\
& & - \frac{1}{4}(\frac{1}{m_{+}} + \frac{1}{m_{-}})F_{\mu\nu}(Q)F^{\mu\nu}(Q)
- \frac{\sqrt{m_{+} + m_{-}}}{2m_{-}}F_{\mu\nu}(A)F^{\mu\nu}(Q) \nonumber\\
& & + \sqrt{m_{+} + m_{-}} \eps_{\mu\nu\lambda} Q^{\mu}\pa^{\nu}A^{\lambda}
 + \frac{1}{2}(m_{+} + m_{-})\eps_{\mu\nu\lambda} A^{\mu}\pa^{\nu}A^{\lambda}
 + A_{\mu}J^{\mu}]
 \label{w} 
\end{eqnarray}
Performing the integral over the $ Q_{\mu}$ variables yields,
\begin{eqnarray}
{\cal Z } & = & \int dA_{\mu} \delta(\pa_{\mu}A^{\mu}) \exp i \int d^{3}x 
[- \frac{1}{4} F_{\mu\nu}(A)F^{\mu\nu}(A) + \frac{1}{2}(m_{+}m_{-}) 
A_{\mu}A^{\mu} \nonumber\\
& & + \frac{1}{2}(m_{-} - m_{+})\eps_{\mu\nu\lambda} A^{\mu}\pa^{\nu}A^{\lambda}
 + A_{\mu}J^{\mu}]
\end{eqnarray}\label{x}
Express the delta function in the measure by an integral over a variable 
$ \alpha$,
\begin{eqnarray}
{\cal Z } & = & \int dA_{\mu} d \alpha \exp i \int d^{3}x [\alpha \pa_{\mu}
A^{\mu} - \frac{1}{4} F_{\mu\nu}(A)F^{\mu\nu}(A) + \frac{1}{2}(m_{+}m_{-})
A_{\mu}A^{\mu} \nonumber\\
& & + \frac{1}{2}(m_{-} - m_{+})\eps_{\mu\nu\lambda} A^{\mu}\pa^{\nu}A^{\lambda}
 + A_{\mu}J^{\mu}]
\end{eqnarray}\label{y}
Introducing a St\"{u}ckelberg transformed field 
$ A_{\mu} \rightarrow A_{\mu} + (m_+m_-)^{-1}\pa_{\mu}\alpha $ 
and using the conservation of the source $(i.e. \pa_{\mu} J^{\mu} = 0)$ it follows that, 
\begin{eqnarray}
{\cal Z } & = & \int dA_{\mu}\exp i \int d^{3}x [- \frac{1}{4} F_{\mu\nu}F^{\mu\nu}
 + \frac{1}{2}(m_{+} m_{-})A_{\mu}A^{\mu}\nonumber\\
& &  + \frac{1}{2}(m_{-} - m_{+})\eps_{\mu\nu\lambda} A^{\mu} \pa^{\nu}A^{\lambda} 
+ A_{\mu}J^{\mu}] \label{z}
\end{eqnarray}
where the integral over $ \alpha$ has been absorbed in the normalisation.

As before, the generating functional for the Maxwell-Chern-Simons theory with an explicit mass term is obtained. The connection of the basic field $A_{\mu}$ with the original doublet,of course, remains the same as in (\ref{r}).

%%%%%%%%%%%%%%%%%%%%%% Hamiltonian Analysis %%%%%%%%%%%%%%%%%%%
\section{Hamiltonian reduction and Canonical Transformations}
\label{canonical}

%%%%%%%%%%5
The results of the previous section were achieved in the lagrangian formulation
by combining the doublet to yield the composite model. A complementary viewpoint
will now be presented in the hamiltonian formulation. By solving the constraint,
the hamiltonian of the model is expressed in term of a reduced set of variables.
Next, by means of a canonical transformation, the hamiltonian gets decomposed
into two distinct pieces, which correspond to the hamiltonians of the Maxwell-
Chern-Simons doublet. Defining a new set of parameters,
\begin{eqnarray}
m_{-} - m_{+} & =  & \theta \nonumber\\
m_{+}m_{-} & = & m^{2} \label{aa} 
\end{eqnarray} 
the lagrangian (\ref{j11}) takes the form,
\be
{\cal L } = - \frac{1}{4} F_{\mu\nu}F^{\mu\nu} + \frac{\theta}{2} \eps_
{\mu\nu\lambda} A^{\mu} \pa^{\nu}A^{\lambda} + \frac{m^{2}}{2} A_{\mu}A^{\mu} 
\label{ab}
\ee 
The canonical momenta are,
\be 
\pi_{i} = \frac{\pa{\cal L}}{\pa \dot{A}^{i}} = - (F_{0i} + \frac{\theta}{2} 
\eps_{ij} A_{j}) \label{ac}
\ee 
while,
\be
\pi_{0} \approx 0 \label{ad} 
\ee 
is the primary constraint. The canonical hamiltonian is given by,
\be
H = \frac{1}{2} \int d^2x [\pi_{i}^{2} + \frac{1}{2}F_{ij}^{2} + (\frac{\theta^
{2}}{4} + m^{2})A_{i}^{2} - \theta  \eps_{ij}A_{i}\pi_{j} + m^{2}A_{0}^{2}]
 + \int d^2x A_{0} \Omega  \label{ae}
\ee  
where,
\be
\Omega = \pa_{i}\pi_{i} - \frac{\theta}{2} \eps_{ij}\pa_{i}A_{j} - m^{2} A_{0}
\approx 0 \label{af} 
\ee 
is the secondary constraint. Eliminating the multiplier $A_{0}$ from (\ref{ae})
by solving the constraint (\ref{af}) we obtain,
\begin{eqnarray}
H & = & \frac{1}{2}\int d^2x [\pi_{i}^{2} + (\frac{1}{2} + \frac{\theta^{2}}
{8m^{2}}) F_{ij}^{2} + \left( \frac{\theta^{2}}{4} + m^{2}\right) A_{i}^{2} - \theta
\eps_{ij}A_{i}\pi_{j}] \nonumber \\ 
& & + \frac{1}{2m^{2}} \int d^2x [( \pa_{i}\pi_{i})^{2} - \theta \pa_{i}\pi_{i}
\eps_{lm}\pa_{l}A_{m}] \label{ag} 
\end{eqnarray} 
Making the canonical transformations in terms of the new canonical pairs
$ (\alpha, \pi_{\alpha}) $ and $ (\beta, \pi_{\beta})$,
\begin{eqnarray}
A_{i} & = & \frac{2m}{\sqrt{4m^{2} + \theta^{2}}} \eps_{ij}\frac{\pa_{j}}{\sqrt
{-\pa^{2}}} (\alpha + \beta) + \frac{1}{2m} \frac{\pa_{i}}{\sqrt{-\pa^{2}}}
(\pi_{\alpha} - \pi_{\beta})\nonumber \\
\pi_{i} & = & - \frac{\sqrt{4m^{2} + \theta^{2}}}{4m} \eps_{ij}\frac{\pa_{j}}
{\sqrt{-\pa^{2}}}(\pi_{\alpha} + \pi_{\beta}) + m \frac{\pa_{i}}{\sqrt{-\pa^
{2}}} (\alpha - \beta) \label{ah}
\end{eqnarray}
the hamiltonian decouples into two independent pieces,
\be
H (A_{i}, \pi_{i}) = H (\alpha, \pi_{\alpha}) + H (\beta, \pi_{\beta})
\ee \label{ai}
where,
\begin{eqnarray}
 H (\alpha, \pi_{\alpha}) & = & \frac{1}{16m^{2}} \sqrt{4m^{2} + \theta^{2}}
(\sqrt{4m^{2} + \theta^{2}} - \theta)\int d^{2}x \pi_{\alpha}^{2} + 
 \frac{\sqrt{4m^{2} + \theta^{2}} + \theta}{\sqrt{4m^{2} + \theta^{2}}} \int 
d^{2}x(\pa_{i}\alpha )^{2} \nonumber\\ 
& & \hskip 1cm  + m^{2}\frac{(\sqrt{4m^{2} + \theta^{2}} - \theta)}{\sqrt
{4m^{2} + \theta^{2}}} \int d^{2}x \alpha^{2} \nonumber \\
H (\beta, \pi_{\beta}) & = & \frac{1}{16m^{2}} \sqrt{4m^{2} + \theta^{2}}
(\sqrt{4m^{2} + \theta^{2}} + \theta)\int d^{2}x \pi_{\beta}^{2} +
 \frac{\sqrt{4m^{2} + \theta^{2}} - \theta}{\sqrt{4m^{2} + \theta^{2}}}
\int d^{2}x(\pa_{i}\beta )^{2} \nonumber \\
& & \hskip 1cm  + m^{2}\frac{(\sqrt{4m^{2} + \theta^{2}} + \theta)}
{\sqrt{4m^{2} + \theta^{2}}} \int d^{2}x \beta^{2} 
\end{eqnarray} \label{aj}
To recast these expressions in a familiar form, a trivial scaling is done,
\begin{eqnarray}
\alpha^{2} &\rightarrow &\frac{1}{2}\frac{\sqrt{4m^{2} + \theta^{2}}}{\sqrt{4m^
{2} + \theta^{2}} + \theta} \alpha^{2},\hskip .2 cm 
{\pi_{\alpha}^{2}}  \rightarrow 2 \frac{\sqrt{4m^{2} + \theta^{2}} + \theta}
{\sqrt{4m^{2} + \theta^{2}}} \pi_{\alpha}^{2} \nonumber\\
\beta^{2} &\rightarrow &\frac{1}{2}\frac{\sqrt{4m^{2} + \theta^{2}}}{\sqrt{4m^
{2} + \theta^{2}} - \theta} \beta^{2},\hskip .2 cm \pi_{\beta}^{2} \rightarrow 
2 \frac{\sqrt{4m^{2} + \theta^{2}} - \theta}{\sqrt{4m^{2} + \theta^{2}}}
\pi_{\beta}^{2} \label{ak}
\end{eqnarray}
so that,
\begin{eqnarray}
H (\alpha, \pi_{\alpha}) & = & \frac{1}{2}\int d^{2}x [(\pa_{i}\alpha)^{2} +
 \pi_{\alpha}^{2} + m_{+}^{2}\alpha^{2}] \nonumber\\
H (\beta, \pi_{\beta}) & = &\frac{1}{2}\int d^{2}x [(\pa_{i}\beta)^{2} +
\pi_{\beta}^{2}+ m_{-}^{2}\beta^{2}] \label{al}
\end{eqnarray}
with, 
\be
m_{\pm} = \sqrt{m^{2} + \frac{\theta^{2}}{4}} \mp  \frac{\theta}{2}\label{am}
\ee 
These relations show that the theory possesses two massive modes with mass
$m_{+}$ and $m_{-}$ which satisfy the Klein Gordon equation.
Furthermore since $ m_{\pm}$ in (\ref{am}) are the solutions to the set 
(\ref{aa}), these can be identified with the corresponding mass parameters 
occurring in the Maxwell-Chern-Simons doublet (\ref{lequiv}) and (\ref{mequiv}).
The above hamiltonians are indeed the reduced expressions obtained
from (\ref{mequiv}) and (\ref{lequiv}), respectively. The canonical reduction of the 
Maxwell-Chern-Simons theory has been done in \cite{DJT} but we present it 
here from our viewpoint for the sake of completeness. Let us, for instance, 
consider the lagrangian (\ref{lequiv})\footnote{The variable P, for convenience,
is now called A}. The multiplier $A_{0}$ enforces the
Gauss constraint,
\be
\Omega = \pa_{i}\pi_{i} - \frac{m_{-}}{2}  \eps_{ij} \pa_{i} A_{j} \approx 0 
\label{an}
\ee 
where $(A_{i}, \pi^{i})$ is a canonical set. The hamiltonian on the constraint
surface is given by,
\be
H = \frac{1}{2} \int d^2x [ \pi_{i}^{2} + \frac{1}{2}F_{ij}^{2} + m_{-} \eps_{ij}\pi_{i} A_{j} + \frac{m_{-}^{2}}{4} A_{i}^{2}] \label{ao}
\ee 
Next, consider the canonical transformation,
\begin{eqnarray}
A_{i} & = & \frac{\pa_{i}}{\sqrt{-\pa^ {2}}} \pi_{\theta} + \eps_{ij}\frac
{\pa_{j}}{\sqrt{-\pa^ {2}}} \beta \nonumber \\
\pi_{i} & = & \frac{\pa_{i}}{\sqrt{-\pa^ {2}}} \theta - \eps_{ij}\frac{\pa_{j}}
{\sqrt{-\pa^ {2}}} \pi_{\beta} \label{ap} 
\end{eqnarray}
where $( \theta, \pi_{\theta})$ and $( \beta, \pi_{\beta})$ form independent
canonical pairs. Since this is a gauge theory, a gauge fixing is imposed.
We take the standard Coulomb gauge,
\be
\pa_{i}A_{i} = 0 \label{aq}
\ee
The presence of the gauge, together with the constraint (\ref{an}),  modifies
the canonical structure of the $(A_{i}, \pi_{i})$ fields; i.e. their brackets
are no longer canonical. The modified algebra can be obtained either by the 
Dirac algorithm \cite{Dirac} or, as done here, by just solving the constraints.
Their solution leads to the following structure,
\begin{eqnarray}
A_{i} & = & \eps_{ij}\frac{\pa_{j}}{\sqrt{-\pa^ {2}}} \beta \nonumber \\
\pi_{i} & = & - \frac{m_{-}}{2} \frac{\pa_{i}}{\sqrt{-\pa^ {2}}} \beta - 
\eps_{ij}\frac{\pa_{j}}{\sqrt{-\pa^ {2}}} \pi_{\beta} \label{ar}
\end{eqnarray}
which satisfies a nontrivial algebra,
\begin{eqnarray}
[ A_{i}(x), \pi_{j}(y)]  &=& i (-\delta_{ij}+\frac{\pa_{i}\pa_{j}}
{\pa^{2}} ) \delta(x-y) \nonumber\\ 
\left [ \pi_i(x),\pi_j(y) \right] &=& -i\frac{m_{-}}{2}\eps_{ij}\delta(x-y)
\label{as}
\end{eqnarray}
%%%%%%%%%%%
The same result follows by replacing the Poisson bracket by the Dirac bracket.
Using (\ref{ar}) the reduced hamiltonian is obtained from (\ref{ao}),
\be
H  =  \frac{1}{2}\int d^{2}x [(\pa_{i}\beta)^{2} + \pi_{\beta}^{2} + 
m_{-}^{2}\beta^{2}] \label{at} 
\ee
which has exactly the same structure as the second relation in (\ref{al}). 
Likewise the other Maxwell-Chern-Simons theory with a coupling $m_{+}$ can 
be reduced to the first relation in (\ref{al}). It might be mentioned
that the two lagrangians (\ref{lequiv}) and (\ref{mequiv}) differ not only in the
respective mass parameters, but also in the signature of the Chern-Simons
term. However a scaling argument shows that, apart from the field dependencies,
these are connected by $m_+ \rightarrow -m_-$. Since the hamiltonian is 
quadratic in the mass term, this sign difference therefore does not affect the result. 

Thus the reduced hamiltonian of the Maxwell-Chern-Simons theory with a mass 
term is the sum of the reduced hamiltonians of a doublet of Maxwell-Chern-
Simons theories with distinct mass parameters $m_{\pm}$. There is a complete
correspondence between the lagrangian and hamiltonian formulations.  
%%%%%%%%%%%

\section{The energy momentum tensor and spin}
\label{eng-moment}

As emphasised in \cite{DJT}, spin in $2 + 1$ dimensions cannot be properly
identified from only the angular momentum operator since it does not conform 
to the conventional algebra. It is essential to consider the complete energy 
momentum tensor. Incidentally, although $\alpha$ and $\beta$ in (\ref{al})
satisfy the Klein-Gordon equation, these cannot be regarded as scalars due
to presence of the factor $ \sqrt{-\pa^{2}}$ in the transformations (\ref{ah}).
A complete analysis of the energy momentum tensor will be done which 
unambiguously determines the spin of the excitations.
The energy momentum tensor following from (\ref{ab}) is given by,
\begin{eqnarray}
\Theta_{\mu\nu} & = & 2 \frac{\pa {\cal L}}{\pa g^{\mu\nu}} - g_{\mu\nu} 
{\cal L} \nonumber \\
 & = & - F_{\mu\alpha}F_{\nu}^{\alpha} + m^{2} A_{\mu}A_{\nu} + \frac{1}{4}
g_{\mu\nu} F_{\alpha\beta}F^{\alpha\beta} - \frac{m^{2}}{2} g_{\mu\nu} 
A_{\alpha}A^{\alpha} \label{au}
\end{eqnarray}
The discussion of the hamiltonian has already been done. The momentum is given by,
\begin{eqnarray}
P_{i} & = & \int d^{2}x \Theta_{0i} \nonumber \\
& = & \int d^{2}x ( - F_{0j}F_{i}^{j} + m^{2} A_{0}A_{i}) \label{av} 
\end{eqnarray}
To pass over to the reduced variables, $A_{0}$ is first eliminated by using
the constraint (\ref{af}). Next, the canonical transformations (\ref{ah})
and (\ref{ak}) are applied. This leads to the diagonal form,
\be
P_{i} = \int d^{2}x [ \pi_{\alpha}\pa_{i}\alpha + \pi_{\beta}\pa_{i}\beta ]
\label{aw}
\ee 
The rotation generator is given by,
\be
M_{ij} = \int d^{2}x [x_{i}\Theta_{0j} - x_{j}\Theta_{0i} ] \label{ax}
\ee
which, following the same techniques, is put in the diagonal form,
\be
M_{ij} = \int d^{2}x [ (x_{i} \pi_{\alpha} \pa_{j}\alpha  -  x_{j}
\pi_{\alpha} \pa_{i}\alpha) + (x_{i} \pi_{\beta} \pa_{j}\beta - x_{j}  
\pi_{\beta} \pa_{i}\beta ) ]\label{ay} 
\ee
Both the translation and rotation generators have their expected forms with
the fields $\alpha$ and $\beta$ transforming normally. Using the inverse
transformation of (\ref{ah}) it is seen that the original field $A_{i}$
also transforms normally,
\begin{eqnarray}
[A_{j}, P_{i}]  & = & i \pa_{i} A_{j} \nonumber \\
\left [A_{k}, M_{ij}\right ] & = & i (x_{i}\pa_{j}A_{k} - x_{j}\pa_{i}A_{k} - \delta
_{ik}A_{j} + \delta_{jk}A_{i}) \label{az} 
%\ee
\end{eqnarray}
Finally, the boosts are considered and it is found that the diagonal form is 
given by,
\begin{eqnarray}
M_{0i} & = & t\int d^2x \Theta_{0i} - \int d^2x  x_{i}\Theta_{00} \nonumber \\  
 & = & t\int  d^2x \pi_{\alpha} \pa_{i}\alpha - \frac{1}{2} \int  d^2x  x_{i} [
(\pa_{j}\alpha)^{2} + \pi_{\alpha}^{2} + m_{+}^{2}\alpha^{2} ] + m_{+}
\eps_{ij} \int  d^2x \pi_{\alpha} (\frac{\pa_{j}}{\pa^{2}}) \alpha  \nonumber\\
&&+ t\int d^2x \pi_{\beta}\pa_{i}\beta - \frac{1}{2} \int  d^2x x_{i} 
[(\pa_{j}\beta)^{2} + \pi_{\beta}^{2} + m_{-}^{2}\beta^{2} ] \nonumber\\
 &&- m_{-}\eps_{ij} \int  d^2x \pi_{\beta}(\frac{\pa_{j}}{\pa^{2}}) \beta 
\label{ba}
\end{eqnarray}
The boost generator has extra factors which clearly show that $\alpha$ and
 $\beta$ do not transform as scalars. These extra pieces are however essential
to correctly reproduce the usual transformation of the  original vector
field $A_{i}$,
\be
[A_{j}, M_{0i}] = i (t \pa_{i}A_{j} - x_{i}\pa_{0}A_{j} + \delta_{ij}A_{0})
\label{bb}
\ee
where recourse has to be taken to the solution of the constraint (\ref{af}) 
to obtain the final structure involving $A_{0}$.

The presence of the abnormal terms in the boost leads to a zero momentum
anomaly in the Poincare algebra,
\be
[ M_{0i}, M_{0j}]  = i( M_{ij} + \eps_{ij} \Delta \label{bc})
\ee
where,
\be
\Delta = \frac{m_{+}^{3}}{4\pi} \left (\int d^{2}x \alpha \right)^{2} + \frac{m_{+}}{4\pi}
\left(\int d^{2}x \pi_{\alpha}\right)^{2} -  \frac{m_{-}^{3}}{4\pi} \left(\int 
d^{2}x \beta \right)^{2} - \frac{m_{-}}{4\pi} \left(\int d^{2}x \pi_{\beta}
\right)^{2}
\ee
Following exactly the same steps as in \cite{DJT} it is possible to remove 
this anomaly, simultaneously fixing the spin of the excitations. Consider
the mode expansions,
\begin{eqnarray}
\alpha(x)& = & \int \frac{d^{2}k}{2\pi\sqrt{2\omega(k)}} [a(k) e^{-ik.x}
 + a^{\dagger}(k)e^{ik.x} ]  \nonumber\\
\beta(x) & = & \int \frac{d^{2}k}{2\pi\sqrt{2\omega(k)}}[b(k) e^{-ik.x}
 + b^{\dagger}(k)e^{ik.x} ] \label{be}
\end{eqnarray}
suitably modified by the phase redefinitions,
\be 
a \rightarrow e^{-i\phi}a, \hskip 1 cm b \rightarrow e^{ i\phi}b \label {bf} 
\ee
where,
\be
\phi  = \tan^{-1}\left(\frac{k_{2}}{k_{1}}\right)
\ee
It leads to the following expressions for the boosts and rotation generator,
\begin{eqnarray}
M_{0i} &=& \frac{i}{2} \int d^2k \omega (k) \vert {a}^\dagger(k)
\stackrel{\leftrightarrow}{\pa}_{i} a(k)\vert + \eps_{ij} \int d^2k
\frac {1}{\omega (k) + m_{+}} k_j {a}^\dagger(k)a(k) \nonumber\\
& & \frac{i}{2} \int d^2k \omega (k) \vert b^\dagger(k) \stackrel{\leftrightarrow }{\pa}_{i} b(k)\vert - \eps_{ij} \int d^2k\frac {1}{\omega (k) + m_{-}} k_j 
{b}^\dagger(k)b(k) \label {bg}
\end{eqnarray}

\begin{eqnarray}
M_{ij} & = &  \eps_{ij}\left(\int d^2k {a}^{\dagger}(k) \frac{1}{i} \frac{\pa}
{{\pa }{\phi}} a(k) - \int d^2k  {a}^{\dagger} (k)a(k)\right) \nonumber\\
&  &  +  \eps_{ij}\left(\int d^2k  b^{\dagger}(k) \frac{1}{i} \frac{\pa}{\pa 
\phi} b(k) + \int d^2k  b^{\dagger}(k)b(k) \right) \label {bg1}
\end{eqnarray} 
which satisfy the Poincare algebra 
\be
[M_{0i}, M_{0j}] = i M_{ij} \label {bh}
\ee
%%%%%%%%%%%%%%%%%%%%%
An inspection of the rotation generator shows that it comprises of two distinct terms denoted by the parentheses. The first factor in each corresponds to the usual orbital part. The additional pieces show that the spin of the excitations associated with $\alpha$ and $\beta$ are, respectively, $-1$ and $+1$. 
This also happens in the case of the Maxwell-Chern-Simons theory \cite{DJT}.
The difference from the spin of the excitations in the Maxwell-Chern-Simons theory is noteworthy.
There the sign of the spin is fixed by the sign of the coefficient of the Chern-Simons parameter. 
In the present case it is seen from (\ref{am}) that, irrespective of the sign of $\theta$, the mass parameters $m_{\pm}$ are always positive. Hence the sign of the spin associated with $\alpha$ and $\beta$ is also uniquely determined. 

Note that for $m_{+} = m_{-}$, the theory becomes parity conserving. This is the case when the Maxwell-Chern-Simons doublet with identical mass yields the Proca model \cite{D,BW}. 
%%%%%%%%%%%%%%%%%%%%%%%%%%%%%%%%%%%%5
\section{Conclusion}
\label{con2}

We have considered the description of a doublet of self dual models with distinct
topological mass parameters, having opposite signs. The difference in sign 
implies that the doublet comprises a self dual and an anti-self dual model. 
Specifically, this was a pair of the gauge invariant Maxwell-Chern- theory
or, equivalently, its dual gauge variant version . 
The effective theory, characterising such a doublet, turned out to be the Maxwell-Chern-Simons 
theory with an explicit mass term, referred as the Maxwell-Chern-Simons-Proca model. The basic field of the effective theory was just the difference of the doublet variables.

A canonical analysis of the effective theory was done. Based on a set of 
canonical transformations, the   was diagonalised into two separate
pieces. The two massive modes were found to be a combination of the topological
and explicit mass parameters. In hamiltonian fact these were identified with the two modes 
of the Maxwell-Chern-Simons doublet that led to the effective theory. In this 
way a correspondence was established between the  lagrangian approach of combining 
the doublet into an effective theory and the hamiltonian approach of decomposing 
the  latter back into the doublet. The spin of the excitations was obtained from
a complete study of the Poincare algebra by adopting the method advocated in
\cite{DJT}.  

When the Maxwell-Chern-Simons doublet has identical topological mass $\pm m$,
parity is conserved since one degree of freedom is just mapped to the other.
The spin carried by the two degrees of freedom is $\mp 1$. This has the same 
kinematical structure as the Proca theory which is a parity conserving theory 
with two massive modes having spin $\mp 1$.  An explicit demonstration of this was 
provided earlier \cite{D,BW}. This result is reproduced here by putting $m_{+} = m_{-}$.
 
For the more general case where the Maxwell-Chern-Simons doublet has different
topological masses $ m_{\pm}$, parity is no longer conserved, although 
the other considerations remain valid. Hence the kinematics of such a doublet
resembles a non gauge parity violating theory with two massive modes having
spin $\mp 1$. This turned out to be the Maxwell-Chern-Simons theory with an 
explicit mass term, as elaborated here in details.
%%%%%%%%%%%%%%%%%%%%%55
%%%%%%%%%%%%%%%%%%%%%
%%%%%%%%%%%%%%%%%
%%%%%%%%%%%%%%%%%%%   chapter 4
\chapter{Coupling with Higher Order Chern-Simons term}
\label{TCS}

The abelian Chern-Simons(CS) topological action represented by, \\
$I_{CS}=\frac{\theta}{2}\int d^3x(\eps^{\mnl}f_{\mu}\pr_{\nu}f_{\lambda})$ has been studied extensively in different aspects in 2+1 dimensions. Recently the higher derivative extensions of the CS action, specially the leading third derivative order (TCS) has been considered. For abelian vector fields the action $I_{TCS}$ can be given by $I_{TCS}=\frac{1}{2m}\int d^3x \eps^{\mnl} \Box f_{\mu}\pr_{\nu}f_{\lambda}$. Such a term, even if it is not present originally , will be introduced as a result of the fermionic loop integration. It is an intriguing term that remains gauge invariant, parity odd but no longer topological like the usual CS term, because of the metric dependence in the additional covariant derivative factor \cite{DJ1}. Many interesting observations come out in coupling this third order Chern-Simons(TCS) term to either pure Maxwell term or usual CS term or to both of these terms \cite{DJ1}. 

The study of higher derivative models poses problems that are usually not encountered in normal cases. Special 
techniques are necessary. In this part we follow an alternative view. 
Instead of directly analysing the higher derivative Chern-Simons model we provide their connection with some
familiar models such as the Maxwell-Chern-Simons, the Proca, or the Maxwell-Chern-Simons-Proca models, which contain quadratic derivative terms at most in the action. 

This chapter is organised as follows.
In Sec.~4.1 we have calculated the polarisation vectors of the higher derivative models considered here. It has been shown that the form of these polarisation vectors are coming identical with that of  conventional models. This analysis has been done by the lagrangian formulation \cite{RBT}. 

In Sec.~4.2 we adopt the hamiltonian analysis by considering a particular model. In this section the problem accounted in quantising these higher derivative models is illustrated \cite{LSZ,GT}.

In Sec.~4.3 we consider the same particular model. Here we have discussed about the gauge transformation property of the system; using the Wigner's Little Group \cite{W,W1,HKS}. This group is shown to act as a gauge generator. 

Conclusive remarks left for the Sec.~4.4

Here metric convention is $g^{\mu\nu}=(+,-,-)$; and $\eps^{012}=+1=\eps_{012}$.
%%%%%%%%%%%%%%%%%%%%%%%%%%%%%%%%%%%%%
%%%%%%%%%%%%%%%%%%%%%%%%%%%%%%%%%%%%%%% 
%%%%%%%%%%%%%%%%%%%%%%%%%%555
\section{Analysis of Polarisation vector}

To start with, let us first consider the pure Maxwell term coupled with the 
third derivative CS term, giving the lagrangian,
\beq
{\cal{L}}_{MTCS} = -\frac{1}{4}F_{\mu\nu}F^{\mu\nu} + \frac{1}{2\theta}
\eps_{\mu\nu\lambda} \Box f^{\mu}\pr^{\nu}f^{\lambda}
\label{a}
\eeq
where the field strength is defined as, 
$$
F_{\mu\nu} = \pr_{\mu}f_{\nu} - \pr_{\nu}f_{\mu}
\nonumber
$$
and $\theta $ has the dimensions of mass. The equation of motion obtained from (\ref{a}) is,
\beq
\pr_{\mu}F^{\mu\nu} + \frac{1}{\theta}\eps^{\nu\lam\sg} \Box \pr_{\lam}f_{\sg}= 0
\label{b}
\eeq
Substituting the solution for the negative energy component in terms of the polarisation vector $\eta_\nu$,
\beq
f_{\mu}(k) = \eta_{\mu}(k)e^{ik.x}
\label{c}
\eeq
we get,
\beq
\eta^{\mu} = \frac{1}{k^2}k^{\mu}(k.\eta) - \frac{i}{\theta}\eps^{\mu\nu\lam}
k_{\nu}\eta_{\lam}
\label{d}
\eeq
Two cases are now possible; $k^2=0$ for massless modes and $k^2\neq0$ for massive modes. Consider first the massless case. We can choose the momentum $k^\mu$ propagating along the second direction so that,
\beq
k^{\mu} = (1\,\, 0\,\, 1)^T ; k_{\mu} = (1\,\, 0\,\, -1)^T
\label{e}
\eeq
Replacing $k^\mu$ and $\eta^{\mu}= (\eta^{0} \eta^{1} \eta^{2})^T$ in 
(\ref{d}), we get
\beq
k.\eta = 0
\nonumber
\eeq
So that $\eta^\mu$ has the form, $\eta^{\mu}=(\eta^{2} \eta^{1}\eta^{2})^T$.
Now using the gauge invariance of the model (\ref{a}), $\eta^{\mu}$ may be further reduced to 
\beq
\eta^{\mu} = (0\,\, a\,\, 0)^T
\label{f}
\eeq
where 'a' is some arbitrary parameter. So solution in (\ref{f}) is valid, can be easily verified from 
(\ref{d}). This shows that the model in (\ref{a}) has one massless excitation \cite{ DJ1}.

Now we investigate for the massive case i.e $k^2\ne0$, where we are allowed to go to a rest frame. 
The momentum can be chosen as $k^{\mu }=(\Lambda,0,0)$. Express $\eta^\mu$ in the rest frame as
$$\eta^{\mu}(0) = (\eta^{0}(0),\eta^{1}(0),\eta^2( 0))^T\nonumber$$
Using these structures for $k^\mu$ and $\eta^\mu$ in (\ref{d}), we get
\begin{eqnarray}
\eta^{1}(0) = -i\frac{\Lambda}{\theta}\eta^{2}(0)\nonumber
\eta^{2}(0)=  i\frac{\Lambda}{\theta}\eta^{1}(0)
\label{g}
\end{eqnarray}
Mutual consistency of the above equations yields,
\beq
\Lambda = |\theta|
\label{h}
\eeq
On the other hand, using the gauge invariance of the model, $\eta^{0}(0)$ can be set equal to zero. 
Thus in the rest frame the required polarisation vector is,
$$
\eta^{\mu}(0) = ( 0 ,\eta^{1}(0), i\frac{|\theta|}{\theta}\eta^{1}(0) )^T
\nonumber
$$
modulo a normalisation factor. This can be fixed from the condition,
$$
\eta^{*\mu}(0)\eta_{\mu}(0) = -1
\nonumber
$$
so that $\eta^\mu$ finally takes the form,
\beq
\eta^{\mu}(0) = \frac{1}{\sqrt2}( 0\,\, 1\,\, i\frac{|\theta|}{\theta} )^T
\label{i}
\eeq
Note that $\eta^\mu$ naturally satisfies the transversality condition $k.\eta = 0$.

We can now compare these results with those in the Maxwell case(for the massless mode) and in the Maxwell-Chern-Simons(MCS) case(for the massive mode). First recall that the Maxwell lagrangian in D=2+1 dimensions,
\beq
{\cal L}_{M}=-\frac{1}{4}F^{\mn}F_{\mn}
\label{j1}
\eeq
yields a single massless mode with a polarisation vector identical to (\ref{f}).
So the massless excitation in (\ref{a}) has similar characteristics of the massless mode of the Maxwell part present in the ${\cal L}_{MTCS}$.

Similarly the MCS lagrangian,
\beq
{\cal L}_{MCS} = -\frac{1}{4}F^{\mn}F_{\mn} - \frac{\theta}{4}\eps^{\mnl}f_{\mu}F_{\nu\lambda}
\label{j2}
\eeq
gives a single massive mode. The polarisation vector in the rest frame is given by \cite{RBT,Gir},
\beq
\eta^{\mu}(0) = \frac{1}{\sqrt2}( 0\,\,  1\,\,  i\frac{|\theta|}{\theta} )^T
\label{j3}
\eeq
which is identical structure to (\ref{i}). Thus the massless(massive) mode of MTCS model behaves, as far as the structure of polarisation vector is concerned, exactly like the corresponding modes in the Maxwell(MCS) theory.

Next we will consider the model containing both CS and TCS terms. The lagrangian is given by,
\beq
{\cal L}_{CSTCS} = \frac{1}{4}\eps^{\mnl} \Box f_{\mu}F_{\nu\lam} +
\frac{m^2}{4}\eps^{\mnl}f_{\mu}F_{\nu\lam}
\label{k}
\eeq
where 'm' has the dimension of mass. The equation of motion following from (\ref{k})is,
\beq
\eps^{\mnl}(\Box + m^2)F_{\nu\lam} = 0
\label{k1}
\eeq
Again substitution of (\ref{c}) in (\ref{k1}) yields,
\beq
\eps^{\mnl}(k^2 - m^2)(k_{\nu}\eta_{\lam} - k_{\lam}\eta_{\nu}) = 0
\label{k2}
\eeq
once again there can be two possibilities, massless modes for $k^2 =0$ and massive modes for $k^2\ne0$. Following the same procedure we get no physical massless excitation. So the case may be omitted. Now for the massive modes, let us take the rest frame configuration i.e $k^\mu = (\Lambda,0,0)^T$.
Now using the equation of motion (\ref{k2}) and the gauge invariance of the model
the polarisation vector comes out as,
\beq
\eta^{\mu} = (0\,\, \alpha\,\, \beta)^T
\label{k3}
\eeq
where $\alpha,\beta$ are some arbitrary parameters. Also the mass mode $\Lambda$
is found from the equation of motion as,
$$
\Lambda = |m|
\nonumber
$$
Therefore the model has a spectrum containing two modes of mass m. Such a spectrum has a close analogy 
with that of the Proca model.

The last model to be considered comprising all three terms, the Maxwell, the CS and the TCS. The lagrangian takes the form as,
\beq
{\cal L} = -\frac{1}{4}F^{\mn}F_{\mn} + \frac{\theta}{4}\eps^{\mnl}f_{\mu}
F_{\nu\lam} + \frac{1}{4m}\eps^{\mnl} \Box f_{\mu}F_{\nu\lam}
\label{l}
\eeq
where $\theta$ and m are the distinct mass parameters. The equation of motion 
following from (\ref{l}) as 
\beq
\pr_\mu F^{\mn} + \frac{\theta}{2}\eps^{\nu\alpha\lambda} F_{\alpha\lambda}
+ \frac{1}{2m}\eps^{\nu\alpha\beta} \Box  F_{\alpha\beta} = 0
\label{l1}
\eeq
As before substituting the gauge field $f_\mu$ defined in (\ref{c}) in Eq.(\ref{l1}) we get,
\beq
\eta^\nu = \frac{1}{k^2}[ k^\nu(k.\eta) + i(\theta - \frac{k^2}{m})\eps^{\nu\alpha\beta}k_{\alpha}\eta_{\beta} ]
\label{l2}
\eeq
Now following identical steps as before one finds that only the massive modes are to be accounted. 
In the rest  frame configuration $k^\mu =(\Lambda,0,0)^T$ the polarisation vector comes out as,
\beq
\eta^{\mu} = \frac{1}{\sqrt2}(0\,\,\,1\,\,\,{\mp}i\frac{|\Lambda|}{\Lambda})^T
\label{l3}
\eeq
Thus $\eta^\mu$ is not only complex but also shows a dual characterisation.
Furthermore, mutual consistency of relations akin to (\ref{g}) show that the mass of these modes are given by,
\beq
\Lambda^2 = \frac{m^2}{2}\left[\, (1+\frac{2\theta}{m}) \pm 
\sqrt{1+\frac{4\theta}{m}} \,\, \right]
\label{l4}
\eeq
Eq.(\ref{l4}) indicates four distinct possibilities for $\Lambda$. 
$$
\Lambda_1 = \frac{m}{2}\left[\,1+\sqrt{1+\frac{4\theta}{m}}\,\,\right]$$
$$\Lambda_2 = \frac{m}{2}\left[\,1-\sqrt{1+\frac{4\theta}{m}}\,\,\right]$$
$$\Lambda_3 = -\frac{m}{2}\left[\,1+\sqrt{1+\frac{4\theta}{m}}\,\,\right]$$
$$\Lambda_4 = -\frac{m}{2}\left[\,1-\sqrt{1+\frac{4\theta}{m}}\,\,\right]$$

Out of these four possibilities, depending upon the sign of $\Lambda$
two will lead to the result 
\beq
\eta^\mu = \frac{1}{\sqrt2}( 0\,\, 1\,\, +{i})^T
\label{l5}
\eeq
and the other two will give just the complex conjugate of (\ref{l5}), i.e
\beq
{\eta^\mu}^* = \frac{1}{\sqrt2}( 0\,\, 1\,\, -{i})^T
\label{l6}
\eeq

Thus only two distinct types of massive polarisations are possible. These findings are quite close to that obtained in the Maxwell-Chern Simons-Proca model. The lagrangian of M-CS-P can be given as,
\beq
{\cal L}_{MCSP} = -\frac{1}{4}F^{\mn}F_{\mn} + \frac{\theta}{4}\eps^{\mnl}
f_{\mu}\pr_{\nu}f_{\lam} + \frac{m^2}{2}f^{\mu}f_{\mu}
\label{m}
\eeq
where the mass modes are \cite{BK2},
\beq
\Lambda_{\pm} = \sqrt{\frac{\theta^2}{4} + m^2} \pm \frac{\theta}{2}
\label{m1}
\eeq
The polarisations for these massive modes in rest frame are exactly identical
\cite{RBT} to (\ref{l5}) and (\ref{l6}).
%%%%%%%%%%%%%%%%%%%%%%%%%%%%%%%%%%%%%%%%%5
%%%%%%%%%%%%%%%%%%%5
\section{Hamiltonian analysis}

So far we have considered the lagrangian formalism. Now we can discuss the hamiltonian structure of such higher derivative models. Due to the presence of second order time derivative either the momenta or hamiltonian will be non-trivial. So to get a better understanding let us consider only the Extended MCS model(i.e MTCS model) for simplicity.

Let us recall the lagrangian ${\cal L}_{MTCS}$ in (\ref{a}). Now following Ostrogradski formalism for higher-order lagrangian \cite{LSZ,GT},  presence of the second order derivative term leads to the following canonical momenta,
\begin{eqnarray}
p_{0} = \frac{\pr {\cal L}}{\pr {\dot{f_0}} } - \frac{d}{dt}\frac{\pr {\cal L}} {\pr {\ddot{f_0}}} = -\frac{1}{2\theta}\eps_{ij}\pr_{i}\dot{f_j}
\label{4a}
\end{eqnarray}
\begin{eqnarray}
p_{i} &=& \frac{\pr {\cal L}}{\pr \dot {f^i}} - \frac{d}{dt}\frac{\pr {\cal L}}
{\pr \ddot{f^i} }\nonumber \\
 &=& - F_{0i} - \frac{1}{2\theta}\eps_{ij}\Box f_{j} +\frac{1}{2\theta}\eps_{ij}\pr_{j}\dot{f_0} 
\label{4c}
\end{eqnarray}
\begin{eqnarray}
{\tilde p_{0}} = \frac{\pr {\cal L}}{\pr \ddot {f^0} }
= \frac{1}{2\theta}\eps_{ij}\pr_{i}f_{j}
\label{4b}
\end{eqnarray}
\begin{eqnarray}
{\tilde p_{i} } = \frac{\pr {\cal L}}{\pr\ddot{f^i}}
= \frac{1}{2\theta}\eps_{ij}F_{0j}
\label{4d}
\end{eqnarray}
where the canonically conjugate pair can be identified as [${f_{\mu},p^{\mu}}$]
and [${\dot {f_\mu},\tilde {p^\mu} }$], thus total twelve phase space variables span the space. Now we can easily identify the three primary constraints,
\beq
\Omega_{0} = p_{0} + \frac{1}{2\theta}\eps_{ij}\pr_{i}\dot{f_{j}}
\label{6a}
\eeq
\beq
\Omega_{i} = \tilde{p_i} - \frac{1}{2\theta}\eps_{ij}F_{0j}\,\,\,\, for \,\,i=1,2
\label{6b}
\eeq
\beq
\Omega_{3} = \tilde{p_0} - \frac{1}{2\theta}\eps_{ij}\pr_{i}f_{j}
\label{6c}
\eeq
The canonical hamiltonian takes the form,,
\begin{eqnarray}
{\cal H} &=& \dot{f_\mu}p^{\mu} + \ddot{f_\mu}\tilde{p^\mu} - {\cal L} \nonumber\\ 
         &=& \dot{f_0}p_{0} + {2\theta}p_{i}\eps_{ik}\tilde{p_k} - 
            p_{k}\pr_{k}f_0 - {2\theta^2}{\tilde p_{i}}^2 -
            \tilde{p_i}{\nabla}^2f_i \nonumber \\ &+& \frac{1}{4}(F_{ij})^2 
+ \frac{1}{2\theta}\eps_{ij}{\nabla}^{2}f_{0}\pr_{i}f_{j}
\label{5}
\end{eqnarray}
Now considering the above hamiltonian, the time conservation of the primary constraints lead to the secondary constraints,
\beq
\Omega_{4} = p_{0} + \pr_{i}\tilde{p_i} 
       = \Omega_{0} + \pr_{i}\Omega_{i}
\label{6d}
\eeq
\beq
\Omega_{5} = \pr_{k}p_{k} + \frac{\theta}{2}\eps_{ij}{\nabla^2}\pr_{i}f_{j}
\label{6e}
\eeq

Therefore above total set of constraints (\ref{6b})-(\ref{6e}) ( including primary as well 
as secondary) denote the independent set of constraints out of which we can    
identify $\Omega_i$'s as only second class and others are the first class 
constraints. With the help of these let us find how the canonical brackets 
are changing. Due to the constrained nature of the system the Poisson brackets will be replaced 
by Dirac brackets \cite{Dirac}
$$
[X,Y]_{D} = \{X,Y\} - \{X,\Omega_{i}\}{C_{ij}}^{-1}\{\Omega_{j},Y\}$$
where ,$C_{ij} =[\Omega_{i},\Omega_{j}]_{PB}=\frac{1}{\theta}\eps_{ij}\delta(x,y)$. For this case $C_{ij}^{-1}=-\theta\eps_{ij}\delta(x,y)$ with $i,j=1,2$.

Now we mention only the non-trivial D'brackets,
$$
[f_{i}(x),p_{j}(y)]_{D} = -\delta_{ij}\delta(x,y)$$
$$
[\dot{f_{i}}(x),\dot{f_j}(y)]_{D} = -\theta\eps_{ij}\delta(x,y)$$
$$
[p_{0}(x),\tilde{p_i}]_{D} = -\frac{1}{4\theta}\eps_{ij}\pr_{j}\delta(x,y)$$
$$
[\til{p_i},\til{p_j}]_{D} = -\frac{1}{4\theta}\eps_{ij}\delta(x,y)$$
\beq
[\dot{f_i},\til{p_j}] = -\frac{1}{2}\delta_{ij}\delta(x,y)
\label{7b}
\eeq

Using these brackets one can easily show that it correctly reproduce the Euler-Lagrange equation of motion and also the relevant constraints from the Hamilton equation of motion. So the equivalence between the lagrangian and hamiltonian formalism is satisfied. It might be observed that the algebra of $\dot {f_i}$ is 
identical to the algebra of the basic field in the usual self dual model,
$$
{\cal L}_{SD} = \frac{1}{2}f^{\mu}f_{\mu} - \frac{1}{2\theta}\eps^{\mnl}
f_{\mu}\pr_{\nu}f_{\lambda}
\nonumber
$$
%%%%%%%%%%%%%%%%%%%%%%%%%%%%%
%%%%%%%%%%%%%%%%%%%%5555
\section{Wigner's Little Group}

Next we will discuss briefly the gauge symmetries of the model concerned. It is clear from the model (\ref{a}), that the extended CS term involves metric dependence, so that it becomes non-topological. Again it has been  
already shown that there exist number of first class constraints which implies this is a gauge model. 
So the extended MCS model stands for a higher derivative massive gauge theory. Now  let us try to find out the exact gauge variation in the model using the concept of Wigner's little group. Wigner's little group E(2), which is a subgroup of the Lorentz group $SO(1,3)$ preserves the four momentum invariant, but the polarisation vector $\eta^\mu$ undergoes the gauge transformation,
$$
\eta^\mu(k) \rightarrow \eta^{'\mu}(k)= \eta^{\mu}(k)+f(k)k^\mu
$$
where $f(k)$ can be identified as the gauge parameter. It has been shown earlier \cite{W1,HKS} that translation like generators of Wigner's little group E(2) can act as generators of gauge transformation in the pure Maxwell theory in (3+1) dimensions where only massless quanta is admitted. It is shown \cite{RBT} that such little group generator can generate gauge transformations in topologically massive gauge theories
like $B{\wedge}F$ theory in (3+1)dimensions as well as in Maxwell Chern-simons theory which is topological and permits topologically  massive quanta in (2+1) dimensions. So our natural question arises whether this same little group can provide the gauge transformations regarding the non-topological extended MCS model where both massless and massive modes are present.

Let us first recall  the polarisation vector for massive excitations. So the rest frame configuration is available, i.e $k^\mu = (|\theta|\,\,0\,\,0)^T$. In particular, for simplicity let us take $\theta>0$.
Following little group representations as in \cite{RBT} it is now straightforward to show that, $\eta^\mu$ undergoes the transformation,
\begin{eqnarray}
\eta^{\mu^\prime} &=& W^{\mu}_{\nu}\eta^\nu  
= \frac{1}{\sqrt2}\left(\begin{array}{ccc} {1}&{\alpha}&{-i\alpha}\\
{0}&{1}&{0}\\ {0}&{0}&{1}  \end{array}\right) \left(\begin{array}{c}
{0}\\{1}\\{i\frac{|\theta|}{\theta} } \end{array} \right) \nonumber \\
&=& \frac{1}{\sqrt2} \left( \begin{array}{c} {\al + \al\frac{|\th|}{\th} }\\
{1}\\ {i\frac{|\th|}{\th} } \end{array} \right) \nonumber \\
&=& \frac{1}{\sqrt2}\left( \begin{array}{c} {0}\\{1}\\{i\frac{|\th}{\th}}
\end{array}\right) + (\frac{\al}{|\th|} +\frac{\al}{\th}) \left(\begin{array}{c}
{|\th|}\\  {0}\\ {0} \end{array} \right)
\nonumber\\
&=& \eta^{\mu} + \frac{2\al}{\theta}k^{\mu}
\label{w1}
\end{eqnarray}

Therefore we get  the gauge transformation on the polarisation vector of the massive extended MCS quanta in its rest frame \cite{W1} where $\al$ is the gauge parameter. It is interesting to note that the same little group representation given by $W^{\mu}_{\nu}$ can generate gauge transformations in both MCS and extended 
MCS cases(provided $\th$ is positive). This is due to the fact that the expression of the polarisation vector for the two cases are similar. 

On the other hand if we consider the polarisation vector $\eta^{\mu}(k)$ (since
rest frame is not available) of the massless mode of the MTCS model and successively operate $\eta^{\mu}(k)$ by the relevant little group $W^{\mu}_{\nu}$ which plays the same role of a gauge generator of gauge transformation in pure Maxwell theory it can be easily seen that gauge transformation obtained  exactly matches with that of Maxwell case. Reason is same as for the massive case mentioned earlier.

\section {Conclusion}
\label{con3}

We have considered the third derivative extension of the topological Chern-Simons term either coupled with the Maxwell term(MTCS) or the Chern-Simons term (CS-TCS) itself or with both of these terms(MTCS-CS). It is shown via lagrangian analysis, that the three models mentioned above, reveal some distinct similarities with more familiar models e.g the Maxwell-Chern-Simons(MCS), the Proca, or the Maxwell-Chern-Simons-Proca (which contain quadratic term at most). It was observed in the MTCS model, which has both massive and massless modes, the structure of polarisation vector for the massive modes is exactly identical to that of the MCS case (and similar to the Maxwell case for the massles mode).  Since the structure of the polarisation vectors is known to yield the spin of the various modes in the usual models \cite{RBT,Gir}, the above mapping between MCS and MTCS cases enables us to specify the helicities of the two massive modes in the latter theory to be $ \pm1 $
\\
Next we discussed the hamiltonian formulation. We have considered only the MTCS model for convenience. Due to the presence of third order time derivative it becomes very non-trivial to get the canonical pairs
specially the momenta. So we adopted the Ostrogradski formalism for higher order lagrangian and successively constructed the momentum as well as the hamiltonian. Here we illustrated the constrained feature of the model and computed the Dirac brackets as well.  
\\
Finally we have investigated the gauge transformation property of that previously mentioned model(i.e MTCS) using the Wigner's Little Group. Due to the presence of mass term there which was neither truly Proca like mass nor a topological one, the origin of gauge invariance was not clear. We studied this problem based on an approach \cite{RBT,HKS} using Wigner's little group. It is observed that the identical representation of the Wigner's Little group which acts as a gauge generator for the MCS case(which allows only the topologically massive quanta) is also able to induce gauge transformation for the higher derivative MTCS case (which contains non-topological massive quanta). This is because of the identical structure of the polarisation vectors.
%%%%%%%%%%%%%%%%%%
%%%%%%%%%%%%%%%%%           Chap-5 Gravity
%%%%%%%%%%%%%%%%%%%%%%%%%%%%%
\chapter{Extrapolation to Gravity}
\label{gravity}

Duality is a fascinating symmetry which keeps appearing in many contexts. It was originally developed for electromagnetism, where duality invariance of the Maxwell equations leads to the introduction of magnetic
sources and the quantization of electric charge. It also has been at the origin of many remarkable developments in Yang-Mills theory. More recently duality has revolutionised the understanding of string theory by providing non-perturbative insight. These latter developments indicate that duality should play a central role in gravitational theory as well. Three dimensional higher derivative theories of gravity have received
considerable attention over the years. The first example of such a higher derivative theory is topologically  massive gravity (TMG)model. The TMG lagrangian consists of the usual Einstein-Hilbert term, which by itself
does not describe any degrees of freedom in three dimensions, a Lorentz Chern-Simons (LCS)term which is parity odd and third order in the derivatives. The two terms together describe a single massive state of helicity 
$ +2 $ or $ -2 $, depending on the relative sign between the EH and LCS terms. Last couple of years 
there has been intense activity on this subject of  higher spin theories in different dimensions and their dual formations.
 
In this part of our work we will consider first order self dual model suggested in \cite{AK} which is the helicity +2 analogue of the helicity +1 self dual model in D=2+1.
Drawing analogy with the spin-1 vector models, here 
we shall implement the notions developed in the previous chapters to illuminate different features of such
rank two tensor models which arise in linearised approximation of gravity models. 
As in Chap. 3 where we illustrated the combination of a doublet of self-dual models with distinct masses and spins $\pm1$ to yield an effective Maxwell-Chern-Simons-Proca model, here we will consider the combination of such SD doublet of helicity $\pm2$ tensor gravity models. Due to the involved algebra we have opted only the method based on the equation of motions technique which is a completely lagrangian formalism.

In Sec.5.1 we will investigate the self dual massive spin-2 tensor model including the source term with identical mass parameters but with opposite sign (signifies opposite spin). We will discuss the relevant factorisation properties. Sec.(5.2) revisits this similar calculation but with the distinct mass parameters.  For sake of simplicity we will drop the source terms in this respective analysis. Section ends with the conclusion.
%%%%%%%%%%%%%%%%%%%%%%%55
%%%%%%%%%%%%%%%%%%%%%%%%%%%%%%%%%%%%%%%%%%%%%%%%%%%%%%%%%%%%%%%%%%%%
\section{Spin-2 self dual tensor model with same mass}

Following similar procedure, let us start with the action of first order self dual massive spin 2 model as suggested in \cite{AK}
\be
S = \int{d^3}x\lbrack \frac{m}{2}\eps^{\mu\nu\lambda}{f_\mu}\,^{\alpha}\pa_{\nu}f_{\lambda\alpha} +
\frac{m^2}{2}(f^2 - f_{\mu\nu}f^{\nu\mu}) \rbrack
\label{3.1a}
\ee
where $f = \eta^{\mu\nu}f_{\mu\nu}$. The metric is flat :$\eta^{\mu\nu}$= diag (-,+,+). 
In this model we use second rank tensor fields, like $ f_{\alpha\beta}$ with no symmetry with respect to
their indices. Replacing $m$  by $-m$ in (\ref{3.1a}) implies helicity change from $+2$ to $-2$. 
The first term in (\ref{3.1a}) is reminiscent of a spin one topological Chern-Simons term which will be
called a Chern-Simons term of first order.
The second term in (\ref{3.1a}) is the Fierz- Pauli(FP) mass term \cite{F} which is the spin two analogue 
of a spin one Proca mass term. Note that FP term breaks the local invariance of the Chern-Simons term.
The above first order self dual massive spin-2 field action can be easily found after writing topologically 
massive gravity in an intrinsically geometric form language and then linearizing it \cite{AK1}.
 
Let us then consider the following doublet of first order lagrangian densities in presence of 
source terms ,

\be
{\cal{L}}(f) =   \frac{m}{2}\eps^{\mu\nu\lambda}{f_\mu}\,^{\alpha}\pa_{\nu}f_{\lambda\alpha} +
\frac{m^2}{2}(f^2 - f_{\mu\nu}f^{\nu\mu}) -\frac{m}{2}f^{\mu\nu}J_{\nu\mu}
\label{3.2}
\ee
\be
{\cal{L}}(g) =  - \frac{m}{2}\eps^{\mu\nu\lambda}{g_\mu}\,^{\alpha}\pa_{\nu}g_{\lambda\alpha} +
\frac{m^2}{2}(g^2 - g_{\mu\nu}g^{\nu\mu}) + \frac{m}{2}g^{\mu\nu}J_{\nu\mu}
\label{3.3}
\ee
where $f^{\mu\nu}$ and $g^{\mu\nu}$ are distinct fields. Note that although the helicites are $\pm2$,
the mass term is identical for both lagrangian (\ref{3.2}) and (\ref{3.3}). The case of different masses
will be dealt in the next section. Now the equations of motion are given by,
%%%%%%%%%%%%%%%%%% Equation of Motion %%%%%%%%%%%%%%%%%
\be
{\eps_{\mu}}\,^{\nu\lambda}\pa_{\nu}f_{\lambda\alpha}  + m\,(f\eta_{\mu\alpha} - f_{\alpha\mu})= \frac{1}{2}J_{\alpha\mu}
\label{3.4}
\ee
\be
{\eps_{\mu}}\,^{\nu\lambda}\pa_{\nu}g_{\lambda\alpha}  - m\,(g\eta_{\mu\alpha} - g_{\alpha\mu})= \frac{1}{2}J_{\alpha\mu}
\label{3.5}
\ee
%%%%%%%%%%%%%% combination of Fields %%%%%%%%%%%%%%%%%%%%%555
Following our previous approach, let us introduce new fields $F$ and $G$ as,
\begin{eqnarray}
F_{\mu\alpha} = f_{\mu\alpha} + g_{\mu\alpha}; \quad\quad G_{\mu\alpha} = f_{\mu\alpha} - g_{\mu\alpha};\nonumber\\
F =\eta^{\mu\alpha}F_{\mu\alpha}= f+g;\quad\quad G=\eta^{\mu\alpha}G_{\mu\alpha}=f-g
\label{3.5a}
\end{eqnarray}
Now adding (\ref{3.4}) and (\ref{3.5}) and substituting old fields by new ones defined in
(\ref{3.5a}) leads to
\be 
{\eps_{\mu}}\,^{\nu\lambda}\pa_{\nu}F_{\lambda\alpha}  - m\,( G_{\alpha\mu} -\eta_{\alpha\mu}G) = J_{\alpha\mu}
\label{3.6}
\ee
Our motivation now is to express the above equation solely in terms of the G-field. To achieve this
we abstract certain results from (\ref{3.4}).

$\bullet$ 
Contraction by $\eta^{\alpha\mu}$ of (\ref{3.4}) yields
\be
{\eps^{\alpha\nu\lambda}}\pa_{\nu}f_{\lambda\alpha} + 2mf =\frac{1}{2}J;\quad\quad\quad 
J=\eta^{\mu\alpha}J_{\mu\alpha}
\label{3.7}
\ee

$\bullet$
 Contraction by $\eps^{\mu\alpha\rho}$ of (\ref{3.4}) leads to 
\be
  -\pa_{\alpha}f^{\rho\alpha} + \pa^{\rho}f -m\,{\eps^{\mu\alpha\rho}}f_{\alpha\mu}=
\,\frac{1}{2}{\eps^{\mu\alpha\rho}}J_{\alpha\mu }
\label{3.8}
\ee

$\bullet$ 
Operating (\ref{3.4}) by $\pa^\mu$ on both sides gives
\be
 -m\,(\pa^{\mu}f_{\alpha\mu} - \pa_{\alpha}f) = \frac{1}{2} \pa^{\mu}J_{\alpha\mu}
\label{3.9}
\ee

Taking the difference of (\ref{3.4}) and (\ref{3.5}) and exploiting (\ref{3.5a}) leads to 
\be
{\eps_{\mu}}\,^{\nu\lambda}\pa_{\nu}G_{\lambda\alpha}  - m\,( F_{\alpha\mu}  -\eta_{\alpha\mu}F) = 0
\label{3.10}
\ee
Taking the trace yields the identity,
\be
 F \,=\,-\,\frac{1}{2m}{\eps^{\mu\nu\lambda}}\pa_{\nu}G_{\lambda\mu} 
\label{3.11}
\ee
Using (\ref{3.11}) in (\ref{3.10}) we obtain 
\be
 F_{\alpha\mu} \,=\, \frac{1}{m}\eps_{\mu}\,^{\nu\lambda}\pa_{\nu}G_{\lambda\alpha} \,-\,
\frac{1}{2m}\eta_{\alpha\mu}[\eps^{\rho\sigma\omega}\pa_{\sigma}G_{\omega\rho}]
\label{3.12}
\ee

Now substituting $F_{\alpha\mu}$ in (\ref{3.6}) gives,
\be
  \frac{1}{m}{\eps_{\mu}}\,^{\nu\lambda}\pa_{\nu} [{\eps_\alpha}\,^{\rho\sigma}\pa_{\rho}G_{\sigma\lambda}\,
-\,\frac{1}{2} \eta_{\lambda\alpha}\eps^{\rho\sigma\omega}\pa_{\sigma}G_{\omega\rho}]
\,-\,m[ G_{\alpha\mu}  -\eta_{\alpha\mu}G] = J_{\alpha\mu }
\label{3.13}
\ee
Combining (\ref{3.7}) and  (\ref{3.8}) we obtain,
\be
\eps^{\mu\alpha}\,_{\rho}f_{\alpha\mu} =\frac{1}{2m^2}\pa^{\mu}J_{\rho\mu }-
\frac{1}{2m}\eps^{\mu\alpha}\,_{\rho}J_{\alpha\mu}
\label{3.13a}
\ee
The corresponding equation for $g$ follows by replacing $m$ by $-m$, 
\be
\eps^{\mu\alpha}\,_{\rho}g_{\alpha\mu} =\frac{1}{2m^2}\pa^{\mu}J_{\rho\mu }+
\frac{1}{2m}\eps^{\mu\alpha}\,_{\rho}J_{\alpha\mu}
\label{3.13b}
\ee
Subtracting (\ref{3.13b}) from (\ref{3.13a}) yields
\be
\eps^{\mu\alpha\rho}G_{\alpha\mu}=-\frac{1}{m}\eps^{\mu\alpha\rho}J_{\alpha\mu}
\label{3.13c}
\ee
Therefore from (\ref{3.13c}) we can conclude,
\be
\eps^{\mu\alpha\rho} \pa_{\rho} G_{\alpha\mu}=-\frac{1}{m}\eps^{\mu\alpha\rho} \pa_{\rho}J_{\alpha\mu}
\label{3.14}
\ee
\be
G_{\alpha\mu} - G_{\mu\alpha} = -\frac{1}{m}(J_{\alpha\mu} - J_{\mu\alpha} )
\label{3.15}
\ee
Substituting (\ref{3.14}) in (\ref{3.13}) we get 

\be
{\eps_{\mu}}\,^{\nu\lambda}\pa_{\nu} [{\eps_\alpha}\,^{\rho\sigma}\pa_{\rho}G_{\sigma\lambda}]-
{m^2}[ G_{\alpha\mu}  -\eta_{\alpha\mu}G] = -\frac{1}{2m}{\eps_{\mu\nu\alpha}}\pa^{\nu}
 [\eps^{\rho\beta\omega}\pa_{\beta}J_{\omega\rho}] + m J_{\alpha\mu}
\label{3.16}
\ee
The symmetrised version of the above equation reads,
\begin{eqnarray}
{\eps_{\mu}}\,^{\nu\lambda}\pa_{\nu} [{\eps_\alpha}\,^{\rho\sigma}\pa_{\rho} G_{\sigma\lambda}]
+{\eps_{\alpha}}\,^{\nu\lambda}\pa_{\nu} [{\eps_\mu}\,^{\rho\sigma}\pa_{\rho}G_{\sigma\lambda}]
-{m^2}[ G_{\alpha\mu}+G_{\mu\alpha}] \cr
 + 2m^{2}\eta_{\alpha\mu}G= m (J_{\alpha\mu}+J_{\mu\alpha})
\label{3.16a}
\end{eqnarray}
Exploiting (\ref{3.15}) we obtain the final effective equation of motion,
\be
 \frac{1}{2}{\eps_{\mu}}\,^{\nu\lambda}\pa_{\nu} [{\eps_\alpha}\,^{\rho\sigma}\pa_{\rho}(G_{\sigma\lambda}+
G_{\lambda\sigma})] - {m^2}[ G_{\alpha\mu}  -\eta_{\alpha\mu}G] = m J_{\mu\alpha}
\label{3.17}
\ee
Let us now discuss, in the absence of sources, the factorisability of the equations of motion. Some
conditions on the tensor field are necessary to achieve this factorisation. It is known \cite{DM1}
from a study of the equations of motion of (\ref{3.1a}) that the tensor field $f_{\mu\nu} $ satisfies
 $a)$ tracelessness $ f^{\mu }_{\mu}=0$, $b)$ transversality $ \pa^{\mu }f_{\mu\nu}=0$ and $c)$ symmetricity
$f_{\mu\nu} = f_{\nu\mu}$. Consequently the composite fields $F_{\mu\nu},G_{\mu\nu} $ in (\ref{3.5a})
should also satisfy these properties. Indeed one may also verify this directly from (\ref{3.17}), in the absence
of sources. Under these conditions (\ref{3.17}) factorises as,
\be
(-\eps^{\mu\rho\beta}\pa_{\rho} + m \eta^{\mu\beta})(\eps_{\mu}\,^{\nu\lambda}\pa_{\nu} - m \eta^{\lambda}_{\mu})
G_{\lambda\alpha}=0
\label{4b.3}
\ee
We observe that the above equation of motion (\ref{3.17}) corresponds to an effective theory whose action is given by
\be
S = \int{d^3}x\lbrack \frac{1}{4}G.d\Omega(G)   + \frac{m^2}{2}(G^2 - G_{\mu\nu}G^{\nu\mu})-\frac{1}{2}m
G_{\mu\nu}J^{\nu\mu}  \rbrack
\label{3.18}
\ee 
where 
$$ G.d\Omega(G) = G^{\mu\alpha}{\eps_{\mu}}\,^{\nu\lambda}\pa_{\nu} [{\eps_\alpha}\,^{\rho\sigma}\pa_{\rho}
(G_{\sigma\lambda}+G_{\lambda\sigma})]\nonumber
$$
\\
Note that the first term in the action (\ref{3.18}) stands for the quadratic Einstein-Hilbert term
while the second one is the Pauli-Fierz mass term applicable for spin 2 particle. In the absence
of source this action corresponds to an effective theory which is the analogue of Proca
model for spin-1 case in vector theory. 

Proceeding in a likewise manner the equation of motion for the $F$-field emerges as, 

\begin{eqnarray}
 \frac{1}{2}{\eps_{\mu}}\,^{\nu\lambda}\pa_{\nu} [{\eps_\alpha}\,^{\rho\sigma}\pa_{\rho}(F_{\sigma\lambda}+
F_{\lambda\sigma})] - {m^2}[ F_{\mu\alpha}  -\eta_{\alpha\mu}F] = 
m [\eps_{\alpha\mu\rho}\pa_{\omega} J^{\rho\omega} + \cr
\eps_{\mu\nu\lambda}\pa^{\nu} J^{\lambda}\,_{\alpha}+\eps_{\alpha\nu\lambda}\pa^{\nu} J^{\lambda}\,_{\mu}]
\label{3.20}
\end{eqnarray}

In the absence of sources this is just a replica of (\ref{3.17}). Thus, as happened for the vector
model, either combination $F$ or $G$ yields an effective theory which has the Einstein-Hilbert term
and the F-P term with differences cropping in the source terms.

The analogue of the map (\ref{2.10}) is now written for the spin $2$ example,

\be
{\cal L}_{2SD}(f,g) \Longleftrightarrow {\cal L}_{EHFP}(f \pm g)
\label{3.20a}
\ee
Here the doublet of self dual models on the left hand side is given by (\ref{3.2},\ref{3.3}) while the composite
Einstein-Hilbert Pauli-Fierz model is defined in (\ref{3.18}).

\section{Tensor fields with distinct mass}

In this section we repeat the analysis for the doublet (\ref{3.2}) and(\ref{3.3})
but with distinct mass parameters. To avoid technical complications we drop the source terms. 
We show that combining this doublet yields an effective theory that has an E-H term, a FP mass term
and a generalised first order CS term. This CS term contains, apart from the standard form
given in  (\ref{3.1a}), two other similar terms with a different orientation of indices.
Consider therefore the lagrangian densities, 

\be
{\cal{L_+}}(f) =   \frac{m_1}{2}\eps^{\mu\nu\lambda}{f_\mu}\,^{\alpha}\pa_{\nu}f_{\lambda\alpha} +
\frac{{m_1}^2}{2}(f^2 - f_{\mu\nu}f^{\nu\mu}) 
\label{3.21}
\ee
\be
{\cal{L_-}}(g) =  - \frac{m_2}{2}\eps^{\mu\nu\lambda}{g_\mu}\,^{\alpha}\pa_{\nu}g_{\lambda\alpha} +
\frac{{m_2}^2}{2}(g^2 - g_{\mu\nu}g^{\nu\mu}) 
\label{3.22}
\ee
Now (\ref{3.21}) and (\ref{3.22}) yield the equations of motion,
\be
{\eps_{\mu}}\,^{\nu\lambda}\pa_{\nu}f_{\lambda\alpha}  + {m_1}\,(f\eta_{\mu\alpha} - f_{\alpha\mu})= 0
\label{3.23}
\ee
\be
{\eps_{\mu}}\,^{\nu\lambda}\pa_{\nu}g_{\lambda\alpha}  - {m_2}\,(g\eta_{\mu\alpha} - g_{\alpha\mu})= 0
\label{3.24}
\ee
Following identical field definitions as (\ref{3.5a}) and analogous steps, 
it can be shown that the final form of the equation of motion for $ G_{\mu\nu} $ is given by,

\begin{eqnarray}
- \frac{1}{2}{\eps_{\mu}}\,^{\nu\lambda} \pa_{\nu} [{\eps_\alpha}\,^{\rho\sigma}\pa_{\rho}(G_{\sigma\lambda}+
G_{\lambda\sigma})] - {m_1}{m_2}[ G_{\alpha\mu}  -\eta_{\alpha\mu}G]  + \cr 
\frac{1}{2}({m_1}-{m_2})({\eps_{\mu}}\,^{\nu\lambda}\pa_{\nu}G_{\lambda\alpha }  
 + {\eps_{\alpha}}\,^{\nu\lambda}\pa_{\nu}G_{\lambda\mu}) =0
\label{3.25}
\end{eqnarray}
A similar equation of motion is also obtained for the other variable $F_{\mu\nu} $. 

The action from which the above equation of motion (\ref{3.25}) follows is given by,

\begin{eqnarray}
S = \int{d^3}x\lbrack \frac{{m_1}{m_2}}{2}(G^2 - G_{\mu\nu}G^{\nu\mu})+
\frac{1}{4}G.d\Omega(G)   
+\frac{1}{8}({m_2}-{m_1}) \cr
\{{\eps_{\mu}}\,^{\nu\lambda}G^{\mu\alpha}\pa_{\nu}G_{\lambda\alpha } 
+ {\eps_{\mu}}\,^{\nu\lambda}G^{\alpha\mu}\pa_{\nu}G_{\alpha\lambda}  
+ \,2 {\eps_{\mu}}\,^{\nu\lambda}G^{\alpha\mu}\pa_{\nu}G_{\lambda\alpha } \} ]
\label{3.26}
\end{eqnarray}

We have thus successfully combined different mass terms in the spin 2 case to yield
the action (\ref{3.26}) of the effective theory.
While the first two terms are the usual FP and E-H terms the last piece, which is a consequence
of different masses, is a generalised form of the CS term. As announced earlier it has, apart
from the usual structure, two other pieces that may be obtained from a reorientation of indices. 
In fact it has all possible orientations of indices leading to a first order Chern-Simons term.
Furthermore if we impose a condition of symmetricity $G^{\mu\alpha}=G^{\alpha\mu} $, then all
pieces become identical and the standard first order C-S term with a coefficient $\frac{1}{2}({m_2}-{m_1})$
is obtained.
The first term in (\ref{3.26}) is the Fierz-Pauli(FP) mass term with mass co-efficient $m=\sqrt{{m_1}{m_2}}$.
The second term involves the usual kinetic term (defined in the previous section) which is
equivalent to linearised Einstein-Hilbert(EH) term upto quadratic order.
Thus the action (\ref{3.26}) for spin-2 particle may be interpreted as an analogue 
of Maxwell-CS-Proca model for spin-1 particle. Incidentally the C-S term for the vector case
has a unique orientation of indices $\eps_{\mu \nu\lambda}f^{\mu}\pa ^{\nu} f ^\lambda$ and any
changes are absorbed in a trivial normalisation of signs.

%%%%%%%%%%%%%%%%%%%%%%%%%%%%%%%%%%%%%%% (General Conclusion) Chap-6%%%%%%%%%%
\chapter{Conclusions}
\label{con4}
%%%%%%%%%%%%%%%%%%
The present analysis depicts the important role of symmetry in understanding various models in odd dimensions.
The dual nature of symmetry manifested in (left-right) chirality or (anti)self duality was responsible for the properties of the final theory. We have observed that 
the quantum mechanical example served as the bedrock from where the more involved examples of field theory
and gravity were studied. More specifically, the similarity in the structures of the 
quantum mechanical model and the other models in field theory/gravity naturally suggested
this possibility of dual composition.
We have also discussed the factorisability of equations of motion of different models. Such a phenomenon
illuminates the dual composition of the models. Specially in case of gravity, this factorisation is 
possible subject to certain conditions following from the equation of motion.
\\

In Chapter.2 we started with the basic oscillator model in two dimensions which was shown 
to be composed of two chiral oscillators moving in opposite directions. Chirality gets hidden in an ordinary
two dimensional oscillator since the opposing effects of chirality in its constituent pieces are cancelled.
There were two approaches to visualize this doublet structures of a composite theory--one based on the lagrangian formulation and the other is canonical formalism or the hamiltonian analysis. The first approach
was through soldering mechanism which was demonstrated in Sec.(2.2). In this method the distinct lagrangians were combined through a contact term.  In Sec.(2.3) we considered topological quantum mechanical model -- such an example was the generalised Landau problem with steady electric and magnetic fields. 
Initially we considered a particle model describing  motion in steady magnetic field but in the absence of electric field. Implementation of the soldering method on the doublet of such model produced a bi-dimensional harmonic oscillator. Expectedly the initial chiral symmetry of the primary models got hidden. Subsequently in Sec.(2.3.1) a hamiltonian analysis was performed. Using canonical transformation the hamiltonian was diagonalized  into independent pieces corresponding to the individual hamiltonian of chiral models. Thus these two results actually were complementary with each other. 
\\

The soldering formalism eventually becomes technically involved requiring arcane field redefinitions.
 So before going into the intricacy
of the analysis we have introduced a method in Sec.(2.3.2) based on equations of motion necessitating very simple field redefinitions and  generic to a wide variety of models. We have considered a variant of the above quantum mechanical model with distinct frequencies and illustrated the features of this process in detail. We have also discussed the factorisability property of the final equation of motion emphasising the chiral nature latent in it. Either of these above techniques led to a new composite model without having symmetry of the basic primary models. 

It is now well known that the measurement of space time coordinates at small scale involves unavoidable
effects of quantum gravity. This effect can be incorporated in a physical theory by making the space time coordinates non-commutative. We have discussed such effects in Sec.(2.4).  The representations of Galilean generators were constructed  on a space where both position and momentum coordinates were non-commutative operators. A simple dynamical model invariant under non commutative(NC) phase space transformations was constructed. Analysing the model via Dirac brackets reproduced the original NC algebra. Also the generators
in terms of NC phase space variables were abstracted in a consistent manner. Finally the role of Jacobi
identities was emphasised to produce the noncommutating structure that usually occurs when an electron is subjected to a constant magnetic field and Berry curvature. 
\\

It is well known that  models in  quantum mechanics can could be interpreted as a field theory in (0+1)dimension and also the result would serve as precursor to genuine field theory in (2+1)dimension.
Chapter.3  provides the complete correspondence between those quantum mechanical models and self dual
field theoretical models in odd dimensions. The analysis with respect to quantum models in (0+1)dimension was directly extended to (2+1)dimensional vector field theory. It was observed that all the results and interpretations found in the quantum mechanical examples had the exact analogues in the corresponding field theory. In Sec.(3.1) we studied self(anti-self)dual  doublet which were analogues  of left(right)chiral oscillators. Here the topological mass parameters were taken to be distinct and with opposite sign $\pm $ signifying spin $ \pm 1 $ for vector fields.
For generalisation of the model the source term was also included. Following our trodden path we  approached the analysis first by the equation of motion technique. Similar field redefinitions and specific algebraic steps
led to the final equation of motion corresponding to the effective form of Maxwell-Chern Simons-Proca model.
Self dual or anti selfdual symmetry which got hidden in the M-CS-Proca model actually did manifest in the final factored form of the equation of motion. It was reassuring to note that the Proca model was reproduced in the case where the masses were identical. This finding was reported earlier in different article. 
\\

The factorisation of equation of motion of this final composite M-CS-Proca model revealed two distinct mass modes with two degrees of freedom as known earlier. On the other hand computations of the correlation functions with respect to vector fields in the final model manifested the contrast between the true selfdual nature of fields of the basic models and of the M-CS-Proca model. We concluded the lagrangian formulations by implementing the soldering mechanism in Sec.(3.2).

So far we  discussed about the self dual (anti selfdual) models. But the equivalence between the SD and MCS model is well known. We have reviewed this feature in Sec.(3.3) first by implementing the soldering method. We have illustrated briefly how a MCS doublet with distinct mass parameters culminated in  the required M-CS-Proca theory. Since the result was similar to that obtained via SD doublet so from this perspective an equivalence is established.
Alternately in Sec.(3.4) this equivalence was interpreted from path integral approach. A generating functional was constructed for the doublet of SD and ASD models yielding M-CS-Proca theory. Similar analysis was then carried out for the doublet of MCS model. 
\\
In Sec.(3.5) we have investigated in details the hamiltonian form  of this M-CS-Proca vector model. Illustrations of the constraints and the Dirac brackets of the phase space coordinates had been explored. 
On the other hand a suitable canonical mapping enabled us to decouple the composite hamiltonian into its constituent pieces $ H_\pm $. In Sec.(3.6) an elaborate study of Energy-Momentum Tensor ensured the the spin components of the vector fields in the final model to be $ \pm 1 $ depending on the sign of the mass term.  

%%%%%%%%%%%%%%%%%%%% chap 4%%%%%%%
So far we discussed the first order abelian topological Chern-Simons (CS)term. But the coupling of higher derivative order (here third order derivative)of CS term with either usual Maxwell term or ordinary CS term or with both of these terms suggests interesting features. We have investigated in Chapter.4 such models in (2+1)dimensions. The polarisation vectors in these models unveiled an identical structure with the corresponding expressions for usual models which contain at most quadratic structures. We also studied hamiltonian structure
of these models and revealed how Wigner's Little group acted as a gauge generator. 

%%%%%% chap 6 %%%%%%%%%%55
In various chapters we exploited dual descriptions where a particular theory is interpreted as a combination 
or a doublet of theories. A typical illustration is the Proca model in (2+1) dimensions. The two massive modes of this model were known to be obtained from a doublet of self-dual models with helicity $ \pm 1 $. 
In Chapter.5  we exploited  similar notions and concepts to study a new version of topologically massive gravity in (2+1)dimensions. The propositions made for spin-1 vector models in (2+1)dimensions were extended to this part. We considered first order self dual massive spin-2 tensor model in terms of tensor fields like $ f_{\alpha\beta} $ 
with no symmetry with respect to their indices. We investigated specially the combination of a doublet of 
spin $ \pm 2 $ models that arise in linearised gravity. To avoid repetition we have chosen only the method based on equations of motion to analyse these doublet. In the first part, we analysed identical mass parameters '$ m $' where replacing $ m $ by $ -m $ implied helicity change from $ +2 $ to $ -2 $. 
Strikingly here also the generic field redefinitions worked properly. 
\\
Systematic steps combined the doublets
into an effective  model containing the quadratic Einstein-Hilbert term with the Pauli-Fierz mass term
applicable for spin-2 particle which could be easily recognised as an analogue of Proca model for spin-1 case
in vector theory. Similar calculations  were repeated by considering doublets with distinct 
mass parameters $ m_1 $ and $ m_2 $. The final equation of motion  yielded 
the composite model  having \-- an Einstein-Hilbert term, a Fierz-Pauli mass term and interestingly
a generalised first order Chern-Simons term. This CS-term contained, apart from the standard CS term , two other similar terms with a different orientation of indices. Thus the action for the effective model with distinct mass parameters could be interpreted as an analogue of Maxwell-Chern-Simons-Proca theory for the spin -1 particle. 

Let us visualise in general terms the obtention
of a new theory from a combination of chiral ones. Chiral theories occur in doublets corresponding to 
the left and right degrees of freedom. We may take the equation of motion approach as symbolic and try to explain this point. The equations of motion following from a doublet are form invariant,
differing only by a sign in the chiral piece. Adding and subtracting these equations naturally leads to
a combination which is either a sum or a difference of the original variables. Renaming this `sum' and `difference' as new fields yields a pair of coupled differential equations. 
It is then possible to eliminate one of these new fields in favour of the other using these differential equations. The final outcome is an equation of motion involving only the new fields. Furthermore, the symmetrical treatment implies that we obtain identical equations of motion for both the new fields. Consequently we are led to a unique new theory obtained by a composition of the chiral degrees of freedom. 
 
The other approaches are more sophisticated leading to fresh insights, nevertheless this basic
idea runs as a common string.
  
%%%%%%%%%%%%%%%%%%%%%%%%  the end %%%%%%%%%%%%%

\end{document}